\documentclass[twocolumn]{aastex631}

\usepackage{appendix}

\newcommand{\Msun}{M$_{\odot}$}

\newcommand{\LIRLsun}{$\rm log_{10} \rm L_{\rm IR}/\rm{L}_\odot$}
\newcommand{\LogM}{$\rm Log_{10} \rm M^*/\rm{M}_\odot$}
\newcommand{\LIR}{L$_{\rm IR}$}
\begin{document}

\title{The galaxies missed by Hubble and ALMA: the contribution of extremely red galaxies to the cosmic census at $3<z<8$ }

\author[0000-0003-2919-7495]{Christina C.\ Williams}
\affiliation{NSF’s National Optical-Infrared Astronomy Research Laboratory, 950 North Cherry Avenue, Tucson, AZ 85719, USA}
\affiliation{Steward Observatory, University of Arizona, 933 North Cherry Avenue, Tucson, AZ 85721, USA}

\author[0000-0002-8909-8782]{Stacey Alberts}
\affiliation{Steward Observatory, University of Arizona, 933 North Cherry Avenue, Tucson, AZ 85721, USA}

\author[0000-0001-7673-2257]{Zhiyuan Ji}
\affiliation{Steward Observatory, University of Arizona, 933 North Cherry Avenue, Tucson, AZ 85721, USA}

\author[0000-0001-9262-9997]{Kevin N. Hainline}
\affiliation{Steward Observatory, University of Arizona, 933 North Cherry Avenue, Tucson, AZ 85721, USA}

\author[0000-0002-6221-1829]{Jianwei Lyu }\affiliation{Steward Observatory, University of Arizona, 933 North Cherry Avenue, Tucson, AZ 85721, USA}

\author[0000-0003-2303-6519]{George Rieke} \affiliation{Steward Observatory and Dept of Planetary Sciences, University of Arizona 933 North Cherry Avenue Tucson AZ 85721 USA}

\author[0000-0003-4564-2771]{Ryan Endsley} \affiliation{Department of Astronomy, University of Texas, Austin, TX 78712 USA}

\author[0000-0002-1714-1905]{Katherine A. Suess}\affiliation{Department of Astronomy and Astrophysics, University of California, Santa Cruz, 1156 High Street, Santa Cruz, CA 95064 USA}\affiliation{Kavli Institute for Particle Astrophysics and Cosmology and Department of Physics, Stanford University, Stanford, CA 94305, USA}

\author[0000-0002-9280-7594]{Benjamin D. Johnson}
\affiliation{Center for Astrophysics $\vert$ Harvard \& Smithsonian, 60 Garden Street, Cambridge, MA 02138, USA}

\author[0000-0001-5097-6755]{Michael Florian}\affiliation{Steward Observatory, University of Arizona, 933 North Cherry Avenue, Tucson, AZ 85721, USA}

\author[0000-0003-4702-7561]{Irene Shivaei} \affiliation{Centro de Astrobiolog\'ia (CAB), CSIC-INTA, Ctra. de Ajalvir km 4, Torrej\'on de Ardoz, E-28850, Madrid, Spain}

\author[0000-0002-0303-499X]{Wiphu Rujopakarn} \affiliation{National Astronomical Research Institute of Thailand, Don Kaeo, Mae Rim, Chiang Mai 50180, Thailand}\affiliation{Department of Physics, Faculty of Science, Chulalongkorn University, 254 Phayathai Road, Pathumwan, Bangkok 10330, Thailand}

\author[0000-0003-0215-1104]{William M.\ Baker} \affiliation{Kavli Institute for Cosmology, University of Cambridge, Madingley Road, Cambridge CB3 0HA, UK} \affiliation{Cavendish Laboratory, University of Cambridge, 19 JJ Thomson Avenue, Cambridge CB3 0HE, UK}

\author[0000-0003-0883-2226]{Rachana Bhatawdekar}\affiliation{European Space Agency (ESA), European Space Astronomy Centre (ESAC), Camino Bajo del Castillo s/n, 28692 Villanueva de la Cañada, Madrid, Spain}

\author[0000-0003-4109-304X]{Kristan Boyett}
\affiliation{School of Physics, University of Melbourne, Parkville 3010, VIC, Australia}
\affiliation{ARC Centre of Excellence for All Sky Astrophysics in 3 Dimensions (ASTRO 3D), Australia}

\author[0000-0002-8651-9879]{Andrew J. Bunker}
\affiliation{Department of Physics, University of Oxford, Denys Wilkinson Building, Keble Road, Oxford OX13RH, U.K.}

\author[0000-0002-6719-380X]{Stefano Carniani}
\affiliation{Scuola Normale Superiore, Piazza dei Cavalieri 7, I-56126 Pisa, Italy}

\author[0000-0003-3458-2275]{Stephane Charlot}
\affiliation{Sorbonne Universit\'e, CNRS, UMR 7095, Institut d'Astrophysique de Paris, 98 bis bd Arago, 75014 Paris, France}

\author[0000-0002-9551-0534]{Emma Curtis-Lake}
\affiliation{Centre for Astrophysics Research, Department of Physics, Astronomy and Mathematics, University of Hertfordshire, Hatfield AL10 9AB, UK}

\author[0000-0002-4781-9078]{Christa DeCoursey} \affiliation{Steward Observatory, University of Arizona, 933 North Cherry Avenue, Tucson AZ 85721 USA}

\author[0000-0002-2380-9801]{Anna de Graaff}
\affiliation{Max-Planck-Institut f\"ur Astronomie, K\"onigstuhl 17, D-69117, Heidelberg, Germany}

\author[0000-0003-1344-9475]{Eiichi Egami}
\affiliation{Steward Observatory, University of Arizona, 933 North Cherry Avenue, Tucson, AZ 85721, USA}

\author[0000-0002-2929-3121]{Daniel J. Eisenstein}
\affiliation{Center for Astrophysics $\vert$ Harvard \& Smithsonian, 60 Garden Street, Cambridge, MA 02138, USA}

\author[0000-0003-1903-9813]{Justus L.\ Gibson}\affiliation{Department for Astrophysical and Planetary Science, University of Colorado, Boulder, CO 80309, USA}

\author[0000-0002-8543-761X]{Ryan Hausen} \affiliation{Department of Physics and Astronomy, The Johns Hopkins University, 3400 N. Charles St., Baltimore, MD 21218}

\author[0000-0003-4337-6211]{Jakob M. Helton}
\affiliation{Steward Observatory, University of Arizona, 933 North Cherry Avenue, Tucson, AZ 85721, USA}

\author[0000-0002-4985-3819]{Roberto Maiolino}
\affiliation{Kavli Institute for Cosmology, University of Cambridge, Madingley Road, Cambridge, CB3 0HA, UK; Cavendish Laboratory, University of Cambridge, 19 JJ Thomson Avenue, Cambridge, CB3 0HE, UK}

\author[0000-0003-0695-4414]{Michael V. Maseda}
\affiliation{Department of Astronomy, University of Wisconsin-Madison, 475 N. Charter St., Madison, WI 53706 USA}

\author[0000-0002-7524-374X]{Erica J. Nelson}\affiliation{Department for Astrophysical and Planetary Science, University of Colorado, Boulder, CO 80309, USA}

\author[0000-0003-4528-5639]{Pablo G. P\'erez-Gonz\'alez}\affiliation{Centro de Astrobiolog\'{\i}a (CAB), CSIC-INTA, Ctra. de Ajalvir km 4, Torrej\'on de Ardoz, E-28850, Madrid, Spain}

\author[0000-0002-7893-6170]{Marcia J. Rieke}\affiliation{Steward Observatory, University of Arizona, 933 North Cherry Avenue, Tucson, AZ 85721, USA}

\author[0000-0002-4271-0364]{Brant E. Robertson}
\affiliation{Department of Astronomy and Astrophysics, University of California, Santa Cruz, 1156 High Street, Santa Cruz, CA 95064, USA}

\author[0000-0002-4622-6617]{Fengwu Sun}\affiliation{Steward Observatory, University of Arizona, 933 North Cherry Avenue, Tucson, AZ 85721, USA}

\author[0000-0002-8224-4505]{Sandro Tacchella}
\affiliation{Kavli Institute for Cosmology, University of Cambridge, Madingley Road, Cambridge, CB3 0HA, UK}
\affiliation{Cavendish Laboratory, University of Cambridge, 19 JJ Thomson Avenue, Cambridge, CB3 0HE, UK}

\author[0000-0001-9262-9997]{Christopher N. A. Willmer}
\affiliation{Steward Observatory, University of Arizona, 933 North Cherry Avenue, Tucson, AZ 85721, USA}

\author[0000-0002-4201-7367]{Chris J. Willott}
\affil{NRC Herzberg, 5071 West Saanich Rd, Victoria, BC V9E 2E7, Canada}

\begin{abstract}

Using deep JWST imaging from JADES, JEMS and SMILES, we characterize optically-faint and extremely red galaxies at $z>3$ that were previously missing from galaxy census estimates.
The data indicate the existence of abundant, dusty and post-starburst-like galaxies down to $10^8$M$_\odot$, below the sensitivity limit of Spitzer and ALMA. 
Modeling the NIRCam and HST photometry of these red sources can result in extreme, high values for both stellar mass and star formation rate (SFR); however, including 7 MIRI filters out to 21$\mu$m results in decreased mass (median 0.6 dex for \LogM$>$10), and SFR (median 10$\times$  for SFR$>$100\Msun/yr). At $z>6$, our sample includes a high fraction of Little Red Dots (LRDs; NIRCam-selected dust-reddened AGN candidates). 
We significantly measure older stellar populations in the LRDs out to rest-frame 3$\mu$m (the stellar bump) and rule out a dominant contribution from hot dust emission, a signature of AGN contamination to stellar population measurements. 
This allows us to measure their contribution to the cosmic 
census at $z>3$,  below the typical detection limits of ALMA ($L_{\rm IR}<10^{12}L_\odot$). We find that these sources, which are overwhelmingly missed by HST and ALMA, 
could effectively double the obscured fraction of the star formation rate density at $4<z<6$ compared to some estimates, showing that prior to JWST, the obscured contribution from fainter sources could be underestimated. Finally, we identify five sources with evidence for Balmer breaks and high stellar masses at $5.5<z<7.7$. While spectroscopy is required to determine their nature, we discuss possible measurement systematics to explore with future data.

\end{abstract}
\keywords{High-redshift galaxies, Active galaxies, AGN host galaxies}

\section{Introduction} \label{sec:intro}

Our picture of galaxy formation and growth during the first 2 billion years of the Universe is, currently, an incomplete one. The Hubble Space Telescope (HST) is only sensitive to the rest-frame ultra-violet (UV) light for galaxies at $z > 3$, which at Cosmic Noon ($1<z<3$), is known to miss the overwhelming majority of star formation activity and energy output from galaxies due to dust obscuration \citep[e.g.][]{MadauDickinson2014}. While the addition of sub-millimeter and radio measurements have greatly improved our understanding of the star formation census at Cosmic Noon \citep[e.g.][]{HodgedaCunha2020}, the difficulty in making these measurements increases dramatically at $z>3$. This is in part due to a decline in the abundance 
of the bright, sub-millimeter galaxies that are detectable by wide-area sub-mm surveys \citep[e.g.][]{Simpson2014, Brisbin2017, Danielson2017, Casey2021}. It is also in part because  of the large uncertainty in the abundance of more common, but less-extreme galaxies, that are below the detection limits of typical multi-wavelength surveys \citep[and difficulties in determining their counterparts and redshifts, e.g.][]{Smail2021}. 
As the abundance of massive galaxies decline, less massive and fainter sources \citep[e.g.][]{Pope2017,Pope2023} could plausibly  harbor an increasing fraction of the obscured star-formation at early times. 
Progress has been made by identifying fainter, dusty galaxies with ever increasing ALMA area \citep[in particular using longer wavelength $>2-3$ mm selections to filter out lower-redshift sources][]{Bethermin2015, Casey2021, Cooper2022}. Despite these advances, recent ALMA-based estimates at $z>4$ still rely on relatively few detections, strong lensing, or extrapolations of the infrared luminosity function \citep[e.g.][]{Zavala2021, Algera2023, Barrufet2023a, Traina2023, Fujimoto2023}.
Uniform and large-area surveys are still needed for unbiased samples, and remain challenging with ALMA's small field of view. 
Thus, the pre-JWST census of galaxies in the first 2 Gyr of cosmic history remained highly biased to relatively unobscured, star-forming sources. 
A complete understanding of the evolution of the early Universe requires accurate accounting of the abundance, the growth rate, and the energy output from a complete and unbiased census of the galaxy population.

Prior to the launch of JWST \citep{Gardner2023}, observations at the limit of Spitzer Space Telescope and ALMA capabilities indicated the existence of surprisingly abundant, massive galaxies that are very red, enough to evade HST selection. Some of these massive galaxies are serendipitiously identified at submillimeter wavelengths to faint limits, thanks to ALMA's extremely sensitive receivers (otherwise lacking counterparts at shorter wavelengths; i.e. optical/near-infrared-faint or ``dark"). Sub-millimeter or radio sources lacking counterparts have been known for decades and hypothesized to be high redshift \citep[e.g.][]{HuRidgway1994, Hughes1998,Dey1999,  Dunlop2004, Frayer2004}. However, in recent years, ALMA's sensitivity, small field of view, and ever deeper surveys at optical/near-infrared suggest that massive, dust-obscured and red galaxies were even more common at high-redshift than previously thought \citep{Williams2019, Umehata2020, Fudamoto2021, Sun2021}. The number densities and redshift distributions of these red sources are still highly uncertain, but imply that previous HST surveys might have missed up to 90\% of massive galaxies ($\rm Log_{10}\rm M/M_{\odot}>$ 10.5) at $z > 3$ \citep{Wang2016, Wang2019}. 
 While the idea that the distant Universe may contain abundant massive galaxies waiting to be discovered is tantalizing, ALMA left many open questions regarding the properties of the stellar populations, morphologies, and potential for these missing galaxies to harbor black holes. These questions remain unanswered without deep near- to mid-infrared data.

The first year of JWST observations has now confirmed the existence of very red galaxies missed by HST. First-look papers indicated that the properties of these sources are consistent with being dusty or quiescent galaxies at $z \sim 3-7$, with some fraction of bluer sources turned red by strong emission lines boosting long-wavelength filters \citep{Barrufet2023, Rodighiero2023, PerezGonzalez2023, Endsley2023, Smail2023, Fujimoto2023dualz, BargerCowie2023}. The new discovery of these red galaxies demonstrated conclusive evidence that HST’s limited wavelength coverage had left us with a critical blind spot. 

Interestingly,  NIRCam imaging is revealing that distant red and optically-faint galaxies are remarkably diverse as a population. In particular, a number of galaxies with peculiar or surprisingly shaped red spectral energy distributions (SEDs) have been discovered. Some SEDs resemble strong Balmer breaks with strong red rest-optical continuum leading to high inferred stellar masses despite their early observation times \citep[e.g.][]{Labbe2023a}. Another class exhibit similarities in SED and morphology to highly reddened QSOs \citep[visibly appearing as ``little red dots"; LRDs;][]{Labbe2023b,Furtak2023a}. A number of such LRDs have now been spectroscopically confirmed to host some form of AGN activity \citep{Kocevski2023, Harikane2023, Greene2023, Matthee2023, Maiolino2023, Furtak2023b, Kokorev2023}. Both subclasses raise important questions about how these galaxies and black holes could grow at such fast rates at early cosmic time, and motivate the identification of larger samples with expanded deep panchromatic data.

Among the broader red and massive $z>3$ galaxy population, some first results have indicated that in fact their abundance has been previously significantly underestimated, with a factor of 20-25\% higher stellar mass density at $3<z<6$ and $>100$\% at $6<z<8$ \citep{Gottumukkala2023}.  This points to an early start to rapid stellar mass growth that is obscured by large dust columns, assuming our extrapolations from $z\sim0$ stellar population modeling are correct \citep[e.g.][]{Steinhardt2023temp, Wang2023b, Woodrum2023}. This could alter the picture of early galaxy growth, requiring increased efficiency of star formation than previously thought \citep{Labbe2023a, BoylanKolchin2023, Xiao2023b}. However, further exploration remains to be done, since stellar mass and photometric redshift estimates at high mass and high redshift can also be sensitive to the data set used,  \citep[e.g. inclusion of near-IR medium-band filters;][]{Desprez2023}, thus motivating new exploration also using expanded wavelength coverage.

With this paper, we provide a detailed look at the SEDs and physical properties of galaxies that were missed by previous surveys using HST, and in many cases even Spitzer and ALMA, primarily due to their extremely red colors and relatively faint fluxes. The goal of this work is to determine what types of galaxies we have been missing, and, what fraction of the cosmic census has been unaccounted for as a result. This work will focus on a uniformly selected sample of extremely red sources (JWST/NIRCam 1.5$\mu$m--4.4$\mu$m colors $>$ 2.2) that were previously unaccounted for by HST (fainter than 27 ABmag at 1.5$\mu$m), a criteria that identifies massive and obscured sources primarily at $z>3$ by bridging the Balmer break at high-redshifts \citep[e.g.][]{Caputi2012, Wang2016, Barrufet2023}.

In Sections \ref{sec:data} and \ref{sec:colorsel} we overview the panchromatic data we use in the characterization of these sources. In Sections \ref{sec:meas} and \ref{sec:results}, we measure the physical properties using SED fitting, revealing both sources with high attenuation, and those with older stellar populations, and dust-reddened AGN candidates. We also highlight how their extreme colors make it difficult to measure their properties using HST+NIRCam alone, and how the deepest MIRI plus ALMA (only available in this field) can alter the interpretation of these sources. Finally, in Section \ref{sec:discussion} we characterize their contribution to the galaxy census, identifying what fraction of star formation and stellar mass density was previously missing. 
We assume a $\Lambda$CDM cosmology with  H$_0$=70 km s$^{-1}$ Mpc$^{-1}$, $\Omega_M$ = 0.3, $\Omega_\Lambda$ = 0.7, and a  \citealt{Kroupa2001} initial mass function (IMF)

\section{Datasets} \label{sec:data}

In this work, we focus on a select sample of optically faint galaxies identified using JWST/NIRCam imaging inside the Great Observatories Origins Deep Survey South \citep[GOODS-S;][]{Giavalisco2004}, overlapping with the Hubble Ultra Deep Field \citep{Beckwith2006}. Given the exceptional ancillary data that is available in this prime region of the sky, we focus on providing a detailed look at this population to give broader insight for larger samples with more limited data. In this section we outline the various datasets we use and our photometric methods to measure fluxes in each.

\subsection{Space-based data and photometric catalog construction}

\begin{figure*}[t]
\includegraphics[width=1\textwidth]{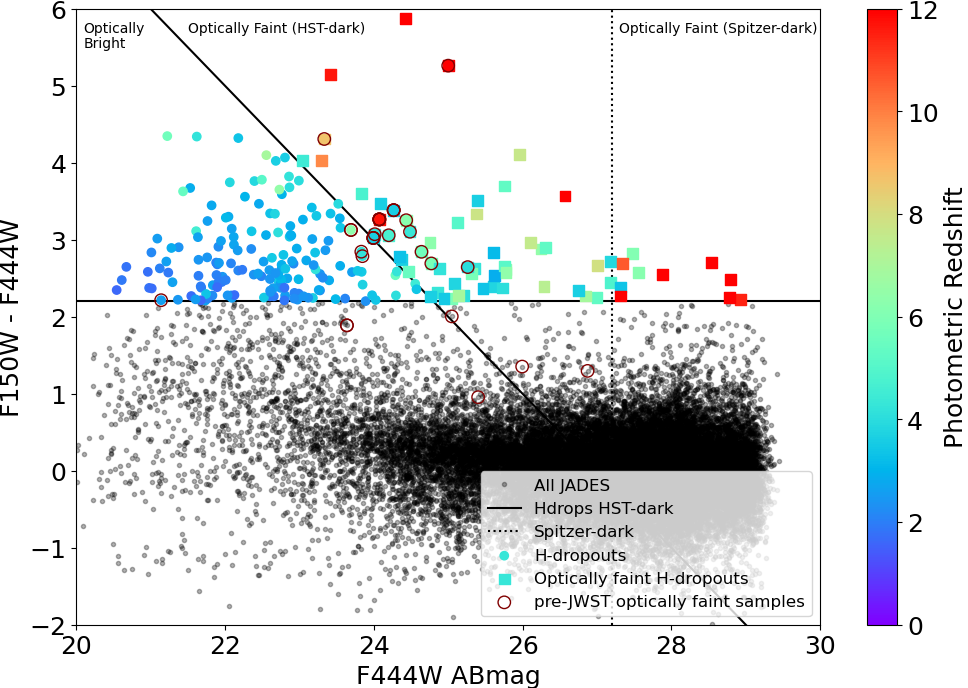}

\caption{F150W--F444W color vs F444W magnitude illustrating our selection from the JADES catalog with S/N $>$ 20 in F444W (black points). We identify very red sources by their F150W--F444W$ > 2.2$ color (all colored points) color coded by their EAZY photometric redshift. We further identify the optically fainter sample whose colors and magnitudes are consistent with $z>3$ massive galaxies that were absent from earlier HST selected samples (i.e. have F150W $<$ 27 ABmag: the black diagonal line). In this work, we focus on the optically faint sources behind the black line marked as colored squares. Existing samples of optically faint objects (maroon open circles) are collected from the literature using pre-JWST datasets \citep{Wang2016,Wang2019,Franco2018,AP2019,Xiao2023}. We note the detection limit of the Spitzer GREATS program (dotted line) indicating newly discovered sources by JWST.}\label{fig:seljades}
\end{figure*}

\subsubsection{ HST Imaging}

In our analysis we include deep optical and near-IR HST imaging from the Advanced Camera for Surveys (ACS) and Wide Field Camera 3 (WFC3) as compiled by use of the Hubble Legacy Field  \citep[HLF;][and references therein]{Illingworth2016, Whitaker2019} imaging in GOODS-S from HST. The HLF represents the deepest composite imaging including nine filters between 0.4-1.6$\mu$m wavelength. We include data from ACS (F435W, F606W, F775W, F814W, F850LP) and WFC3/IR  (F105W, F125W, F140W, F160W). 

\subsubsection{JWST NIRCam Imaging}
 
We identify our sample using data from the JWST Advanced Deep Extragalactic Survey \citep[JADES;][]{Eisenstein2023}. 
JADES imaging with NIRCam \citep{Rieke2023a} covers a deep 27 square arcminute area that 
includes nine filters from 0.9-5 micron (F090W, F115W, F150W, F200W, F277W, F335M, F356W, F410M, and F444W) and a medium-depth region covering an additional 40 square arcminute area with eight filters (F090W, F115W, F150W, F200W, F277W, F356W, F410M, and F444W). We include another 10 square arcminutes of NIRCam imaging from the JADES program ID (PID) 1286 (PI: Lutzgendorf) that additionally includes F070W. 
The deep area also includes the field targeted by the JWST Extragalactic Medium-band Survey \citep[JEMS;][PID: 1963; PIs Williams, Tacchella, Maseda]{Williams2023}.  JEMS obtained imaging in two $2.2^{\prime} \times 2.2^{\prime}$ regions with five medium-band filters (F182M, F210M, F430M, F460M, and F480M). We also combine our F182M, F210M and F444W imaging with that obtained as part of the public First Reionization Epoch Spectroscopic COmplete Survey \citep[FRESCO; PI: Oesch; PID: 1895;][]{Oesch2023}. 
Our data reduction procedure follows the methods outlined in \citet{Rieke2023b} for the JADES data release 1.

\subsection{Photometric catalogs}

Also following the methods outlined in the JADES first data release \citep{Rieke2023b} we measure photometry jointly for the HST and NIRCam imaging based on a detection image constructed as an inverse variance-weighted stack of the NIRCam longwave filters we have across the entire GOODS-S footprint: F277W, F335M, F356W, F410M, and F444W. Photometry is measured via a series of steps that perform source detection with advanced deblending algorithms \citep[see][ Robertson, in prep]{Rieke2023b} that utilize {\tt photutils}, {\tt astropy}, {\tt scipy}, {\tt cupy} and {\tt sextractor} packages.  
We use the {\tt photutils} package to identify source detections, and measure  
forced circular aperture photometry (convolved to the resolution of the F444W filter) for all detected objects. For compact or unresolved sources we use a 0.5" diameter aperture, which we validate is the appropriate size by visual inspection. However, by eye we identified a fraction of sources are more extended than this aperture, and in those cases we instead use an 0.7" diameter aperture.  To estimate photometric uncertainties, we combine in quadrature  the Poisson noise and the noise estimated from the root mean squared (RMS) of 100,000 random apertures  \citep[following methodology outlined in][]{Labbe2005, Quadri2007, Whitaker2011}. 
We generate aperture corrections to account for flux lost from the fixed apertures using encircled energy curves constructed using WebbPSF \citep{Perrin2014} as described in \citet{Ji2023}.

\subsection{Forced photometry } 

Once we identify our sample (as will be discussed in Section \ref{sec:colorsel}), we also include forced photometry on the following data measured based on the NIRCam positions of our galaxies.

\subsubsection{JWST MIRI Imaging}

We include wide-field JWST/MIRI \citep{Wright2023} imaging at 5--25.5$\mu$m from the Systematic Mid-infrared Instrument Legacy Extragalactic Survey program (SMILES; PID 1207; PI Rieke), which covers 34 square arcminutes of the GOODS-S/HUDF region with a nearly complete overlap with JADES and JEMS imaging.  For each of the 15 pointings, $\sim2.2$ hr of science time was spread between 8 MIRI filters\footnote{F2550W is relatively shallow compared to the other bands and so it is omitted from source photometry for lack of constraining power.}, reaching $5\sigma$ point source sensitivities of 0.20, 0.19, 0.38, 0.59, 0.68, 1.7, 2.8 and 16 $\mu$Jy for F560W, F770W, F1000W, F1280W, F1500W, F1800W, F2100W, and F2550W respectively, in apertures containing $65\%$ of the encircled energy for each PSF (see below).  Survey design, data reduction, and source extraction are described in \citet{Lyu2023}\footnote{Point source sensitivities differ slightly from Lyu et al. submitted due to the updated flux calibrations used here.}.  Briefly, data reduction was performed using the JWST Calibration Pipeline v1.10.0 \citep{Bushouse2023} with custom external background subtraction \citep[see][]{AlvarezMarquez2023} and astrometry correction matched to the JADES astrometric solution.  Flux calibration is updated based on the new reference files released in CRDS jwst\_1130.pmap, which include the time-dependent count rate loss (Gordon, in prep.).

The blind MIRI photometric catalog is made using the same pipeline as the NIRCam catalog \citep{Rieke2023b} with the differences that F560W and F770W are stacked for the detection image and the source detection and deblending are optimized for MIRI, given its noise properties and source density. 
Aperture corrections are applied based on model PSFs from WebbPSF \citep{Perrin2014} for F1000W--F2550W and with empirical PSFs for F560W and F770W to account for the ``cruciform" detector artifact \citep[][see \citealt{Lyu2023} for more details]{Gaspar2021}.  
The JADES NIRCam sources that also have a blind SMILES MIRI counterpart are then matched within 0.15$\arcsec$.  Given MIRI's remarkable sensitivity and well-behaved noise, we also include forced photometric measurements using seven MIRI filters out to 21$\mu$m at the NIRCam positions for sources that are not formally detected in the blind MIRI catalog, in order to enable better constraints on the observed mid-IR SEDs (restframe near-infrared). To perform forced photometry, we use {\tt photutils} to measure flux in circular apertures of size 0.7'' arcsecond diameter (comparable to the FWHM at F2100W, which balances the higher resolution MIRI shortwave bands against minimizing the aperture corrections in the longwave bands).  Our forced photometry is verified by repeating this process on detected sources and comparing to the catalog fluxes, which are found to be in good agreement at the few percent level.  
To assess the uncertainties, we measured the median flux in boxes at random sourceless locations across the map to get the RMS in MJy/sr and then to estimate the RMS in the given aperture, we scale by the square root of the aperture area. 

\subsubsection{ALMA imaging}

Extensive ALMA $\sim$1 mm imaging exists within the JADES region of the GOODS-S field, including some of the deepest contiguous ALMA maps on the sky to date. We take advantage of this ancillary data in order to constrain the far-infrared emission. This critical constraint allows a more robust estimate of the dust-obscured star formation, given the relatively featureless power-law SEDs in the rest optical and faintness in the rest-UV. In cases where galaxies are significantly detected at 1-mm in either the deep ASAGAO imaging \citep{Hatsukade2018} or the wider-area  GOODS-ALMA program \citep{Franco2018} we use the published total fluxes in our SED modeling.

For cases where galaxies are not detected in any of the available ALMA imaging, the deep data still  provides strong constraints on the limiting far-infrared luminosity associated with galaxies. To incorporate these deep ALMA limits as constraints in our SED-modeling, we perform forced photometry at the locations of our NIRCam detections using the deepest available map at the sources' position. The data we use for our analysis include the deep combined map built by the ASAGAO team \citep{Hatsukade2018} combined with data from the deep HUDF ALMA footprint \citep{Dunlop2017} and overlapping coverage from the shallower, but much wider GOODS-ALMA program \citep{Franco2018, GomezGuijarro2022}. We prioritize the combined map from \citet{Hatsukade2018} if objects appear in multiple datasets, and provide GOODS-ALMA limits 
(map $\sigma_{rms}\sim180\mu$Jy/beam) 
for the objects outside the ASAGAO coverage. 

The resolution of the combined ALMA data is relatively high (beam size $\sim0.5"\times0.5"$ FWHM) similar in size to our NIRCam photometric aperture \citep{Hatsukade2018}. Without knowledge of the actual size of our sources at far-infrared wavelengths, to ensure we include any extended flux, we perform forced aperture photometry at the location of the NIRCam centroid, using a 1.0" diameter. 
To perform forced photometry on ALMA, we broadly follow the methodology of \citet{Sun2021, Shivaei2022, Betti2019}. In short, we multiply the map units (flux/beam) by the number of pixels per beam, and use \texttt{photutils} \citep[][]{photutils}  to measure the flux within a circular aperture of 1.0" diameter from the primary beam corrected map. We assess the uncertainty as the standard deviation of the fluxes within 100 randomly placed apertures within 30" distance from the NIRCam source. Where the source is within 30" of the edge of the mosaic, we decrease this distance to 15", and do not assess forced photometry for sources closer than 15" to the edge of the map.

\subsubsection{JWST spectroscopy}\label{sec:specz}

For this work, we will cross-match the samples we select in Section \ref{sec:colorsel} with available spectroscopy from JWST. We make use of the public spectroscopic confirmations released by the JADES-NIRSpec program \citep{Bunker2023, Eisenstein2023b}. We also use data from the First Reionization Epoch Spectroscopic COmplete Survey \citep[FRESCO; PI: Oesch; PID: 1895;][]{Oesch2023} that obtained 4$\mu$m NIRCam wide-field slitless spectroscopy. We will discuss these further in Section \ref{sec:eazy}.

\section{Color selection of targets}\label{sec:colorsel}
In this paper we target extremely red sources at $z>3$ that would have been missed by typical pre-JWST surveys given wavelength coverage and detection limits. To build an inclusive sample of red sources, we adopt a selection based on the HST/WFC3 F160W - Spitzer 4.5$\mu$m colors  (H-[4.5] $>$ 2.3;  e.g. \citealt{Caputi2012, Wang2016, Wang2019}). Due to the redshifting of the Balmer break beyond 1.5$\mu$m at $z>3$, this selection targets massive and red (including both dust obscured and quiescent) high-redshift galaxy candidates.  At typical detection limits of HST, such galaxies were referred to as ``H-band dropouts" (i.e. because they are red, and often not detected in HST; see \citealt{Wang2019}).  For this work, we adopt a comparable selection based on JWST/NIRCam colors F150W--F444W $>$ 2.2, slightly less restrictive in order to include existing objects that were identified pre-JWST based on data with larger photometric scatter. Since the first year of GOODS-S imaging from JADES yielded smaller area coverage from F150W than in F444W, we additionally check for additional H-dropouts using the F160W imaging for which we have the other NIRCam filters. We identify two additional sources which happen to fall off of our F150W footprint, but that otherwise have nearly complete NIRCam and MIRI imaging.

We present our red sources in F150W--F444W vs F444W magnitude space in Figure \ref{fig:seljades}, color coded by their initial photometric redshifts measured with EAZY (see Section \ref{sec:eazy}), along with  all significant JADES NIRCam detections (black). This figure highlights the relative rarity of galaxies with this extreme red color compared to the full JADES sample, especially at the fainter F444W magnitudes (which can be seen to roughly correspond to higher-redshift sources based on the EAZY color coding). After visual inspection to remove imaging artifacts and hot pixels, our selection identifies a parent sample of 240 red galaxies accross redshifts, that are both F150W--F444W $>$ 2.2 and F444W S/N $>$ 20 (all colored points in Figure \ref{fig:seljades}). 

Following \citet{Barrufet2023}, we 
retain the F150W-fainter sample for the primary analysis of this paper, which roughly corresponds to a redshift selection (see their Figure 1) to identify the primary targets of study: those with F150W magnitude fainter than 27 (the dashed diagonal line), which are sources that are below the typical detection limit of HST/F160W in GOODS-S. We will refer to this primary sample following convention in the literature as ``HST-dark", of which we identify 66 candidate objects (colored squares in Figure \ref{fig:seljades}). Additionally, we note that $\sim$20\% of those sources are also below the detection limit of the deepest Spitzer 3.6 and 4.5 $\mu$m imaging to date from the GREATS program \citep{Labbe2015} which reached a 3$\sigma$ limiting depth of 27.2 (AB magnitude). H-dropouts rightward of this limit (dotted vertical line) represent galaxies that are newly discovered by JWST.

Our HST-dark sources are listed in Tables \ref{tab:MIRI} and \ref{tab:NIRCAM}. NIRCam cutouts of our sample, along with the aperture used to measure their photometry, are shown in Figures \ref{fig:miricutouts}. The corresponding HST/F160W imaging is also shown.

The sample we target are very red in the restframe optical, exhibiting significant overlap in color space with very late-type stars. Thus, we check whether any of our  objects are both unresolved, and have F115W, F150W, F277W and F444W colors that look like red stellar sources (e.g. brown dwarfs with effective temperatures $<1500$K) and are poorly fit with stellar population synthesis models, yet are well fit by Sonora (2018) substellar atmosphere models \citep[][]{Marley2021, Hainline2023b}. Only one source (ID 190413) is a highly probable brown dwarf candidate based on its colors. This source also has a 0.62\"\ positional offset measured between GREATS 3.6$\mu$m Spitzer imaging and our JWST data \citep[2$\sigma$ confidence, given its low S/N in the Spitzer imaging of 5$\sigma$, suggesting the positional uncertainty is $\sim$1.2" / 5 = 0.2-0.3"; see][]{Hainline2023b}.  This source is additionally poorly fit by stellar population models with a high redshift and improbably high mass solution. Thus, based on the evidence for proper motion, we catalog its properties in Table \ref{tab:MIRI} but remove it from our analysis and census measurements in Sections \ref{sec:results} and \ref{sec:discussion}.
The color similarity indicates the uncertainty involved in distinguishing red galaxies at these redshifts from late-type stars \citep[see e.g.][]{Hainline2023b, Burgasser2023}. 

Our sample has some overlap with a number of red optically faint galaxy populations studied to date with JWST. For context, we note what fraction of objects in JADES identified by their criteria also meet our criteria of F150W--F444W$>$2.2 and F150W$<$27 ABmag.  We find that only 12\% of massive double-break candidates as in \citet{Labbe2023a} overlap with our HST-dark sample, likely because of their requirement of a relatively blue SED between F150W--F277W$<$0.7. We will discuss this overlap further in Section \ref{sec:massdens}. Our sources are also redder than 50\% of sources selected based on F150W--F356W$>$1.5 in \citet{PerezGonzalez2023}, owing primarily to the bluer color chosen to include sources at $z>2$. Thus in comparison to these other samples, our objects tend to be those with consistently red SEDs from the rest-frame UV to near-IR, and, our sources are typically redder across the NIRCam wavelength baseline.

\begin{figure*}[t]
\includegraphics[width=0.33\textwidth]{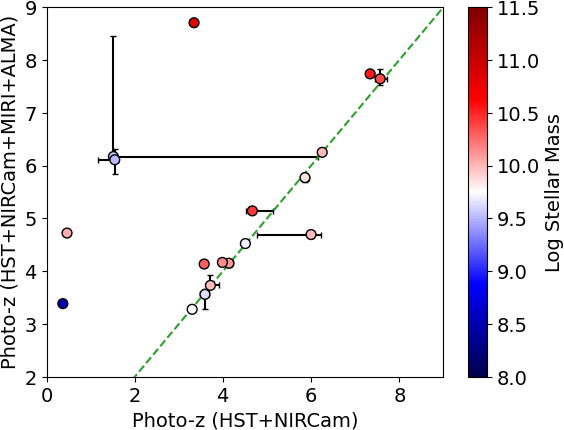}
\includegraphics[width=0.33\textwidth]{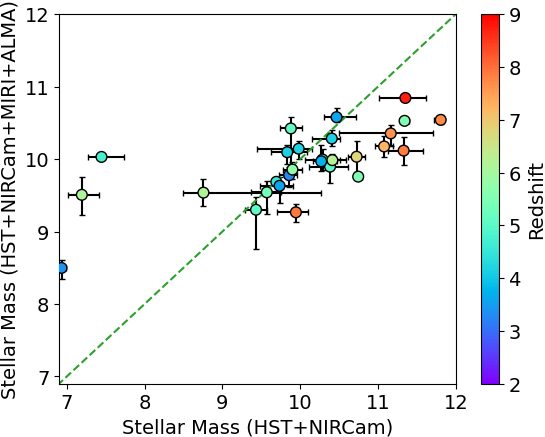}
\includegraphics[width=0.33\textwidth]{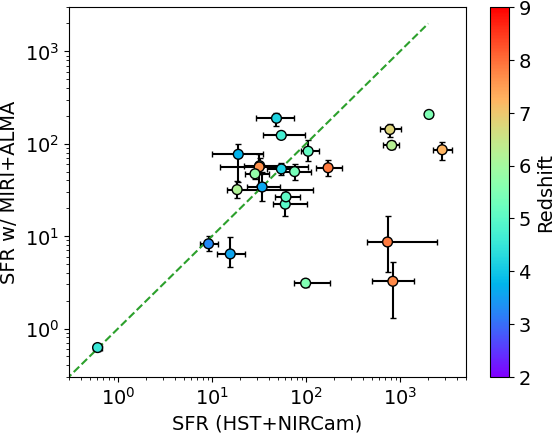}
\caption{For sources with MIRI and ALMA coverage, we compare inferred photometric redshift, stellar mass and SFR from our {\tt prospector} modeling of HST+NIRCam+MIRI+ALMA vs just the HST+NIRCam photometry. Left panel shows the inferred photometric redshift (sources with spectroscopic redshifts are excluded from this panel). Middle panel shows the stellar mass, and right panel shows the SFR. At the very high mass and SFR end, we find that including the MIRI+ALMA data lowers the extreme masses and SFRs that are otherwise inferred (median decrease of 0.6 dex for sources above \LogM$>$10, and median 10$\times$ less for SFR $>$ 100 \Msun/yr). This figure shows that stellar masses and SFRs can be wrong for extreme and red sources using NIRCam data without longer wavelength constraints.}\label{fig:paramcompare}
\end{figure*}

\subsection{Comparison with previously identified optically faint galaxies in GOODS-S}

Our sample includes the majority of known optically faint or HST-dark galaxies that were identified in previous searches in this field \citep[see open circles in Figure \ref{fig:seljades}][]{Wang2016, Wang2019, AP2019, Franco2018, Xiao2023}.
Five out of seven of the H-dropouts identified by \citet{Wang2016,Wang2019} that fully overlap with our imaging  are identified by our parent sample H-dropout color selection, \citep[all of which also overlap with][]{AP2019}. Five optically dark sources identified by pre-JWST data were not red enough to meet our selection (which were identified using different methods \citealt{Franco2018, Wang2019, Xiao2023}) and a further four are brighter than our F150W magnitude limit.

We note that our sample includes the optically dark object AGS11 identified in 
\citet[][our ID 279678]{Franco2018} which had been identified based on a blind ALMA detection with no optical counterpart. JWST now demonstrates that it has one of the most extreme red colors among our sample (F150W--F444W$>$4), remains still undetected in all shortwave filters (despite our ultra deep imaging), and is also one of the largest, reaching nearly one arcsecond in extent.  This source was previously hypothesized to reside in a confirmed overdensity using ALMA spectroscopy to be at $z\sim3.4$ \citep{Zhou2020}, although the source itself was not detected in the ALMA spectroscopy.  Based on our photometry, we find a higher redshift of $z\sim4.7$ for this source (which likely explains their finding of no spectroscopic confirmation). Based on this comparison, our selection identifies a relatively unexplored sample in this field, below the detection limits of other multi-wavelength selections (including ALMA).

\section{Measuring physical properties} \label{sec:meas}

\subsection{Redshift estimations} \label{sec:eazy}

Our sample exhibits very red and sometimes featureless SEDs. To assist in our more detailed SED modeling to infer the physical properties, we first measure preliminary photometric redshifts based on the HST and JWST/NIRCam photometry using EAZY  \citep[][]{Brammer2008} as presented in \citet{Hainline2023a}. We measure the redshift as the probability weighted average peak of the photometric redshift distribution 
without any priors.
This preliminary photometric redshift measurement from EAZY is used to set a prior on redshift for our more detailed photometric modeling that we discuss in Section \ref{sec:prospector}.

A small number of our sample also have spectroscopy. These include 9 sources with one emission line with S/N $>$ 3 in FRESCO data \citep{Oesch2023}, using spectral extractions presented in Sun et al. (in preparation) combining JADES and FRESCO. Two sources that received slits from the JADES NIRSpec program also have one emission line with S/N$>$3 \citep{Bunker2023, Eisenstein2023b}. 
Below we describe our procedure for visually inspecting the spectroscopic redshifts in conjunction with the photometric PDF(z) measured using EAZY, and our process for deciding how to set the redshift priors that we use in our SED modeling in the case of uncertain spectroscopic redshift solutions.

For the case of the FRESCO sources, all nine sources have only one line, resulting in ambiguity in the redshift solution. For four out of the nine, the one line is weakly detected with $3< \rm{S/N} <5$. The five brighter, significantly detected FRESCO sources also only show a single emission line. Thus for all, the solution is heavily dependent on the most probable photometric redshift measured with EAZY. We find that this often leads to a degeneracy between H$\alpha$ at $z\sim5$--6 and [OIII]$\lambda5007$ at $z\sim7$--8 (assumed to be the detected line, given its 3$\times$ brighter than the other [OIII] doublet). This is because our sources are all very red, and the Lyman break is often faint and poorly constrained, thus photometric boosting by strong rest-optical line emission tends to have a strong influence on the photometric redshift solution. Thus, EAZY sometimes yields comparable probability for both H$\alpha$ or H$\beta$+[OIII] solutions. In these cases, we visually inspect the EAZY $\chi^{2}$ surface in conjunction with the SED shape, the detected line's wavelength, and we also consider whether the physical parameters derived from the SED modeling are reasonable (see next Section).

After this iterative process, we find that for four sources with confidently detected lines (219000, 154428, 184838, 204851) the photometric evidence clearly agrees with the FRESCO redshift solutions. 
For a fifth confidently detected single line source, 217926, we find that in fact the photometry (primarily near the Lyman break) supports an altered redshift solution at $z=5.04$ rather than $z=7.6$, assuming the securely detected line is H$\alpha$, not [OIII]$\lambda$5007. 
For three of the less-securely identified objects, 
(IDs 90354, 120484, 104849) the marginal emission lines are more difficult to interpret so close to the limiting signal to noise, although the candidate lines do have solutions that are consistent with the EAZY photometric redshifts. We decide to explore the SED-modeling solutions that are retrieved for both the case where redshift is a free parameter, and also while fixing to the tentative redshift. We get consistent results within the uncertainties either using the tentative spectroscopic redshifts or leaving the redshift free. We therefore consider these spectroscopic redshifts as robust, but in any case this choice does not impact our results.

For the last marginal case (ID 126594) we find a low-confidence marginal emission line which, assuming it is the stronger of the [OIII] doublet, puts this source at $z=7.9$.  In this case, the redshift lines up with one of 3 plausible (narrow) photometric redshift solutions based on the presence of emission line boosting to the photometry, although this redshift solution is not the one most favored by the photometry (which prefers a higher photometric redshift of $z=9.9$). However, at the redshift preferred by EAZY (which is also the redshift preferred by our SED-modeling when redshift is left as a free parameter), the galaxy is quite bright and in excess of expected stellar masses given our small survey area (based on our SED-modeling procedure outlined in the next Section). Thus, we take the approach of comparing stellar masses measured for both cases. We find that at the $z=7.9$ redshift solution, which is in agreement with the tentative line detection in FRESCO data, the SED modeling yields a more realistic stellar mass given our small area (see Section \ref{sec:massdens}). Thus, we opt for this more conservative redshift constraint and fix to the marginal spectroscopic redshift.

In addition, two of our sources were observed as part of the JADES NIRSpec campaign \citep[][]{Bunker2023, Eisenstein2023b}. ID 198459 was spectroscopically confirmed at $z=3.588$, which is consistent with the photometric redshift we measured using EAZY ($z=3.65$). For the second source ID 132229 however, the solution is less obvious. While the redshift is tentative (based on detection of the [OIII]$\lambda$5007 line at $z=7.247$), and consistent with the EAZY redshift ($z\sim7.5$), we found that leaving the redshift as a free parameter yielded an inconsistently high photometric redshift of $z\sim8.1$. If modeled at such a high redshift, we find that the inferred stellar mass is \LogM $\sim$10.7, unphysically high for a $z\sim8$ source in a small area (which we will discuss further in Section \ref{sec:massdens}), lending some credibility to the lower redshift solution. Therefore, for this source we chose to accept the tentative spectroscopic redshift at $z=7.247$ in the modeling and results.

\begin{deluxetable*}{llllllllllll}[!htbp]
\caption{Properties of H-dropouts that are identified inside the MIRI SMILES footprint.   }\label{tab:MIRI}
\tablehead{\colhead{ID$^{a}$}   & \colhead{RA} &  \colhead{Dec }& \colhead{F444W} & \colhead{Photo-z$^{b}$} & \colhead{Spec-z}  & \colhead{Stellar Mass$^{c}$} & \colhead{SFR}$^{d}$ & \colhead{Av$^{e}$}& \colhead{Age$^{f}$} &\colhead{f$_{\rm AGN}$$^{g}$} & \colhead{LRD$^{h}$}  }
\startdata
57356 & 53.115317 & -27.859217 & 25.4 & 5.14 & \nodata & 10.4  $^{+ 0.2 }_{- 0.6 }$ & 82.90  $^{+ 26.6 }_{- 17.8 }$ & 2.57  $^{+ 0.35 }_{- 0.58 }$ & 0.64  $^{+ 0.23 }_{- 0.32 }$ & 0.1  $^{+ 0.3 }_{- 0.0 }$ & 0 \\ 
81400 & 53.125402 & -27.839967 & 26.6 & 6.17 & \nodata & 9.5  $^{+ 0.2 }_{- 0.2 }$ & 31.84  $^{+ 8.1 }_{- 6.2 }$ & 1.76  $^{+ 0.23 }_{- 0.30 }$ & 0.12  $^{+ 0.03 }_{- 0.04 }$ & 0.0  $^{+ 0.1 }_{- 0.0 }$ & 0 \\ 
88481 & 53.105475 & -27.830682 & 24.6 & 4.69 & \nodata & 10.0  $^{+ 0.1 }_{- 0.1 }$ & 123.05  $^{+ 8.8 }_{- 8.6 }$ & 1.92  $^{+ 0.22 }_{- 0.20 }$ & 0.15  $^{+ 0.12 }_{- 0.08 }$ & 0.0  $^{+ 0.0 }_{- 0.0 }$ & 0 \\ 
90354 & 53.133825 & -27.828256 & 27.0 & 7.91 & 7.87$^{k}$ & 10.1  $^{+ 0.2 }_{- 0.2 }$ & 8.64  $^{+ 8.0 }_{- 4.5 }$ & 1.74  $^{+ 0.45 }_{- 0.37 }$ & 0.39  $^{+ 0.10 }_{- 0.09 }$ & 0.0  $^{+ 0.1 }_{- 0.0 }$ & 1 \\ 
104849 & 53.101881 & -27.810949 & 26.3 & 5.29 & 5.12$^{k}$ & 9.1  $^{+ 0.3 }_{- 0.2 }$ & 23.95  $^{+ 5.4 }_{- 4.4 }$ & 1.90  $^{+ 0.36 }_{- 0.34 }$ & 0.15  $^{+ 0.13 }_{- 0.09 }$ & 0.0  $^{+ 0.0 }_{- 0.0 }$ & 0 \\ 
106502 & 53.145292 & -27.808668 & 25.6 & 3.28 & \nodata & 9.8  $^{+ 0.1 }_{- 0.2 }$ & 8.26  $^{+ 1.7 }_{- 1.3 }$ & 1.82  $^{+ 0.25 }_{- 0.31 }$ & 1.41  $^{+ 0.18 }_{- 0.16 }$ & 0.0  $^{+ 0.1 }_{- 0.0 }$ & 0 \\ 
108239 & 53.160381 & -27.806011 & 25.7 & 4.52 & \nodata & 9.7  $^{+ 0.0 }_{- 0.0 }$ & 0.62  $^{+ 0.1 }_{- 0.1 }$ & 0.02  $^{+ 0.02 }_{- 0.02 }$ & 1.28  $^{+ 0.03 }_{- 0.04 }$ & 0.0  $^{+ 0.3 }_{- 0.0 }$ & 0 \\ 
120484 & 53.125426 & -27.787438 & 25.8 & 5.34 & 5.00$^{k}$ & 9.9  $^{+ 0.1 }_{- 0.2 }$ & 22.18  $^{+ 7.5 }_{- 5.5 }$ & 2.00  $^{+ 0.17 }_{- 0.22 }$ & 0.51  $^{+ 0.15 }_{- 0.24 }$ & 0.0  $^{+ 0.0 }_{- 0.0 }$ & 0 \\ 
121710 & 53.126897 & -27.786153 & 25.4 & 7.74 & \nodata & 10.5  $^{+ 0.1 }_{- 0.1 }$ & 3.24  $^{+ 2.0 }_{- 1.9 }$ & 1.26  $^{+ 0.19 }_{- 0.19 }$ & 0.49  $^{+ 0.06 }_{- 0.07 }$ & 0.0  $^{+ 0.0 }_{- 0.0 }$ & 1 \\ 
126594 & 53.144299 & -27.779856 & 27.3 & 10.64 & 7.95$^{k}$ & 9.3  $^{+ 0.1 }_{- 0.1 }$ & 54.94  $^{+ 11.6 }_{- 10.0 }$ & 2.03  $^{+ 0.22 }_{- 0.24 }$ & 0.03  $^{+ 0.03 }_{- 0.02 }$ & 0.0  $^{+ 0.3 }_{- 0.0 }$ & 1 \\ 
132229 & 53.203996 & -27.772097 & 26.1 & 7.56 & 7.25$^{j}$ & 10.2  $^{+ 0.1 }_{- 0.1 }$ & 86.00  $^{+ 17.1 }_{- 19.7 }$ & 2.41  $^{+ 0.24 }_{- 0.22 }$ & 0.34  $^{+ 0.06 }_{- 0.07 }$ & 0.0  $^{+ 0.0 }_{- 0.0 }$ & 1 \\ 
143133 & 53.147950 & -27.759931 & 27.6 & 6.11 & \nodata & 9.5  $^{+ 0.2 }_{- 0.3 }$ & 53.46  $^{+ 19.4 }_{- 14.3 }$ & 3.05  $^{+ 0.40 }_{- 0.51 }$ & 0.14  $^{+ 0.13 }_{- 0.08 }$ & 0.0  $^{+ 0.1 }_{- 0.0 }$ & 1 \\ 
154428 & 53.158173 & -27.739130 & 25.8 & 6.38 & 6.25$^{k}$ & 10.0  $^{+ 0.1 }_{- 0.1 }$ & 95.98  $^{+ 10.6 }_{- 8.9 }$ & 2.41  $^{+ 0.20 }_{- 0.15 }$ & 0.31  $^{+ 0.07 }_{- 0.10 }$ & 0.0  $^{+ 0.0 }_{- 0.0 }$ & 1 \\ 
179755 & 53.082131 & -27.859873 & 24.4 & 3.73 & \nodata & 10.0  $^{+ 0.2 }_{- 0.1 }$ & 76.81  $^{+ 21.3 }_{- 38.6 }$ & 1.80  $^{+ 0.24 }_{- 0.62 }$ & 0.45  $^{+ 0.22 }_{- 0.24 }$ & 0.0  $^{+ 0.1 }_{- 0.0 }$ & 0 \\ 
183348 & 53.082937 & -27.855632 & 27.3 & 3.38 & \nodata & 8.5  $^{+ 0.1 }_{- 0.1 }$ & 1.67  $^{+ 0.3 }_{- 0.3 }$ & 0.77  $^{+ 0.09 }_{- 0.12 }$ & 0.13  $^{+ 0.04 }_{- 0.04 }$ & 0.0  $^{+ 0.1 }_{- 0.0 }$ & 0 \\ 
184838 & 53.096419 & -27.853090 & 27.0 & 5.33 & 5.38$^{k}$ & 9.5  $^{+ 0.2 }_{- 0.3 }$ & 49.91  $^{+ 11.0 }_{- 8.9 }$ & 2.82  $^{+ 0.72 }_{- 0.35 }$ & 0.17  $^{+ 0.17 }_{- 0.12 }$ & 0.0  $^{+ 0.0 }_{- 0.0 }$ & 0 \\ 
189489 & 53.173662 & -27.841913 & 24.9 & 4.15 & \nodata & 10.1  $^{+ 0.1 }_{- 0.2 }$ & 57.72  $^{+ 11.8 }_{- 10.1 }$ & 1.92  $^{+ 0.17 }_{- 0.18 }$ & 0.60  $^{+ 0.18 }_{- 0.19 }$ & 0.0  $^{+ 0.0 }_{- 0.0 }$ & 0 \\ 
189494 & 53.162966 & -27.841946 & 24.5 & 4.50 & \nodata & 10.4  $^{+ 0.1 }_{- 0.3 }$ & 19.31  $^{+ 10.3 }_{- 6.0 }$ & 1.49  $^{+ 0.24 }_{- 0.27 }$ & 0.65  $^{+ 0.28 }_{- 0.28 }$ & 0.2  $^{+ 0.4 }_{- 0.1 }$ & 0 \\ 
189775 & 53.087378 & -27.840274 & 25.4 & 3.56 & \nodata & 9.6  $^{+ 0.1 }_{- 0.2 }$ & 6.41  $^{+ 3.3 }_{- 1.8 }$ & 1.75  $^{+ 0.24 }_{- 0.28 }$ & 0.83  $^{+ 0.22 }_{- 0.26 }$ & 0.0  $^{+ 0.0 }_{- 0.0 }$ & 0 \\ 
190413 & 53.084040 & -27.839348 & 24.4 & 8.70 & \nodata & 10.8  $^{+ 0.0 }_{- 0.0 }$ & 0.08  $^{+ 0.7 }_{- 0.1 }$ & 1.50  $^{+ 0.04 }_{- 0.04 }$ & 0.19  $^{+ 0.00 }_{- 0.00 }$ & 0.0  $^{+ 0.1 }_{- 0.0 }$ & 1 \\ 
198459 & 53.119122 & -27.814039 & 24.3 & 3.58 & 3.59$^{j}$ & 10.6  $^{+ 0.1 }_{- 0.1 }$ & 34.02  $^{+ 15.7 }_{- 9.9 }$ & 2.62  $^{+ 0.38 }_{- 0.19 }$ & 0.57  $^{+ 0.16 }_{- 0.15 }$ & 0.0  $^{+ 0.0 }_{- 0.0 }$ & 0 \\ 
200576 & 53.154771 & -27.806522 & 24.4 & 5.50 & \nodata & 10.5  $^{+ 0.0 }_{- 0.0 }$ & 3.10  $^{+ 0.2 }_{- 0.3 }$ & 0.72  $^{+ 0.03 }_{- 0.03 }$ & 0.77  $^{+ 0.02 }_{- 0.01 }$ & 0.0  $^{+ 0.0 }_{- 0.0 }$ & 0 \\ 
201793 & 53.188275 & -27.801938 & 25.3 & 4.17 & \nodata & 10.1  $^{+ 0.1 }_{- 0.1 }$ & 188.68  $^{+ 27.9 }_{- 32.1 }$ & 2.84  $^{+ 0.17 }_{- 0.21 }$ & 0.19  $^{+ 0.13 }_{- 0.09 }$ & 0.1  $^{+ 0.2 }_{- 0.1 }$ & 0 \\ 
203749 & 53.121420 & -27.794912 & 26.0 & 7.64 & \nodata & 10.4  $^{+ 0.1 }_{- 0.2 }$ & 56.08  $^{+ 20.7 }_{- 20.6 }$ & 1.64  $^{+ 0.17 }_{- 0.17 }$ & 0.33  $^{+ 0.10 }_{- 0.12 }$ & 0.0  $^{+ 0.0 }_{- 0.0 }$ & 1 \\ 
204851 & 53.138593 & -27.790253 & 24.5 & 5.46 & 5.48$^{k}$ & 9.8  $^{+ 0.0 }_{- 0.0 }$ & 207.39  $^{+ 6.9 }_{- 6.7 }$ & 1.83  $^{+ 0.05 }_{- 0.06 }$ & 0.02  $^{+ 0.05 }_{- 0.01 }$ & 0.0  $^{+ 0.0 }_{- 0.0 }$ & 1 \\ 
209303 & 53.206314 & -27.775716 & 25.0 & 5.77 & \nodata & 9.9  $^{+ 0.1 }_{- 0.1 }$ & 47.25  $^{+ 5.3 }_{- 5.4 }$ & 1.13  $^{+ 0.12 }_{- 0.11 }$ & 0.24  $^{+ 0.13 }_{- 0.11 }$ & 0.0  $^{+ 0.1 }_{- 0.0 }$ & 0 \\ 
214839 & 53.196571 & -27.757063 & 24.0 & 4.13 & \nodata & 10.3  $^{+ 0.1 }_{- 0.1 }$ & 53.32  $^{+ 8.7 }_{- 7.9 }$ & 1.61  $^{+ 0.12 }_{- 0.09 }$ & 0.25  $^{+ 0.14 }_{- 0.08 }$ & 0.2  $^{+ 0.3 }_{- 0.1 }$ & 0 \\ 
217926 & 53.184783 & -27.744047 & 26.9 & 6.93 & 5.04$^{k}$ & 9.3  $^{+ 0.2 }_{- 0.5 }$ & 26.59  $^{+ 5.1 }_{- 4.1 }$ & 2.34  $^{+ 0.31 }_{- 0.74 }$ & 0.06  $^{+ 0.04 }_{- 0.04 }$ & 0.0  $^{+ 0.0 }_{- 0.0 }$ & 0 \\ 
219000 & 53.161375 & -27.737665 & 25.1 & 6.92 & 6.81$^{k}$ & 10.0  $^{+ 0.2 }_{- 0.2 }$ & 142.53  $^{+ 20.1 }_{- 24.9 }$ & 1.73  $^{+ 0.17 }_{- 0.14 }$ & 0.19  $^{+ 0.15 }_{- 0.10 }$ & 0.0  $^{+ 0.0 }_{- 0.0 }$ & 1 \\ 
279678 & 53.108795 & -27.869027 & 25.0 & 4.72 & \nodata & 10.0  $^{+ 0.0 }_{- 0.0 }$ & 393.50  $^{+ 16.1 }_{- 14.8 }$ & 3.09  $^{+ 0.10 }_{- 0.09 }$ & 0.01  $^{+ 0.00 }_{- 0.00 }$ & 0.0  $^{+ 0.0 }_{- 0.0 }$ & 0 \\ 
\enddata 
\tablecomments{Properties of our H-dropout sample, measured including MIRI+ALMA data: (a) JADES DR1 ID (b) Photometric redshift measured by {\tt prospector}, or from EAZY in the case of a spectroscopic redshift  (c) Log$_{10}$ of stellar mass  in units M$_{\odot}$ (d) SFR measured by {\tt prospector} using the most recent 30 Myr time bin of the SFH in units M$_{\odot}$/yr (e) V-band Attenuation (f) Mass weighted age in units Gyr (g) the ratio of bolometric luminosity from the galaxy divided by that from the AGN (h) flag (1) indicating the source meets our LRD color selection (j) confirmed in JADES NIRSpec data (k) confirmed in FRESCO data.  }
\end{deluxetable*}

\subsection{SED Modeling}\label{sec:prospector}

To measure the more detailed physical properties, we use the {\tt prospector} Bayesian code to model the SEDs \citep{Johnson2021} using the Flexible Stellar Population Synthesis (FSPS) models \citep{Conroy2009}, MIST stellar isochrone libraries \citep{Choi2016,Dotter2016} and the stellar spectral libraries MILES \citep{Falcon-Barroso2011}. We use the MCMC sampling code {\sc dynesty} \citep{Speagle2020}, adopting the nested sampling procedure \citep{Skilling2004}. Our fiducial {\tt prospector} setup broadly follows that outlined in \citet{Ji2023} with a few alterations.  We adopt the \citet{Madau1995} IGM absorption model. We briefly summarize the other model assumptions and priors used here.

We use a non-parametric star formation history (SFH) composed of nine time bins with a constant star formation rate in each bin. We fix the first two bins to be at $0-30$ and $30-100$ Myr. Throughout this work, we will refer to the SFR as modeled by our SED-fitting as that inferred in the most recent 30 Myr (the latest time bin). The last time bin is assumed to be $\rm{0.85t_H - t_H}$ where $\rm{t_{H}}$ is the Hubble Time at the time of observation; and the remaining 6 bins are evenly spaced in logarithmic space between $\rm{100\, Myr - 0.85t_H}$. It has been shown that the recovered physical properties are largely insensitive to the number of bins used, when it is greater than 5 \citep{Leja2019}. We further adopt the continuity prior (to weight for physically plausible SFH forms, thus mitigating overfitting the data), which has been demonstrated to work well across various galaxy types \citep{Leja2019}. 

We adopt the \citet{Byler2017} nebular continuum and line emission model. We set both the stellar metallicity and gas phase metallicities as free parameters and assume flat priors in logarithmic space (with $\log(Z_*/Z_\sun) \in (-2, 0.19)$ and $\log(Z_{\rm{gas}}/Z_\sun)\in(-2,0.5)$).  Ionization parameter $U$ is also left as free parameter using a flat prior with $\log U\in(-4,1)$.  

We adopt the \citep{DraineLi2007} dust emission model with priors as defined in \citep{Williams2019} to allow for more flexibly hotter dust temperatures which may be prevalent at higher redshift \citep{daCunha2013}. These include flat priors on the starlight intensity on dust grains $U_{\rm{min}}\in(1,25)$, and the faction of stars at $U_{\rm{min}}$, $\gamma\in(0.01,0.99)$. These parameters are related to T$_{\rm dust}\sim 18 \times <U>^{1/6}$ K as in \citet{Draine2014}. We also adopt flat priors on the polycyclic aromatic hydrocarbon (PAH) mass fraction, $qpah\in(0.5,4)$.

We assume a two-component dust attenuation model where the dust attenuation of nebular emission and young stellar populations, and of old stellar populations, are treated differently \citep{Charlot2000}. For stellar populations older than 10 Myr, we assume the dust attenuation using the parametrization from \citealt{Noll2009} (i.e. a modified \citet{Calzetti2000} dust attenuation law). Stellar populations younger than 10 Myr are assumed to have the same dust attenuation law as for the nebular emission \citep[for further details on the various dust parameter priors, dependencies, and prior ranges, see][]{Tacchella2022a, Ji2023}. The dust model priors are set such that the V-band attenuation (Av) can vary between $A_{\rm{V}}\in(0,10)$ with a flat prior.

We also include AGN dust torus templates from \citet{Nenkova2008a} and \citet{Nenkova2008b}, with flat priors in logarithmic space for both the ratio of bolometric luminosity from the galaxy divided by that from the AGN ($f_{\rm{AGN}}\in(10^{-5},3)$), and the optical depth of clumps in AGN dust torus at 5500 \AA\ ($\tau_{\rm{AGN}}\in(5,150)$).

We use the photometric redshift measured using EAZY in the last section as a photometric redshift prior for {\tt prospector} modeling (the mean of a Gaussian prior width $\pm$ 0.5). For the cases where the EAZY fit resulted in a redshift $z>8$, we instead use a flat prior on the redshift to allow the possibility of lower-redshift solutions. For the sources with spectroscopic redshift constraints, we fix to the spectroscopic redshift that was identified in the last Section. We further limit the S/N of any photometric point, which is capped at 20 (minimum 5\% uncertainty, reflecting uncertainties in relative photometric calibration between filters).

\section{Results}\label{sec:results}

\subsection{Impact of MIRI+ALMA data}\label{sec:results:mirialma}

To date, photometric studies of similarly red galaxies at $z>3$ have been restricted to NIRCam and HST data, with some limited wavelength coverage from 1-2 MIRI bands \citep{Barrufet2023, PerezGonzalez2023, Labbe2023a, Labbe2023b, Barro2023, Rodighiero2023, Akins2023, McKinney2023, Endsley2023}. To explore the impact on recovered properties when using this more limited wavelength coverage, we run our \texttt{prospector} modeling for the 29 sources inside the MIRI footprint using only the HST+NIRCam data, and, again including also the MIRI+ALMA data.

In Figure \ref{fig:paramcompare} we present a comparison of the inferred best-fit parameters: photometric redshift, stellar mass, and SFR using HST+NIRCam only, with the result obtained when including the MIRI+ALMA data. 
We find that including photometry from both the seven MIRI filters and ALMA 1-mm data overall results in consistent redshifts as HST+NIRCam (with a few outliers; left panel of Figure \ref{fig:paramcompare}). However, including MIRI+ALMA significantly alters other key parameters recovered using SED modeling. We find that the addition of MIRI+ALMA data serves to lower both the stellar masses and SFRs of galaxies compared with using HST+NIRCam alone. In particular, we find a systematic reduction in stellar mass of (median decrease of 0.6 dex, and as large as 1 dex) for galaxies with HST+NIRCam-measured \LogM$>$10. We similarly find that for SFRs in excess of $\sim$100 M$_{\odot}$/yr as measured by HST+NIRCam alone we calculate a median factor of 10 decrease in SFR inferred when including MIRI+ALMA. These findings indicate that studies based on the more limited data sets are likely to overestimate both the SFRD and the stellar mass density, at high redshift significantly, in particular for galaxies where HST+NIRCam infer high masses and SFRs. This seems to be the case even despite the excellent medium band coverage of JADES.

We note that this finding may not be representative of the general galaxy population, since our sample has much more extreme red colors than typical galaxies at these redshifts. A systematic assessment of the impact of multiple MIRI bands on mass and SFR across redshifts has not yet been undertaken \citep[although see simulations based on mock galaxies in][]{Kauffmann2020, Kemp2019, Bisigello2019}. However, analyses with more limited filters (F560W and/or F770W) suggest mixed results. 
A similar impact to ours was noted in \citet{Papovich2023} using a more representative sample of galaxies at similar redshifts, (although comparing to modeling with a much more limited set of only HST plus IRAC data). Those authors find that, on average, the stellar mass decreases by $\langle {\rm{\Delta }}\mathrm{log}{M}_{* }\rangle =0.25$ dex at $4 < z < 6$ and 0.37 dex at $6 < z < 9$. Those authors also find a systematic reduction in SFRs by $\langle {\rm{\Delta }}\mathrm{log}SFR\rangle$= 0.14 dex at $4 < z < 6$ and 0.27 dex at $6 < z < 9$. MIRI likely has a large impact in this case owing to a lack of NIRCam photometry, which can break degeneracies between continuum and emission lines, which translates to a less robust constraint on rest-optical continuum at high-redshifts. While their result is qualitatively in line with what we find for the most extreme of our red sources, accumulating evidence indicates that including some MIRI data (in particular, F770W, covering rest-frame $\sim$1--2$\mu$m) may not be essential to achieving  more accurate estimates for typical galaxies. Helton et al (in preparation) finds that for $7<z<9$ galaxies (using our same NIRCam filter set), that the stellar population models with and without MIRI data are similar. Additionally, Alberts et al (in preparation) find that stellar masses of \LogM$>$9 galaxies at $3<z<6$ also demonstrate no systematic improvement when F770W is included. 

Regardless, since in this study we have both multiple NIRCam medium bands plus multiple MIRI filters (mitigating uncertainties in constraining the stellar mass from both angles), our measurements are
likely to be relatively reliable despite the exceptional character of our very red sources. However, this exercise demonstrates that  the MIRI and ALMA data are of particular importance for sources with extreme red colors. Thus, caution should be exercised when interpreting the SED modeling of sources with red (and often featureless) SEDs without mid and far infrared wavelength coverage. Further, the uncertainties that arise from lacking the fuller wavelength coverage are not reflected in errorbars of mass and SFR measured with HST+NIRCam alone.

\subsection{Properties of Optically faint (HST-dark) galaxies}

In this section, we characterize the properties of our sample of very red sources, in particular highlighting the diversity of our sample. Our {\tt prospector} modeling indicates that our sources range from $3<z<8$, revising all of the higher-redshift solutions that were measured using EAZY (owing to non-detections in the observed optical and near-infrared; Figure \ref{fig:seljades}).  We also find that our sample includes galaxies at a range of stellar masses from \LogM$\sim8.2-10.8$, with moderate median SFR ($\sim50$ M$_{\odot}$/yr), high attenuation (A$_{\rm V}\sim2$), moderately evolved stellar populations (mass-weighted age $\sim250$ Myr). These are similar to initial JWST explorations \citep{Barrufet2023, Rodighiero2023, PerezGonzalez2023}.

As shown in Figure \ref{fig:miricutouts}, our sample of H-dropouts have incredibly diverse morphologies, ranging from large extended disks \citep[e.g. Gibson et al submitted;][]{Nelson2023}, to potentially merging clumps or groups, while a large fraction are remarkably compact, close to the resolution limit of F444W. This diversity is in line with expectations from simulations that 1) whether dusty galaxies are ``dark" is a strong function of viewing angle for a range of galaxy types, and 2) the most massive galaxies will pass through this phase early in the Universe's history due to prodigious dust production \citep{Cochrane2023}. In the subsections below, we review the sub-categories of objects that we identify in our sample.

\begin{figure*}[t]
\includegraphics[width=1\textwidth, trim=90 120 0 85, clip]{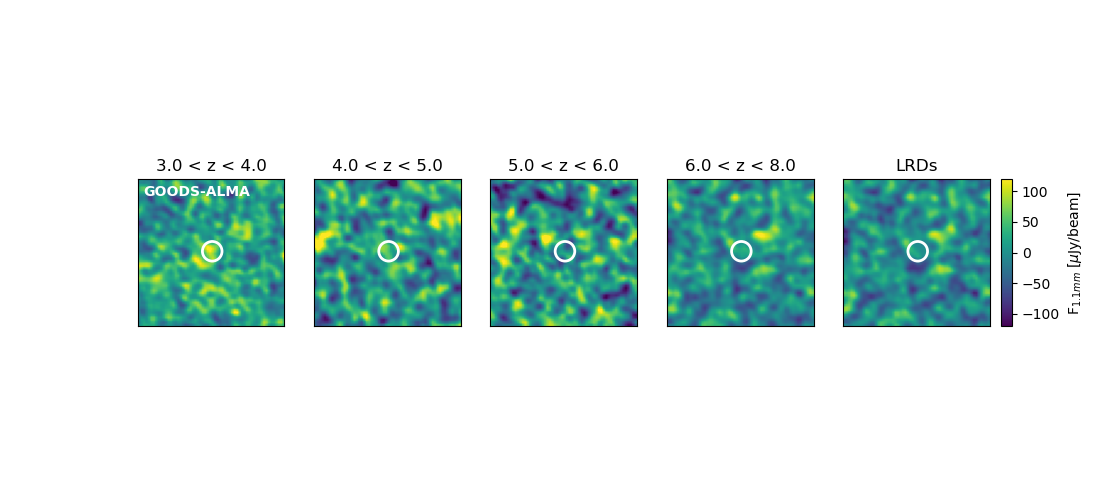}
\includegraphics[width=0.9\textwidth, trim=0 100 0 85, clip]{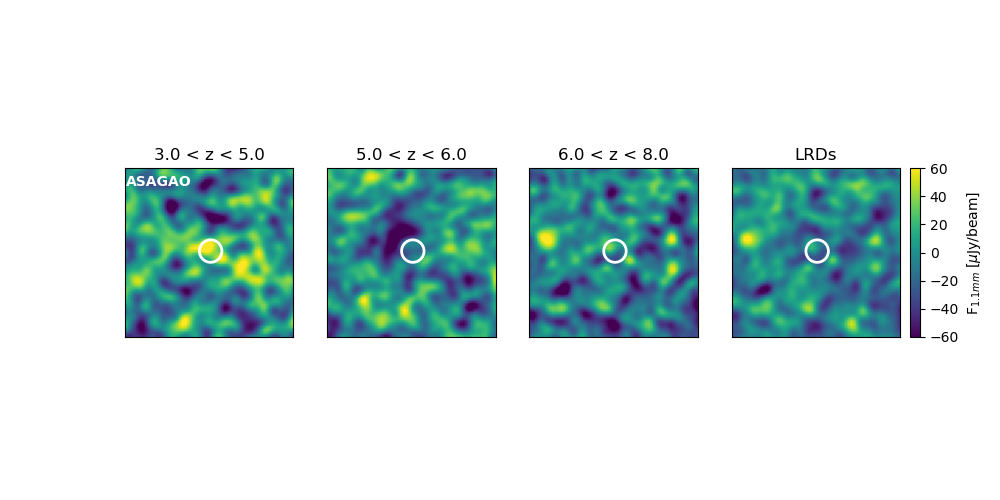}
\caption{Top panel: ALMA 1.1 mm stacks of our entire sample galaxies (including NIRCam-only sources, with individual detections removed) using the GOODS-ALMA data from \citep{Franco2018}, which covers our entire sample. 
Bottom panel: ALMA 1.1 mm stacks of a subset of our galaxies that are inside the footprint of the deeper ASAGAO imaging (individual detections removed). Only one source is inside the footprint at $3<z<4$, so we instead show a combined $3<z<5$ redshift bin. Our sample below $z<5$ exhibit possible cold dust emission ($2.2-3\sigma$ for both maps respectively), consistent with low luminosity DSFGs. Sources at $z>5$ (including all LRDs) do not show detectable dust emission. Far right panels shows the ALMA stack of all 9 LRDs, which reaches a non-detection upper limit of 32$\mu$Jy/beam.}\label{fig:stack}
\end{figure*}

\subsubsection{Dusty star forming galaxies in our sample}\label{sec:dsfg}

A minority of our sources exhibit detectable ALMA emission, indicative of dust obscured star formation at the level typical of dust obscured galaxies  \citep[DSFGs; SFR$>$100 M$_\odot$/yr;][]{Casey2014}. Four sources between $3.6<z<5$ show significant ALMA detections ($>4\sigma$) in the range 0.5--1.0 mJy/beam, well below those of prototypical submillimeter galaxies  \citep[their ALMA properties have been studied elsewhere;][]{Franco2018, Hatsukade2018, Xiao2023}. 
The NIRCam and MIRI cutouts (in Figure \ref{fig:miricutouts} of the Appendix) show that 3/4 ALMA sources are the most extended disk-like objects in our sample, while the 4th (201793) is compact but clearly resolved in the shortwave filters. These sources are qualitatively similar to ALMA-only objects that have been identified by 2--3-mm ALMA imaging with $z>4$, total infrared luminosity  \LIRLsun$\sim12-12.5$, \LogM$\sim10.5-11$, SFR$\sim200-300$ and $\sim25$ ABmag at 4$\mu$m \citep[e.g.][among others]{Williams2019, Manning2022}. Noting that our MIRI area is quite small, we find a similar abundance of $\sim0.1$ square arcmin$^{-1}$.

However, the overwhelming majority of our sources are not detected in any of the 1-mm ALMA data (25 out of 29) and thus have only upper limits to their total infrared luminosity. These upper limits are \LIRLsun$\lesssim$12, based on integrating the 8-1100$\mu$m (restframe) best-fit SED from {\tt prospector} (see Figure \ref{fig:LIR}). While this means we can't robustly constrain how much lower is the exact level of obscured star formation individually, we stack the 1.1mm ALMA images for all sources that are not individually detected (calculating the inverse variance weighted average) from the GOODS-ALMA program \citep[$\sigma_{rms}\sim180 \mu$Jy/beam;][]{Franco2018} for which we have coverage of our entire MIRI footprint.  
We do not find a significant detection in the stacked data at $3<z<4$, or $4<z<5$, and only a 2-sigma detection from the combined redshift range. However, a majority of the sources sit inside the deeper ASAGAO footprint. We perform the stack again for sources inside ASAGAO, finding again a marginal detection for $3<z<5$ sources in our sample with 55$\pm$24 $\mu$Jy/beam. The ASAGAO results from inside the MIRI footprint are shown in Figure \ref{fig:stack}.

\begin{figure}[t]
\includegraphics[width=0.5\textwidth]{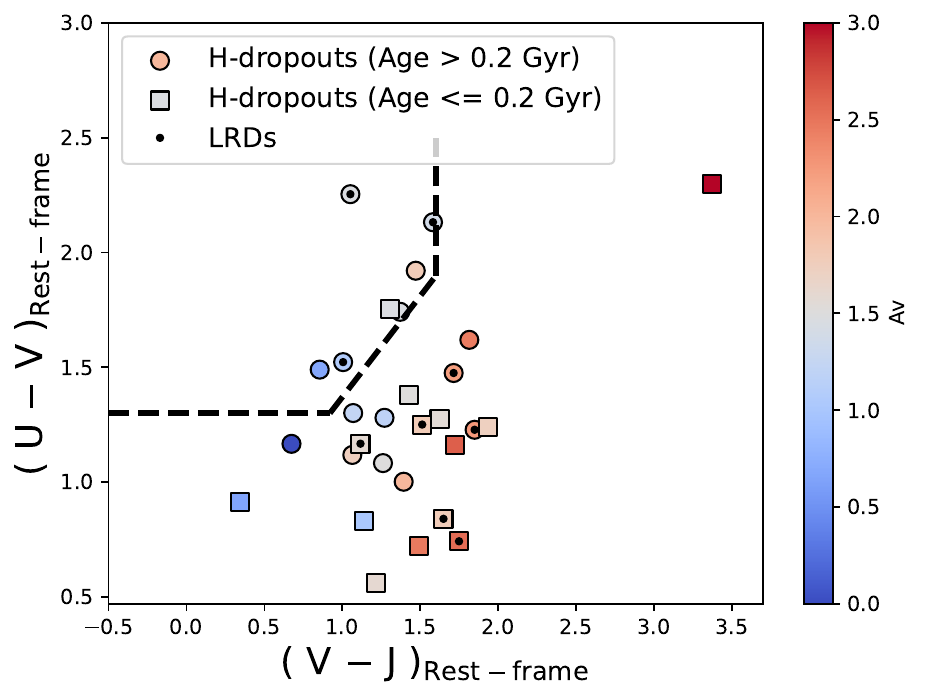}
\caption{UVJ diagram  of our MIRI sample. Points are color coded by their inferred Av and symbol shape indicates old vs young mass weighted ages. LRDs are flagged with black dots. We omit the 1-$\sigma$ uncertainties from the modeling posteriors for clarity, noting the uncertainties are large (sources have low S/N near both the rest frame U and J) enough to scatter away from the UVJ box. 
SEDs are consistent with our sample including primarily star forming sources, some with extreme dust obscuration (red in V-J), and some candidate older/post-starburst-like SEDs (Balmer breaks are also visible in the SEDs).}\label{fig:UVJ}
\end{figure}

It is possible that our limited sample size is too small for a significant detection. Therefore, we repeat the stacking while including our broader NIRCam-only sample within the GOODS-ALMA footprint, which roughly doubles the sample size. We also show the GOODS-ALMA stack from the broader NIRCam-only sample in Figure \ref{fig:stack}. We find that this test does yield a more significant (3$\sigma$) stacked detection in ALMA in the lowest redshift bin, $3<z<4$. Collectively, we interpret these stacking experiments to indicate that our sample is likely dominated by faint sources with some small level of dust obscured star formation that are (individually) below the detection limit of ALMA at the low redshift end, $z<5$.

The NIRCam imaging is sufficiently deep to have detected lower-luminosity analogs of DSFGs at 
even higher-redshifts $z\sim6-7$ \citep[e.g. the serendipitious sources identified in][]{Fudamoto2021}. These ALMA-only sources were thought to have $\sim$25-26 ABmag at $4\mu$m, with obscured SFRs in the range $\sim$40-70 \Msun/yr (based on $f_{1mm}\sim110-190\mu$Jy) and \LogM $\lesssim10.3$. We identify nine sources with inferred properties that are consistent with these, noting that the stellar population modeling for three of those indicate substantially lower sSFRs to \citet{Fudamoto2021} 
To determine whether on average our sources at $6<z<8$ may be comparable, we repeat the ALMA stacking for sources in this redshift range, and we find no stacked detection in GOODS-ALMA to a limit of $f_{\rm1.1mm}=38\pm41$ $\mu$Jy/beam (Figure \ref{fig:stack}; in ASAGAO, we find $f_{\rm1.1mm}=-9\pm28 \mu$Jy/beam, $1\sigma$ upper limits that are a factor of 4--7 below the 1-mm continuum detections in that work). This result suggests that our $6<z<8$ sample is not actually dominated by similar lower luminosity DSFGs.

In Figure~\ref{fig:UVJ} we explore further whether galaxies are red due to age, dust, or both in the restframe U-V and V-J  colors of our sources.   All sources exhibit substantial dust obscuration (Av$>$0.7) except one. Although we do not see a strong dependence in Av with redshift, we note the most extreme sources with Av$>$2.5 are all at $z<6.5$. We also note that by stacking the ALMA data in redshift bins, we find that at $3<z<4$ and $4<z<5$, the stacks show weak but more significant detections. This is consistent with a majority of the sources we identified at $z<5$ being faint dust obscured sources.

While the stacked non-detection in ALMA at $z>6$ could mean that the level of obscured star formation traced by cold dust is relatively low, (i.e. below the ALMA detection limit, \LIRLsun$<$12), this is not definitive, and we unfortunately have limited data to further constrain this on an individual-galaxy basis.
A lack of ALMA detection could also imply that our sources either 1) contain obscured star formation, but the dust is hotter than is typically assumed at lower redshift \citep[see e.g. ][]{deRossi2018}, 2) contain hot dust heated by AGN activity, or, 3) the galaxies have primarily evolved or older stellar populations. In the following sub-sections we now explore these possible scenarios.

\subsubsection{Possible evidence of AGN among red galaxies}\label{sec:AGN}

To look for AGN evidence, we have matched our sample to the pre-JWST AGN catalog built by \citet{Lyu2022} that has integrated the {\it Chandra} X-ray, {\it HST} optical to near-IR, {\it Spitzer} mid-IR and JVLA radio data for a comprehensive search 
of AGN in the GOODS-S field. In total, we found only two matches among the more extended NIRCam-only sample (Table \ref{tab:NIRCAM}): JADES 171973 and 284527; both of them are identified as AGN by their high X-ray luminosity and the X-ray to radio ratio \citep[see details in][]{Lyu2022}. As pointed out in \citealt{Lyu2022}, the AGN selection is complicated by the survey depth, wavelength range and object variations and many AGNs are still likely missed. 

With the improved sensitivity and wavelength coverage of JWST data, significant progress has been made to identify AGN.
Based on semi-empirical SED analysis of MIRI-detected sources with JADES NIRCam and SMILES MIRI photometry, \citet{Lyu2023} has drastically improved the AGN census in the central regions of GOODS-S. 
For our sample,
three new AGN candidates have been revealed from that study: JADES 57356, 106502 and 204851. Notably, 204851 has 
been confirmed to be a broad-line AGN at $z$=5.48 in FRESCO data \citep{Matthee2023}.

Meanwhile, several groups have demonstrated
the existence of fainter broad-line AGNs by selecting sources that feature as Little Red Dots \citep[LRDs;][among other references]{Matthee2023,Labbe2023b,Greene2023} --- objects with strong red continuum and compact morphologies in the NIRCam bands. Although the nature of these objects is still debated \citep[e.g.,][]{Barro2023}, the success rate of AGN search by this selection has been high (e.g., 9/12 in the sample of \citet{Greene2023} show evidence of broad H$\alpha$). 
While confirmed broad-line AGN may be prevalent among LRDs, it remains unclear whether the AGN dominates the host galaxy, and what its contribution is to the restframe UV and restframe optical continua.
We now apply such selections to our sample and discuss their nature via the color/SED analysis with the addition of MIRI data points at longer wavelengths.

\citet{Matthee2023} describe the LRD selection criteria as relatively flat or blue at observed 1 -- 2 micron, with a (very)
red continuum from 2 -- 4 micron.  We cross-check our sample with the following selection criteria for LRD:  $-0.5<$F115W--F200W$<1$ and F277W--F444W$>$1.6 \citep{Greene2023}. Those criteria, based on spectroscopic confirmation of broad line AGN, 
should contain an estimated 80\% AGN fraction \citep{Greene2023}. 
We find that 13 out of 66 objects in the NIRCam footprint are candidate LRDs by this criterion (9 of which lie in our MIRI coverage, excluding the probable brown dwarf candidate 190413). Two of the 13 include sources with evidence of an AGN, including 204851 \citep{Lyu2023, Matthee2023}, as well as 154428, which show possible evidence of broadened H$\alpha$ and a weak narrow component in FRESCO data (Sun et al. in prep). While we do not fold in the explicit compactness cut of \citet{Greene2023} to identify LRDs, we note that all candidates identified by the LRD colors are visibly unresolved, or consistent with point sources, in F444W (including 204851, which in the single-band cutouts of Figure \ref{fig:miricutouts} appears blended with 2 neighbors).  In general, our F150W--F444W selection does not pick up LRDs systematically, or, similarly, the extremely red object (ERO) selection that was used in \citealt{Barro2023} (F277W--F444W$>$1.5). This is because while the LRDs and EROs are red in the long-wave NIRCam filters, some fraction are bluer in the shorter wavelength ones due to the rising rest-frame UV SEDs. The net result is that the bluer, more ``V" SED-shaped sources are preferentially excluded by our H-dropout selection unless combined with a very red rest-optical continuum.

\begin{figure*}[t]
\includegraphics[width=1\textwidth,trim=40 130 40 100, clip]{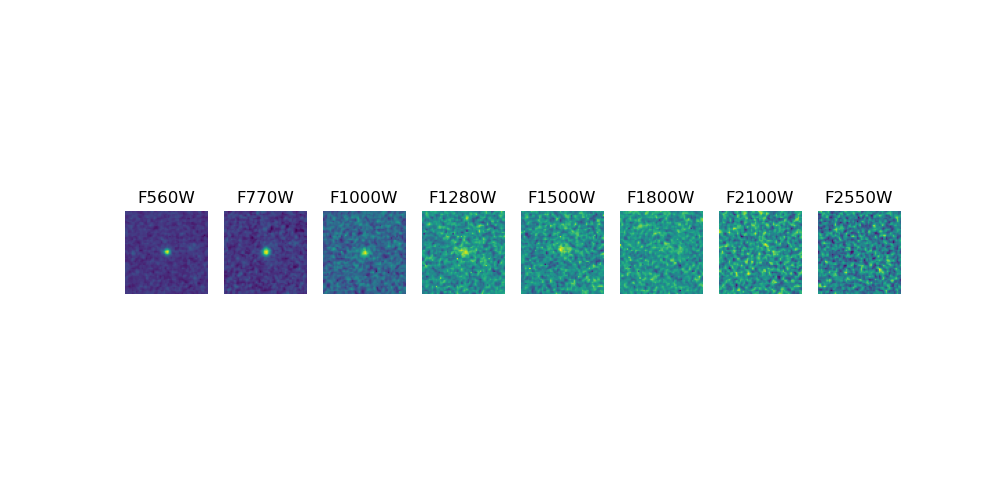}
\includegraphics[width=1\textwidth,trim=10 0 10 40, clip]{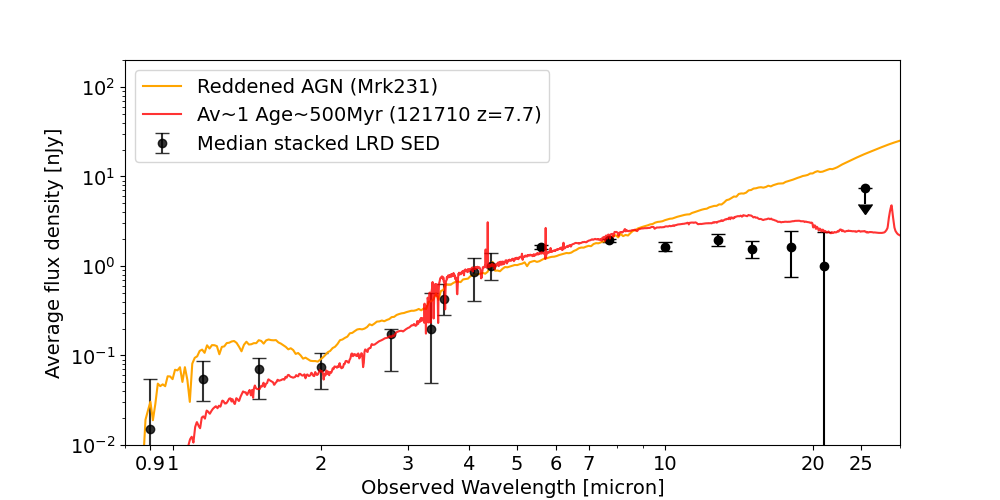}
\caption{Top panel: Stacked MIRI images from 5.6-25$\mu$m of the LRD subset of H-drops (3.6" on a side). Bottom panel: stacked observed SED of the LRD subset, where NIRCam points are the median and interquartile range of the measured photometry, and MIRI points are measured from the median stacked imaging in top panel. The stacks are consistent with the flattening of the restframe near-infrared SED that was observed in the forced photometry of individual sources. They are not consistent with rising mid-infrared flux from a dust obscured AGN \citep[e.g. Mrk 231][]{Polletta2007} and are a better match to stellar models of moderate dust and age (red).  }\label{fig:miristack}
\end{figure*}

\begin{figure*}[t]
\includegraphics[width=0.5\textwidth]{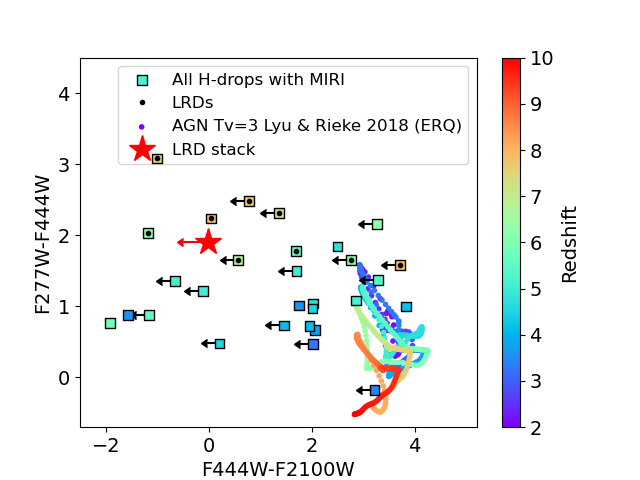}
\includegraphics[width=0.5\textwidth,trim=0 0 0 0, clip]{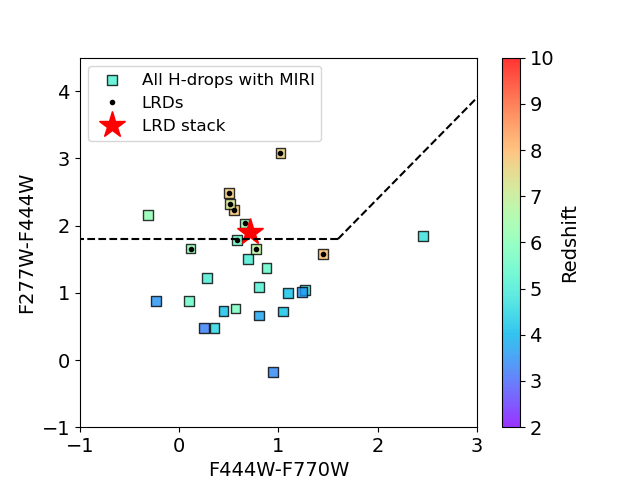}
\caption{NIRCam-MIRI color color diagrams for our sample (squares) with the LRD flagged (black dots). Objects are color-coded by redshift. Red star indicates the average colors obtained from a median-stacked LRD SED in Figure \ref{fig:miristack}.
Left: F277W--F444W vs F444W--F2100W colors of our sample. For comparison we plot a heavily obscured AGN template from \citet[][excluding stellar host]{LyuRieke2018} shown to match the red SED of so-called Extremely Red Quasars (ERQ) which at $z\sim3-8$ are extremely red in the restframe near-infrared. In comparison, LRDs have bluer or flat colors in F444W--F2100W than SEDs dominated by obscured AGN, 
more similar to SEDs dominated by stellar emission (tracing the stellar bump). 
Right: F277W--F444W vs F444W--F770W colors of our sample. Most LRDs are red or flat in F444W--F770W color, inconsistent with an interpretation that emission line boosting in F444W produces the red in F277W--F444W color (which would be blue in F444W--F770W). These colors are more consistent with flat or moderately-rising red continuum driven by starlight (the dashed-line box, specifically diagonal line, defines the color region at F444W---F770W$>$1.5 that excludes heavily reddened AGN continuum at $z>3$; \citealt{Akins2023}). 
The NIRCam-MIRI colors in both plots favor SEDs that are stellar in origin. }\label{fig:NMcolor}
\end{figure*}

\subsubsection{Nature of the LRDs in Our Sample}

Our LRD subset has distinct SEDs that appear different from most of the H-dropouts. Some LRDs are also poorly fit by e.g. single component dusty or quiescent stellar populations. In the literature, this has prompted a number of explorations into differing origins for their blue rest-UV SEDs \citep{Barro2023, Labbe2023b, Endsley2023, Furtak2023a, Greene2023}.  On the one hand, the rest-frame UV could be unobscured star formation. Alternatively, it could be scattered light from either hot stars or from an AGN accretion disk (depending on geometry). Similar ambiguity exists over the origin of their very red rest-optical SEDs, which could be driven by either obscured AGN continuum, or dust obscured stellar emission. Recently, deep ALMA non-detections for LRDs have provided evidence against red continuum produced by dust-obscured star formation, because the expected amount of re-processed dust emission in the far-IR under typical assumptions (e.g. dust temperatures in the range T$_{\rm dust}\sim20-60$ K) 
would be dramatically in excess of the deep ALMA limits \citep[][]{Labbe2023b}. We find similar results for our LRD sample (see right panels in Figure \ref{fig:stack}). This could point to an AGN-dominant SED with hot dust emission. However, recent studies of compact dusty star forming galaxies at high redshift indicate that the dust temperature can be significantly higher due to the higher density of star formation \citep{deRossi2018, Sommovigo2020}. So, this argument in favor of AGN dominance is not definitive.

Our MIRI photometry enables us to explore whether the LRD/AGN candidates from our sample are plausibly sources whose red rest-optical emission is dominated by a heavily reddened AGN \citep[e.g. A$_{\rm V}\sim1-4$, based on the shape of the rest-optical continuum][]{Greene2023}. 
To explore this hypothesis, we plot our sources with MIRI coverage on two NIRCam-MIRI color-color diagrams in Figure \ref{fig:NMcolor}  along with the redshift evolution in colors for a template dominated by an obscured AGN  \citep[i.e. excludes contribution from stars][]{LyuRieke2018}. 
From the empirical AGN SED library built in \citet{LyuRieke2018}, we choose a model template for a typical (normal) AGN obscured by an extended dust distribution featuring large grains with an optical depth $\tau_V=3$. This template matches the typical SEDs of lower-redshift ``extremely red quasars" (ERQ; \citealt{Ross2015, Hamann2017}) that have similarly red F277W--F444W color as our LRD/ERO sample.

Looking at the observed colors, we find that all of our LRD candidates exhibit a 
turnover in their SED redward of F444W, between restframe 0.5--3$\mu$m. This turnover results in blue or flat F444W--F2100W color that is inconsistent with obscured AGN templates, which show a steeply rising shape at longer wavelengths \citep[e.g.][]{LyuRieke2022}. 
These colors instead favor a scenario where the rest-optical emission between $0.5-3\mu$m restframe is dominated by continuum from the stellar populations. This is potentially the spectral signature of the stellar bump at restframe 1.6$\mu$m (caused by a minimum in the H- opacity in the atmospheres of cool and low-mass stars that dominate the near-infrared spectrum of galaxies at age $>$10 Myr; e.g. \citealt{Sawicki2002}). Reddened AGN, in contrast, typically exhibit a steeply rising red continuum redward of rest-frame 1.6$\mu$m, tracing hot dust emission from a torus \citep{AlexanderHickox2012,LyuRieke2022}.

While the MIRI data is deep enough to rule out rising red continuum from a dominant AGN for individual sources, the majority of the LRD subset are not detected in the longer-wavelength filters. To obtain a stronger constraint on the MIRI colors (on average) we  
median stack the MIRI imaging for the LRD subset. For this experiment we also include the 25$\mu$m band (F2550W) which otherwise is shallow relative to the other filters, but on average, can provide a meaningful constraint at restframe $>$3$\mu$m. The MIRI stacks are shown in the top panel Figure \ref{fig:miristack}. 

We measure average aperture photometry using the stacked MIRI images following the same procedure as Section \ref{sec:meas}. On average, the sample remains undetected at wavelengths $>$18$\mu$m, even in the deeper stacked image. For an accurate comparison to the average NIRCam SED of LRDs, we also calculate the median and inter-quartile ranges of the LRD photometric points (in lieu of stacking the NIRCam images, since all sources are already strongly detected in the NIRCam F444W). The full median-stacked SED for the sample of LRDs is shown in the bottom panel of Figure \ref{fig:miristack}. For comparison, we also plot two example SEDs: one for a reddened quasar and its host galaxy, placed at the median redshift of our LRD sample \citep[Mrk 231;][]{Polletta2007}, and, a {\tt prospector} best fit model  that is representative in shape of a number of our LRDs among the H-dropout sample (ID 121710). It is immediately clear that the turnover in the median MIRI SED strongly disfavors a reddened AGN shape, and instead agrees better with the moderately dusty and moderately old stellar model describing 121710 (mass weighted age $\sim$500 Myr, Av$\sim$1). Notably, we find that the limits in the color from the median stack indicate that the SEDs of our sources must be quite flat between F444W and F2100W, with color F444W--F2100W$\sim0$ (and, even consistent with blue color, i.e. a turnover, given the lack of significant stacked detection in F2100W).

Emission line boosting in F444W could also generate red F277W--F444W with blue F444W--F2100W colors, either from [OIII]+H$\beta$ at $z>7$ or H$\alpha$ at z$\sim$5. To explore this possibility, we also plot the rest-optical colors vs F444W--F770W (right panel of Figure \ref{fig:NMcolor}) which should be very blue if F444W is boosted by emission lines (note that we choose F770W because F560W can also be contaminated by H$\alpha$ emission when [OIII]+H$\beta$ is in F444W). We find that the majority of the LRDs have weakly red or flat F444W--F770W colors, suggesting that the F277W--F444W color is not red  due to significant line boosting in F444W.  We additionally find that the median SED color of F444W--F770W=0.7, which is not consistent with the idea that the F444W--F2100W is blue or flat in spite of a red continuum because of strong emission line boosting F444W.
These colors also suggest that the red rest-optical colors are dominated by stellar emission and not obscured AGN emission (which would predict a continual red rise into the mid-IR from AGN continuum rather than turnover from the stellar bump). However, the colors cannot rule out an obscured mid-infrared-dominant AGN whose continuum begins to dominate the SED at restframe wavelengths $>$3$\mu$m.

We note that the theoretical AGN templates provided with {\tt prospector} may not correspond fully to reality; an alternative approach uses empirically based templates \citep[see review by][]{LyuRieke2022}.
To explore this approach, we run {\tt prospector} on our full sample while using the modified and more realistic mid-infrared AGN template set \citep{LyuRieke2018}. We follow the {\tt prospector} setup as outlined in \citealt{Lyu2023}. 
We find that, in line with our exploration of the NIRCam-MIRI colors, the contribution to the restframe 0.5--3$\mu$m continuum of LRDs is not obviously dominated by an obscured AGN. In reality, individual sources may exhibit a broad variety of behavior near restframe 3$\mu$m (where the presence of a mid-IR AGN is expected to be most obvious among typical star forming galaxies) and analysis is complicated by the effect of strong emission lines and low S/N of the longest wavelength MIRI data.

Further, we do a similar exploration using CIGALE \citep{Boquien2019}, which for samples in the literature has found that the steep slopes of the rest-optical continuum for LRDs has a preferred origin from AGN continuum, based on HST+NIRCam photometry alone. For this experiment we use the {\tt SKIRTOR} AGN model, a clumpy two-phase torus model from \citealt{Stalevski2012, Stalevski2016}. We find that, in nearly all cases (for the LRDs) CIGALE prefers fits where AGN continuum dominates the SED between restframe 0.5-0.8$\mu$m, but these AGN dominant models significantly overpredict the MIRI observed SED at F1800W--F2100W by a median/typical factor of 25-40 (well in excess of the photometric uncertainties). We also find that  CIGALE over-predicts the MIRI photometry for the parent H-dropout sample, although to a lesser degree for non-LRDs (factor 2-3). This is likely driven by the redder rest-optical (F277W--F444W) color of the our subset of LRDs compared to the parent sample of H-dropouts (see left panel of Figure \ref{fig:NMcolor}).

Thus, we would find similar conclusions based on HST+NIRCam photometry alone as the AGN-dominant solution identified by other studies in the literature. This highlights that while LRDs may be redder in the rest-frame optical than the non-LRD H-dropout sample, the MIRI data demonstrates that these SEDs are in fact mostly quite similar in the restframe near-infrared, suggesting a stellar origin for both subsets of H-dropouts (see right panel of Figure \ref{fig:NMcolor}).

Given the consistency of the rest-optical and near-infrared SED with dust obscured stellar emission, it  remains puzzling why this does not translate to brighter far-infrared emission from re-processed energy by dust. To explore this further, we stack the 1.1mm ALMA imaging covering our 9 LRDs from the GOODS-ALMA program \citep{Franco2018} to obtain an average far-infrared flux (see Figure \ref{fig:stack}). We find that the 1.1mm flux is not detected (3$\pm37\mu$Jy/beam). Similarly low stacked ALMA limits disfavoring dust-obscured star formation were found in \citet{Labbe2023b}. It remains plausible that compact star formation at high redshift and low metallicity could heat dust well above typical expectations, and we discuss this scenario further in Section \ref{sec:sfrd}.

\subsubsection{Evidence for older stellar populations in our sample}\label{sec:psb}

Such a low far-infrared flux measured in the ALMA stacks is consistent with a primary result from the SED-modeling, which is that a large fraction of LRDs in particular are preferred to have older stellar populations over high levels of dust-obscured star formation. This is presumably, at least in part, a result of {\tt prospector} trying to account for these deep upper limits from ALMA (given the fixed low dust temperatures assumed by the modeling). This possibility of quiescent SED solutions for LRDs was also explored in \citet{Labbe2023b}, however that work determined that the extremely red rest-optical continuum slope disfavored a purely quiescent stellar population. However, we find that an older and evolved stellar population (mass weighted age $>200$Myr), in combination with significant dust attenuation (Av $>$ 1), adequately fits the SEDs of a number of H-dropouts, as well as a large fraction of our LRD subset. This is demonstrated in Figure \ref{fig:UVJ}, which shows that a number of our LRDs reside near the UVJ quiescent box \citep{Williams2009, Muzzin2013mf}. The typical level of dust attenuation that we infer is relatively high. 
However, owing to their faintness, we note that the  uncertainties in the restframe colors  near both the restframe U and J part of the SED are quite large. Thus, we caution against overinterpretation of the location of our sources in the UVJ diagram. 

Regardless of the large errors in rest-frame U and J, we note that a number of the LRD SEDs with red U-V and V-J colors also exhibit clear evidence at high significance for strong Balmer breaks (e.g. IDs 121710, 132229, 219000 and perhaps also 154428) as does one non-LRD, 200576. The  Balmer break evidence is clear in part due to the NIRCam medium band images, which at F182M, F210M and F335M are finely sampling the spectral region near 4000\AA\ across our entire redshift range. These strong Balmer breaks (all at early cosmic time $z>5.5$) are remarkable, in particular because we simultaneously can rule out that the breaks are degenerate with strong line emission or AGN contamination with the 4$\mu$m medium bands and MIRI data. We will discuss their implication later in Section \ref{sec:massdens}.  

We note that for 154428 in particular, in addition to having evidence of a Balmer break, the FRESCO detection indicates possible evidence of broadened H$\alpha$ with a weak narrow component, which could indicate a weak optical AGN \citep[e.g. similar to the broad H$\alpha$ in the quiescent galaxy confirmed at $z=4.6$ in][]{Carnall2023}. Unfortunately, we are unable to robustly rule out the presence of an broad-line region in the other sources with FRESCO coverage, since any broad components may be dust obscured and below the FRESCO detection limit.

To summarize, the MIRI colors of AGN-dominated objects should be red, however, since we don't find evidence for that, we suspect that the light is instead dominated by stars.
However, the conclusion that the LRDs in our sample are dominated by stellar emission is not definitive. In Section \ref{sec:discussion}, we will show results including and excluding these sources from the sample, and discuss these results in the context of the assumption that these galaxies are dominated by stellar emission.

\subsubsection{Excess UV emission}\label{sec:uvex}
As a final note, we discuss the apparent presence of a restframe UV ``excess" in a handful of our sources: flux which is not easily modeled without a secondary SED component (either by unobscured SF, or, scattered UV light from an AGN). We find clear evidence that composite SEDs are needed to explain the flat UV slopes in a number of objects (90354, 120484, 203749; and potentially also 81400, 132229, 183348). Our selection based on very red F150W--F444W colors may have rejected a number of LRDs with more obvious needs for composite SEDs.  The slopes of the UV SEDs are at low S/N, but nonetheless consistent with typical slopes of either AGN or  young stars \citep[see discussion in][]{Greene2023}. Thus we cannot rule out models that have been proposed for similar sources with our data \citep{Labbe2023b, Matthee2023, Barro2023}. Spectroscopy has now confirmed that the continuum slope alone cannot easily differentiate between a SF or AGN origin using similar samples \citep{Greene2023} nor their intrinsic luminosity  (which is degenerate with fraction of light scattered). Thus we can't hope to do better with photometry, and take the presence of excess UV emission as an indication that scattered light from either an AGN or SF may contribute. 
We note that, if the origin of the UV flux is indeed from a frosting of unobscured star formation, the amount of SF is very small (i.e. these objects are barely detected in the very deep JADES imaging, and we measured the typical implied SFR(UV) based on the flux is $<$1 M$_{\odot}$/yr). Thus any assumption about the origin of the UV flux will not impact our results in Section \ref{sec:sfrd}. Pursuing an explanation for the UV emission is outside the scope of this paper.

\begin{figure}[t]
\includegraphics[width=0.5\textwidth]{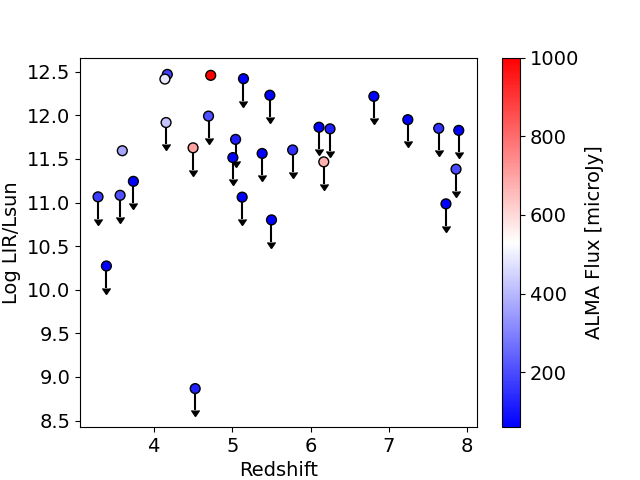}
\caption{\LIRLsun vs redshift of our sources with MIRI coverage that go into our estimate of the cosmic SFRD. For comparison, the average RMS limit of the ASAGAO imaging used is 60 $\mu$Jy and GOODS-ALMA is 180 $\mu$Jy. The majority of sources are inferred to be upper limits to the \LIRLsun based on their non-detection upper limits from the ALMA imaging. }\label{fig:LIR}
\end{figure}

\begin{figure*}
\includegraphics[width=1\textwidth]{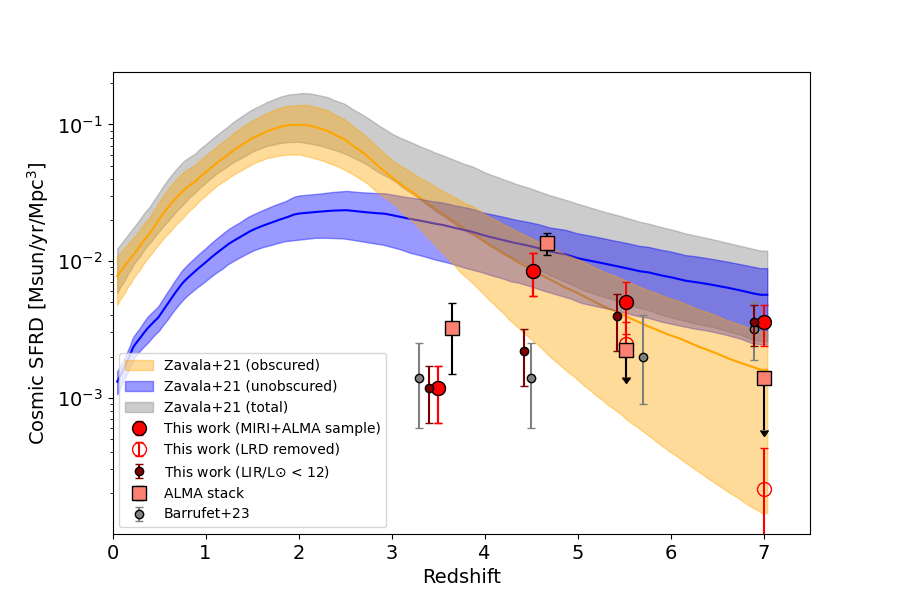}
\caption{The cosmic SFRD of our full H-dropout sample in the MIRI footprint (red points). For comparison we show far-infrared measurements (obscured SFRD; orange shaded region) based on \citep{Zavala2021}, along with the UV-based compilation in that work (unobscured, uncorrected for dust; blue shaded region). 
We also plot the SFRD contribution from our sources with \LIRLsun$<$12 (maroon points). 
Although our MIRI data suggests our LRD subset are stellar-dominated, we show the SFRD when removing them since their nature is uncertain (open red circles). 
We also include an estimate based on the average ALMA flux from stacking the 1.1 mm imaging scaling with a hotter dust template (peach squares). }\label{fig:CSFRD}
\end{figure*}

\section{Discussion: What fraction of the galaxy census was missed due to "dark" galaxies?}\label{sec:discussion}

In this section, we explore how these previously-missed galaxies may contribute to the star formation and stellar mass budget of the early Universe from $3<z<8$, a regime that was previously incomplete among infrared measurements, and which could not be probed uniformly with earlier data. Given the results of Section \ref{sec:results:mirialma}, for this section we only consider sources for which we have MIRI and ALMA constraints, since we find major uncertainties that can change the mass and SFR estimates by up to 1 dex if we have only  HST+NIRCam photometry.

\subsection{The cosmic star formation rate density}\label{sec:sfrd}

Here we estimate the cosmic star formation rate density (SFRD) contribution from these sources (a previous estimate has been made for a similar sample by  \citealt{Barrufet2023}; we now include sub-mm and MIRI data that improves SFR constraints compared to HST+NIRCam data; see Section \ref{sec:results:mirialma}). 
Given that our sources are relatively bright in the detection band compared to the very deep JADES imaging  (F444W S/N $>$ 20) and the relative depths of our F444W and F150W imaging \citep[][]{Eisenstein2023} we find we are sensitive to colors F150W--F444W$<$3.2 even for the fainter sources that are not detected at F150W.  Thus, we expect that selection of H-dropouts is complete for sources brighter than our limiting F444W AB magnitude of 29.4.

We do a simple estimate of the total SFRD by summing the total SFR among our sample (i.e. the SFR during the most recent 30 Myr, as derived from our SED modeling) divided by the cosmic volume within several redshift bins: 
$z\sim$3.5, 4.5, 5.5, and 7 (with $\Delta z\sim1$, except for the highest redshift bin which has width $\Delta z\sim2$). Without completeness our measurements should be considered lower limits, although we pick a high S/N where we are likely complete in both magnitude and color. 

These results are shown as red circles in Figure \ref{fig:CSFRD}, and are compared to the obscured SFRD and unobscured (and un-dust-corrected) SFRD from the MORA survey \citep[orange and blue curves;][]{Zavala2021, Casey2021}. First, we find that at $3<z<4$, optically faint galaxies make up a relatively small fraction of the obscured contribution to the SFRD. This is similar to cosmic noon where the SFRD is still dominated by brighter sources such as sub-millimeter galaxies \citep[e.g.][]{Dudzeviciute2021}, and is consistent with earlier findings \citep[e.g.][]{Wang2019, Sun2021}. 
However, at $4<z<6$, we find that the missing population identified by our H-dropout selection likely contributes non-negligibly to the obscured fraction of the SFRD. We find that in these two redshift bins, 
our full sample is comparable to the total obscured fraction of cosmic SFRD characterized using existing ALMA and far-infrared observations (orange line). 

Since ALMA and far-infrared detections become sparse at $z>3$, the orange region representing the obscured SFRD is measured by combining individual bright detections along with an extrapolation of the infrared luminosity function to faint infrared luminosities \citep[][]{Zavala2021}. 
Since the shape of the infrared luminosity function at such early times is relatively uncertain \citep[in particular at the faint end, \LIRLsun$<$11, where a number of studies report significant differences, e.g.][]{Koprowski2017, Gruppioni2020, Fujimoto2023, Barrufet2023a, Zavala2021, Traina2023}, our data provides an opportunity to compare the contribution of populations below ALMA detection limits to that typically extrapolated from the luminosity function.

To make this comparison, we put our sample in context of existing far-infrared measurements at $z>3$ and estimate their total infrared luminosity  by integrating the maximum likelihood {\tt prospector} model between restframe 8-1100$\mu$m. We then convert this to SFR using the indicator based on total infrared luminosity of \citet{KennicuttEvans2012}. We find that almost all of our sources are below the detection limits of earlier sub-mm surveys at $z>3$, 
with the majority of our sample having upper limits to \LIR\ of $9<$\LIRLsun$<11.8$ (see Figure \ref{fig:LIR}). We have only 4 detected sources with confirmed ALMA flux consistent with \LIRLsun$>$12.  Since many of our objects are non-detections (and thus upper limits), we are likely reaching the ``extrapolation" regime ($9<$\LIRLsun$<11$ Lsun) of the dust-obscured SFRD at $z>3$ \citep{Zavala2021}.

We focus more specifically on sources with \LIRLsun $<$12, to make a direct comparison with the contribution from similar-luminosity populations in \citet{Zavala2021}. That work determined that from $4<z<7$, the fraction of the obscured SFRD contributed by \LIRLsun$<$12 sources was relatively small (20\%) and this fraction was flat with redshift. 
To compare, we also plot the SFRD contribution from our subset of sources with \LIRLsun $<$12  (maroon points in Figure \ref{fig:CSFRD}). 
We find that our measurements indicate that \LIRLsun$<$12 sources make up a higher fraction of the obscured SFRD (26\%) at $4<z<5$, compared to 20\% estimated by Zavala et al (a relatively minor factor of 1.3 increase). However, at $z>5$, the fractional contribution from sources with \LIRLsun $<$12 is much larger, and is comparable with the total obscured SFRD previously estimated. Compared to the earlier estimate of flat 20\% fraction from \LIRLsun$<$12, this suggests that the contribution of \LIRLsun$<12$ sources could be underestimated a factor of $\sim5$ at $z>5$. In comparison to the total obscured contribution from \citet[][orange curve]{Zavala2021}, a factor of 5 increase in the contribution from \LIRLsun$<12$ sources could double the obscured fraction of the SFRD at these redshifts.

While we note that major uncertainties exist in our measurements (see next Section), our data indicates that the overall census of dust obscured SF from earlier studies could be underestimated. Our findings are consistent with and similar to the estimates based on pre-JWST datasets that find 
higher SFRD at early times \citep{Fujimoto2023, Algera2023, Traina2023}, in particular when including estimates based on existence of optically faint or or various ``dark" sources \citep{Williams2019, Gruppioni2020, Talia2021}.

Based on our data, it may also be the case that the dust obscured star formation is underestimated at $z>6$. We find that at $z\sim7$ the SFRD derived using the SED-modeling is log$_{10}\rho_{SFR}=-2.44^{+0.12}_{- 0.18}$ \Msun yr$^{-1}$ Mpc$^{-3}$. This is consistent with \citealt{Algera2023}, but in excess of estimates in both \citealt{Zavala2021} and \citealt{Barrufet2023a}. However, the fraction of our sources which are classified as candidate LRDs at these redshifts by color selection is high. While our analysis in Section \ref{sec:AGN} indicates that the NIRCam and MIRI photometry used to measure their stellar populations is not dominated by the light from the AGN, as a conservative estimate we also calculate the SFRD assuming the star formation contribution of these candidate AGN cannot be robustly determined. Thus we plot a second estimate with these LRD sources removed (open red circles). We find that in doing so, at $z>6$ the obscured SFRD is substantially lower than implied by previous studies using NIRCam and HST data alone \citep{Barrufet2023}, and that the obscured SFRD contributed by our own sample of dark galaxies would be substantially lower. We thus caution against overinterpretation of this measurement, and note that a complete assessment of the SFRD at $z>6$ from dust obscured star formation using JWST selected samples will  likely remain uncertain until we understand the true nature of LRDs.

\subsubsection{Impact of modeling assumptions on the cosmic SFRD}

Despite the panchromatic data and limits used in our SED-fitting, our {\tt prospector}-based SFRs may still be underestimating the intrinsic amount of SFR due to the model assumptions used to infer obscured star formation (namely, that the dust emission model assumed by {\tt prospector} inherently prefer cold dust temperatures at the default prior settings). However, a wealth of evidence now points to hotter dust temperatures among compact, low metallicity systems that make up a higher fraction of the population at high-redshift \citep[T$_{dust}\sim$ 40--60 K; e.g. ][]{deRossi2018, Sommovigo2020, Sommovigo2022, Bakx2020, Schreiber2018dust, Faisst2017, Behrens2018}. The effect of assuming colder dust temperature is to underestimate the SFR, since at fixed 1.1mm flux, the obscured SFR can be higher if dust is hotter. While we have adjusted our priors to allow higher dust temperatures (our {\tt prospector} fits yielded a typical T$_{\rm dust}\sim37$K), this is still below some empirical constraints from dusty galaxies at high redshifts. 

Thus, to test the impact of hotter dust on our estimates, we return to our ALMA image stacking that we used to estimate the average 1.1mm flux in the same redshift bins  (see Section \ref{sec:dsfg}).  Now including the detected sources in the stacks in order to assess the true average flux per bin, we find that in the above redshift bins, the stacked fluxes are 83$\pm$44, 213$\pm$39,  2$\pm$48, and 25$\pm$38 $\mu$Jy/beam. 
However, to interpret the stacked 1.1 mm flux, we now assume a hotter dust SED template. To estimate the corresponding average \LIR, we scale a far-infrared template for Haro-11, an analog for a high-redshift dusty SFG that factors in elevated dust temperatures that more realistically describe compact SFGs at $z>4$ \citep[T$_{dust}\sim47$K, emissivity index $\beta\sim1.9$;][]{Lyu2016}. To obtain the \LIR\ and SFR we integrate the template scaled to the ALMA stacked flux and integrate between restframe 8--1100$\mu$m and convert to SFR as earlier.

We find that our stacked ALMA fluxes are consistent with \LIRLsun= 11.6, 12, $<10$, $<11$ for redshifts $z\sim$3.5, 4.5, 5.5, 7, which correspond to average SFRs that are 
69$\pm$37, 152$\pm$28, 1$\pm$31, 15$\pm$23 M$_{\odot}$/yr
per redshift bin. For comparison, the average SFR per bin based on the {\tt prospector} SED modeling is similar within 2$\sigma$ (although systematically underestimated at $z<5$). 
Using the number of galaxies per redshift bin, we translate these average estimates derived from the ALMA stack to an estimate of the total SFR per redshift bin contributed by this sample, scaled to the same signal-to-noise as reflected in the image stacked flux (see peach squares in Figure \ref{fig:CSFRD}). For the highest two redshift bins where stacked flux is not detected, we instead plot 1$\sigma$ upper limits. Generally, we find that at $z<5$ the hotter dust template does increase our measured SFRD, while at $z>5$ the stacks point to lower SFRD than found with the SED-modeling. More data sampling the far infrared SED to constrain the dust temperature would be needed to resolve the discrepancies between the different measurements.

One potential downstream impact of 
{\tt prospector} not allowing for hotter dust in the modeling is that it likely forces a more quiescent (redder) stellar population solution (which, due to higher M/L ratios of older stars, could result in higher mass solutions at fixed luminosity). We note that redder stars is an easy solution to justify for the sources where we identified clear Balmer breaks indicative of older stellar populations. However, in cases without clear breaks, we may need dust emission modeling with hotter dust more typical of dusty galaxies observed at similar redshifts. However, we note the dust temperatures required may be even hotter \citep[e.g.][]{deRossi2018}, given the extremely compact nature of the objects in our $z>6$ sample.

In fact this parameter space (young compact star forming regions driving hot dust temperature) is seen in the nearby Universe, and could potentially be analogous to these sources \citep[e.g.][]{Hainline2016}. Thus, there is likely to be a significant bias at high redshift (where our sizes are also the most compact) that would mean that these estimates of SFRs are biased low for $z >5$. This is because galaxies are more difficult to detect based on their 1.1 mm emission than they would be if local (colder) far infrared SEDs are used \citep[see also][]{Shivaei2022}. Thus, our estimated contributions to the cosmic SFRD may also additionally be underestimated. Additional higher-frequency dust continuum imaging that would constrain the dust temperature would be needed to investigate this possibility further.

\subsection{Stellar mass census: the abundance of massive galaxies} \label{sec:massdens}

\begin{figure}[t]
\includegraphics[width=0.5\textwidth]{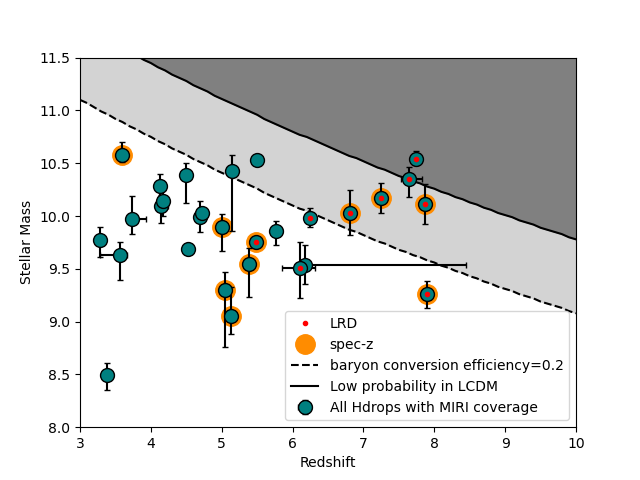}
\caption{The stellar masses vs redshift of our sample of galaxies with MIRI coverage (34 sq arcmin area). For context we include the limiting stellar mass expected for 100\% baryon conversion efficiency (solid line) and 20\% efficiency (dashed line). For six galaxies between $z\sim5.5-8$ we measure stellar masses that likely exceed expectations for a small area, 
even after the stellar masses have decreased by up to an order of magnitude by including MIRI+ALMA data (see Figure \ref{fig:paramcompare}). This comparison to the expectation curves from halo abundance and star formation efficiency show these measurements are likely still overestimated. }\label{fig:mass}
\end{figure}

Despite the findings in Section \ref{sec:meas} that MIRI+ALMA overall result in lower stellar masses, we still identify a remarkably large number of galaxies (16) above $z>3.5$ with \LogM$>$10. The MIRI data also covers a relatively small area of 34 sq arcmin, making the identification of so many massive galaxies unlikely. In fact, 80\% of our objects above $z>7$ have \LogM$>$10.  It has already been pointed out by \citet{Narayanan2023} that stellar masses are essentially unconstrained above $z>7$ (and can dramatically over or underestimate stellar mass) owing to outshining of older stars by young low metallicity stars complicating the reconstruction of the SFH. However, we note that our high-redshift sources are not particularly young (as {\tt prospector} preferred to model them with older stellar populations, either to allow low dust content to fit the ALMA data, or to accomodate clear Balmer breaks in a number of sources). However, even at redshifts below this problematic epoch noted by \citet{Narayanan2023}, the number of high mass objects are also surprisingly, and problematically, high.

To demonstrate this, we show the stellar mass vs redshift for our MIRI+ALMA sample in Figure \ref{fig:mass}. For context, we also plot the expected stellar mass limit (i.e. mass where we only expect one halo, given our MIRI survey area), based on the halo mass function evolution with redshift. To estimate this we use the halo mass function calculator {\tt HMF} published by \citet{Murray2013} and assume the halo mass function of
\citet{Behroozi2013}. We use the limiting halo mass 
to convert to a limiting expected stellar mass by assuming a fiducial baryon conversion efficiency into stars (0.2, dashed line), and for the limit where 100\% of baryons converted in to stars (solid line). While our sample size is small, these curves help to indicate the number of sources whose mass is probably physically unlikely, under the typical assumptions of $\Lambda$CDM. We find one source at $z=5.5$ and five sources at $z\gtrsim6.5$ that are either improbably high mass, or, imply extremely efficient baryon conversion ($>$20\%) compared to typical assumptions. 

In Sections \ref{sec:specz} and \ref{sec:meas} we noted that a number of the photometric and spectroscopic solutions allowed revision to lower redshift upon detailed inspection (thus also allowing further reduction in the stellar masses even after inclusion of MIRI+ALMA data). However, in particular for high mass objects 90354, 121710, 200576 and 219000, we were unable to justify the possibility of lower redshift solutions. Further, while we identified a tentative emission line for 132229 (where the spectroscopic solution is $\Delta z\sim1$ lower in redshift than the photometric redshift inferred with {\tt prospector}), this galaxy still remains at a problematically high mass. We note that the last photometric candidate with unprobably high mass, 203749, has an uncertain redshift solution (we identified two comparable solutions at z=2.41 and the $z\sim7$ one we use for the analysis; we note that its LRD colors and unresolved morphology would make it an outlier among known $z<3$ galaxies). Both redshift solutions are poor fits to the restframe UV photometry due to the presence of UV excess (Section \ref{sec:uvex}). While we do not understand the nature of this object, we conclude it is likely to have alternative explanations besides a high-redshift massive object and do not consider it in the following discussion.

\begin{figure}[t]
\includegraphics[width=0.5\textwidth]{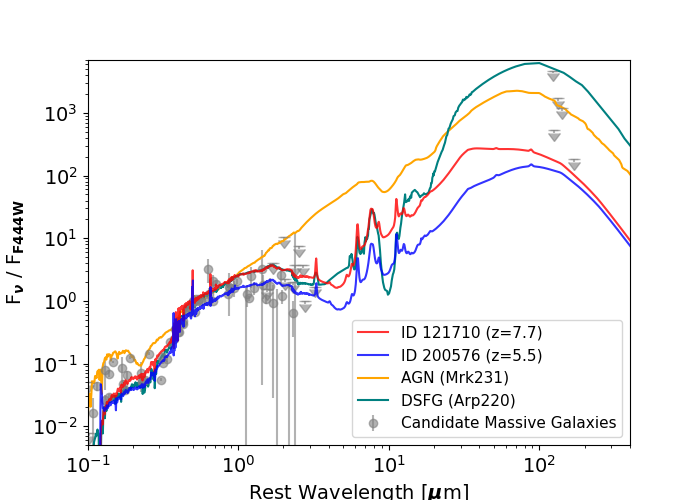}
\caption{The normalized restframe SEDs of the five most confident candidate massive galaxies in Figure \ref{fig:mass} which include four LRDs, plus a Balmer break non-LRD, at $z\sim5.5-8$ (gray points; 203749 is excluded).  All sources exhibit similar double-break SEDs exhibiting well constrained Lyman and Balmer breaks, evidence for a turnover at 1.6$\mu$m, and are consistent with high mass weighted age ($\sim500$ Myr) and moderate dust (Av$\sim$1).
For comparison are a dust-reddened AGN (yellow), a dusty star forming galaxy \citep[teal;][]{Polletta2007}.  While all SED types look very similar in the rest-frame optical (probed by HST+NIRCam) the SEDs diverge at $\lambda_{rest}>1\mu$m, where MIRI shows a clear flattening due to the stellar bump, and ALMA rules out cold dust emission.
}\label{fig:redsed}
\end{figure}

For the other 5 objects with less redshift ambiguity however (all are LRDs except 200576, and are relatively red) we find that under typical SED-modeling assumptions, these galaxies remain well above stellar mass expectations for their redshifts. None of them are well fit with the Sonora-Cholla brown dwarf atmospheric models, as their 2 - 3$\mu$m colors are redder than what is observed in ultracool dwarfs. Three of these sources also meet the double-break criteria used to identify extremely massive candidates in \citet{Labbe2023a}, and exhibit clear evidence of Balmer breaks (as discussed in Section \ref{sec:psb}). In fact, they all have very similar restframe SED shapes, with a strong balmer break, clear turnover at 1.6$\mu$m due to the stellar bump, and with deep ALMA limits, are best fit by moderately aged (mass weighted age $\sim$500 Gyr) and dusty (Av$\sim$1) SEDs (see Figure \ref{fig:redsed}). Due to our excellent wavelength sampling including at least four NIRCam medium bands and seven MIRI bands, it is extremely unlikely we would not be able to account for emission line boosting. All except 90543 exhibit significant MIRI detections even out to 10$\mu$m, making it less possible that we still overestimated the restframe near-infrared continuum due to having only upper limits from MIRI.

While there exists the possibility that these stellar masses are accurate, and these objects trace a population of galaxies that demonstrate highly efficient mass growth  
\citep[e.g.][]{Labbe2023a, BoylanKolchin2023, Xiao2023b}, we offer a few other possible explanations that would require additional data to disentangle.  We note that five out of the six high mass sources are LRDs, and there remains large ambiguity about the nature of such objects. However, despite the evidence presented in Section \ref{sec:AGN} that the rest-optical SED is not dominated by a rising AGN continuum, it remains possible that the redder restframe optical SEDs of LRDs are (still) poorly described by stars alone.  There is also the obvious possibility that various stellar population assumptions are incorrect at higher redshift (i.e. locally-calibrated models do not apply to high redshift phenomena where the stellar properties and environmental conditions may be dramatically different). Below we outline a few hypotheses based on the potential impact of AGN, and modeling assumptions, that could be tested with future high resolution spectroscopy.

\subsubsection{Sub-dominant AGN contamination}

One possibility is that some sub-dominant (but perhaps still impactful) fraction of AGN flux in the rest-optical and near-infrared is still driving up the stellar mass estimates of these four LRDs (ignoring for now ID 203749 with ambiguous redshift). We showed in Section \ref{sec:AGN} that the CIGALE modeling of the rest-optical SED of LRDs predicted much larger discrepancies with the MIRI data, primarily due to their redder NIRCam longwave photometry. The AGN model that overpredicted the MIRI flux included two components: the black body emission from the accretion disk in the rest-optical (e.g. Big Blue Bump), plus hot+warm dust emission in the near and mid infrared. While we have shown that the strongly rising continuum from the hot+warm dust emission component does not agree with our MIRI data, this does not directly constrain the contribution from an accretion disk, unless the relative contributions are physically linked. However, some examples where the relative contribution of an accretion disk could be larger than we assumed \citep[e.g. hot or warm dust deficient AGN;][]{LyuRieke2017}. In that case, the black body emission from an accretion disk could still contribute a small fraction of light to the rest-optical without being so visible in the restframe near-to-mid-infrared. This could be plausible \citep[and could be an explanation given the high confirmation fraction of broad line AGN among LRDs with these colors;][]{Greene2023} but we do not have the data or evidence to determine whether this is the case. AGN contribution from an accretion disk is not definitively ruled out by our relatively flat MIRI SEDs. This could in part explain the large inferred stellar masses. We are not aware of analogous sources at lower-redshifts, except for dust-deficient quasars.
We note that while strong rest-optical emission lines driven by AGN 
can impact the interpretation of high stellar mass \citep{Endsley2023, Kocevski2023}, our inclusion of at least four NIRCam medium bands plus the longer-wavelength MIRI data to anchor the SED beyond the wavelength range with the most contamination is a strong mitigator of this uncertainty.

Future spectroscopy of this sample (in particular to measure restframe optical H$\alpha$, H$\beta$, [OIII] equivalent widths and line profile shapes) could potentially reveal the origin of the rest-optical continuum and validate the photometric measurements. In addition to confirming the likely presence of an AGN accretion disk via the broad lines, \citet{Greene2023} use the relatively low H$\alpha$ EWs to argue that the continuum is not likely dominated by dust-obscured young stars.  However, low EWs could also be explained by older stars and low level star formation over the past 10 Myr (which is consistent with our SED modeling results). Given that these sources typically have high attenuation (Av$\sim1-3$), deeper spectroscopy than was obtained with FRESCO is likely required to adequately measure both broad and narrow line components (if they exist). High spectral resolution will also improve the differentiation between emission line broadening due to outflowing gas vs AGN.

\subsubsection{Potentially errant modeling assumptions}

We note that we use mostly conservative assumptions in our modeling, including a flexible-slope attenuation curve that allows extra attenuation in the UV to avoid ``hiding" stellar mass with a fixed flat slope model \citep[noting that the assumed attenuation curve slope can impact the recovered mass by almost an order of magnitude;][]{LoFaro2017, Williams2019}. 
Additionally, while non-parametric SFHs have been shown to raise stellar mass by 0.3 dex on average based on representative galaxies at lower redshifts, \citep[e.g.][]{Leja2019} these objects would still remain systematically too high (unless the typical systematic offset is not representative for such extreme and red galaxies). 

A number of works now report that the choice of prior on the SFH can significantly impact the ages and masses inferred by the non-parametric SFH \citep[e.g.][]{Leja2019, Lower2020, Tacchella2022b, JiGiavalisco2022}. Therefore, we test whether our choice of continuity prior for the SFH has weighted against bursty SFH solutions, which could result in lower mass solutions by enabling us to explain the SEDs with a larger fraction of stars at younger stellar ages. We re-run our {\tt prospector} modeling for these six high mass sources instead using the Dirichlet prior \citep{Leja2017}, which allows for sudden and extreme changes in SFR in adjacent time bins (and is a weaker prior on the inferred shape of the SFH compared to the continuity). We find that the differences in stellar mass measured with the burstier Dirichlet prior does not cause a systematic shift in mass of our high mass sample, and further, are all consistent within the uncertainties of the stellar masses based on the continuity prior. We do find that the best SFH shapes do change with the Dirichlet solution and appear more stochastic, (indicating that robust stellar ages will require spectroscopy). This typical difference to the SFR in the most recent 30 Myr of only $\sim$1 M$_{\odot}$, but we note that our most active source among the massive sample (219000) sees a 30\% decrease to its recent SFR with the Dirichlet. 
However, our stellar masses, and thus primary conclusions, do not rely heavily on our prior choice. 

Importantly, while the stellar masses from the Dirichlet prior are still consistent within their uncertainties with those inferred using the continuity prior (typical difference is $<0.1$dex) we note that the uncertainties in mass derived from the posteriors of each set of modeling does not marginalize over these assumptions.  
Unfortunately, this is also the case for a number of assumptions that have gone into our modeling, including IMF shape \citep[see e.g.][]{Wang2023b, Woodrum2023}. Thus, the true uncertainty in stellar mass is larger than implied in Figure \ref{fig:mass}. We conclude that more advanced priors or an alternative IMF, among other assumptions, may be able to help lower the stellar masses and account for this tension. Otherwise, these targets are likely excellent testbeds for studying modeling systematics, or the potential for more exotic explanations in the future with near-infrared spectroscopy.

\section{Conclusions}

We study a sample of optically faint sources at $z>3$ that are below the detection limits of the deepest HST and ALMA surveys to date, and have previously been missed from the galaxy census at $3<z<8$. We find that these sources are relatively abundant within the JADES survey (66) and study a subset of those (29) for which deep multi-wavelength MIRI is also available. Our findings include:

\begin{itemize}
\item The population of red optically faint galaxies are diverse in morphology (including both extremely extended sources as well as compact unresolved sources) and in SED shape, including sources resembling dust obscured star forming galaxies, post starburst galaxies, some objects exhibiting evidence of strong Balmer breaks despite high redshifts.
\item We find that stellar population modeling for sources using HST+NIRCam data alone can result large masses and SFRs. When MIRI+ALMA are included, we find a median decrease in 0.6 dex in stellar mass and median decrease of 10$\times$, for sources  sources where HST+NIRCAM alone infer \LogM$>$10 and SFR$>$100 M$_{\odot}$/yr, respectively.  Thus, caution should be exercised when interpreting the SEDs of very red sources from HST+NIRCam data alone.
\item Our sample includes $\sim$30\% candidate AGN (selected as LRDs) and the fraction of LRDs is 100\% among red galaxies above $z>6.5$. Novel measurements with MIRI out to 25$\mu$m for this population confidently rule out that their very red rest-optical continuum primarily originates as obscured AGN continuum. Instead, evidence for a turnover in the SED between restframe 1-3$\mu$m suggests we are seeing the stellar bump, and the red rest-optical continuum is stellar in origin. We cannot rule out the presence of an AGN that becomes dominant in the restframe mid-infrared SED, which requires longer wavelength data.
\item Noting that AGN are not likely dominating the restframe optical emission, we use our stellar population modeling to assess the contribution to the galaxy census of this previously-hidden population of galaxies. We estimate lower limits to the cosmic SFRD and find that galaxies at \LIRLsun$<$12 and $4<z<6$ may contribute 5$\times$ more to the obscured fraction of the SFR than previously estimated based on extrapolation of the infrared luminosity function, which could effectively double the obscured SFRD in this redshift range.
\item We also assess the stellar masses we measure in the context of the limited area we probe in our survey, finding that 
five sources between $z\sim5.5-8$ have very high mass for our small survey area, 
despite the revision to lower stellar mass provided by the MIRI data. These sources have strong Balmer breaks and SED turnovers consistent with the stellar bump in the restframe near infrared, well described by moderately old and dusty SEDs (age$\sim$500 Myr, Av$\sim$1). We discuss plausible reasons for overestimated stellar masses, based on existing assumptions in stellar population and AGN modeling, motivating future work to characterize the physical properties of very red high-redshift galaxies.
\end{itemize}

\begin{figure*}[t]
\includegraphics[width=1\textwidth]{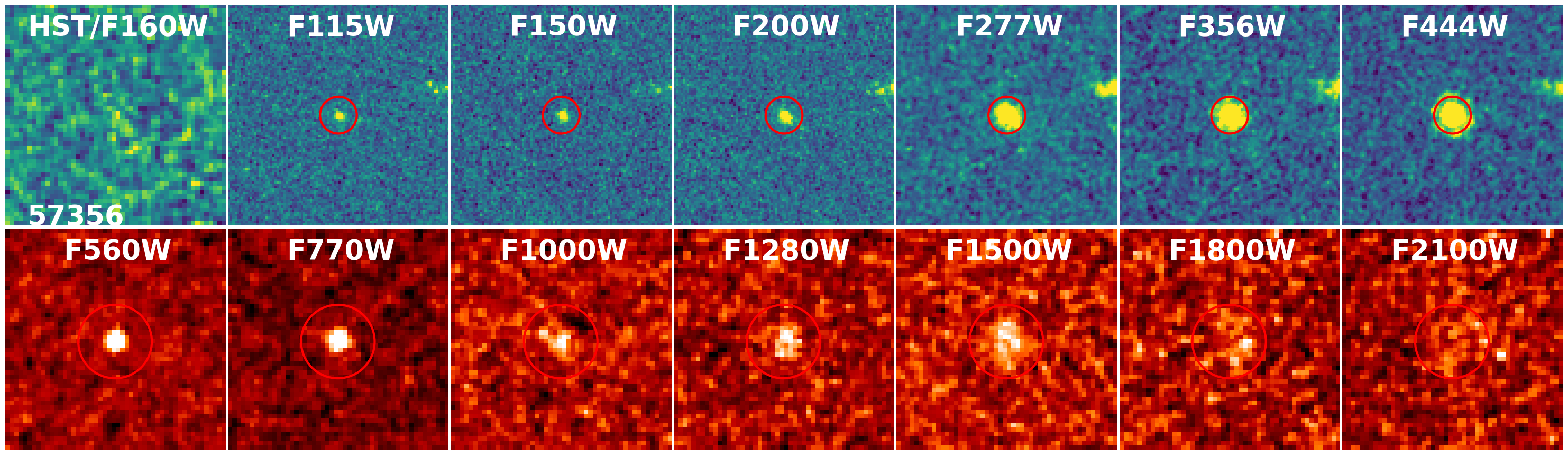}
\includegraphics[width=1\textwidth,trim=10 0 10 10, clip]{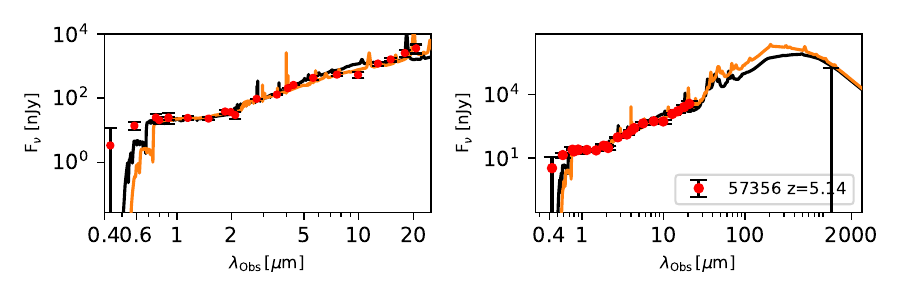}
\includegraphics[width=1\textwidth,trim=10 35 10 10, clip]{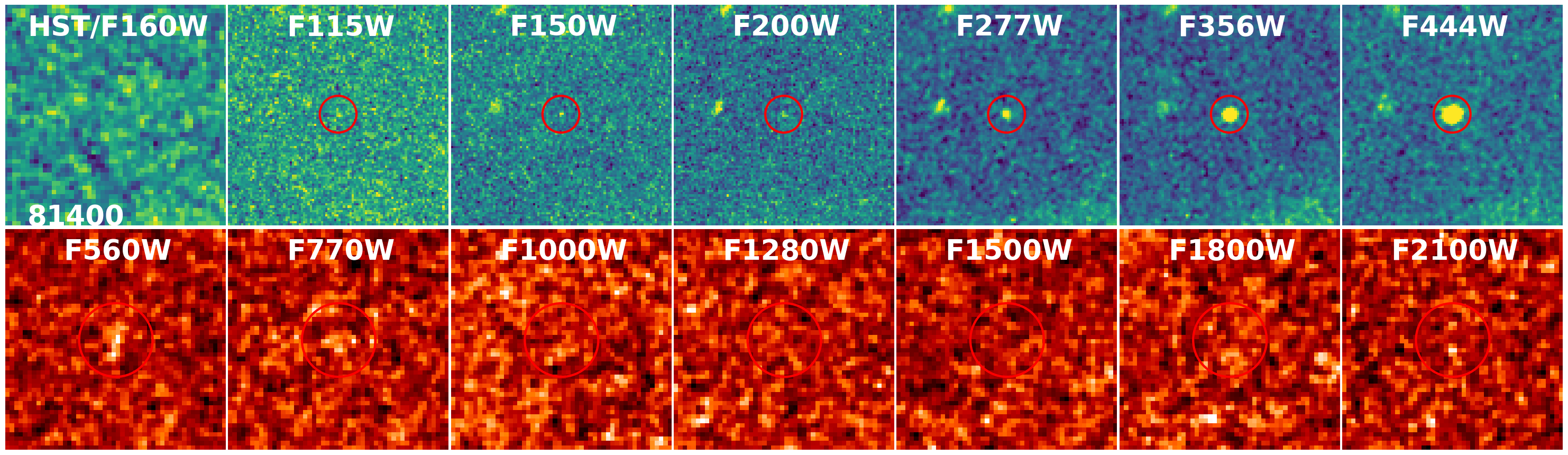}
\includegraphics[width=1\textwidth,trim=10 0 10 10, clip]{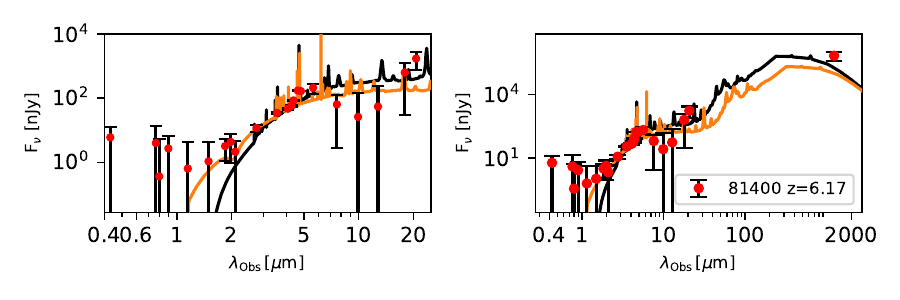}
\end{figure*}
\begin{figure*}[t]
\includegraphics[width=1\textwidth,trim=10 35 10 10, clip]{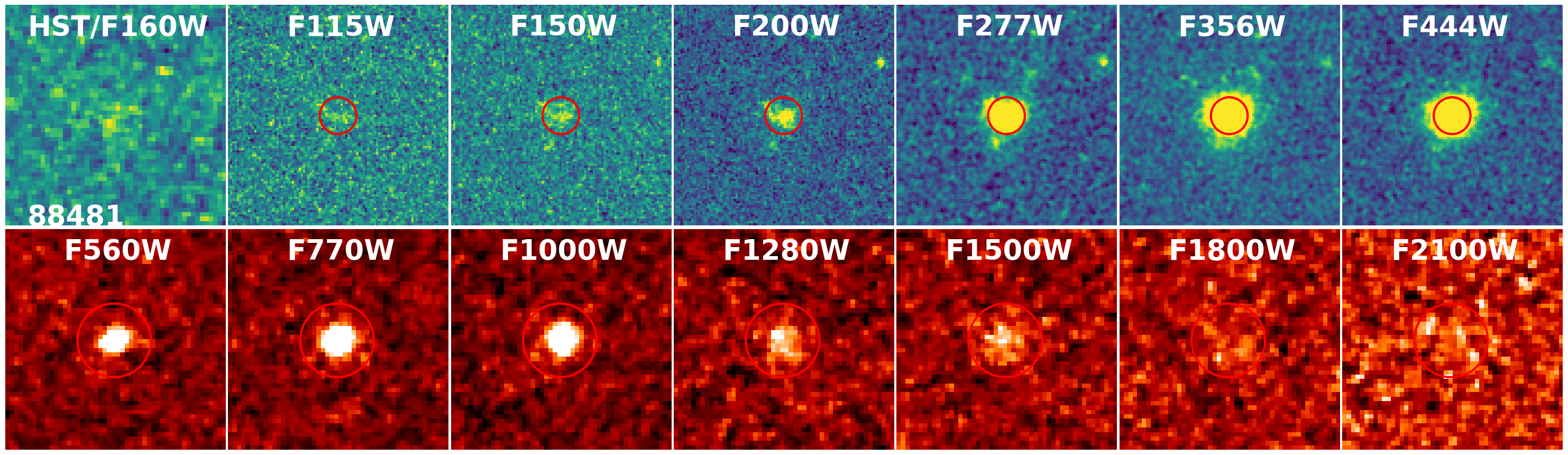}
\includegraphics[width=1\textwidth,trim=10 0 10 10, clip]{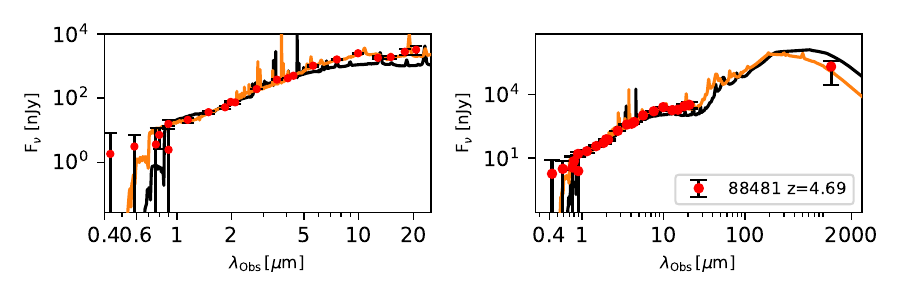}
\includegraphics[width=1\textwidth,trim=10 35 10 10, clip]{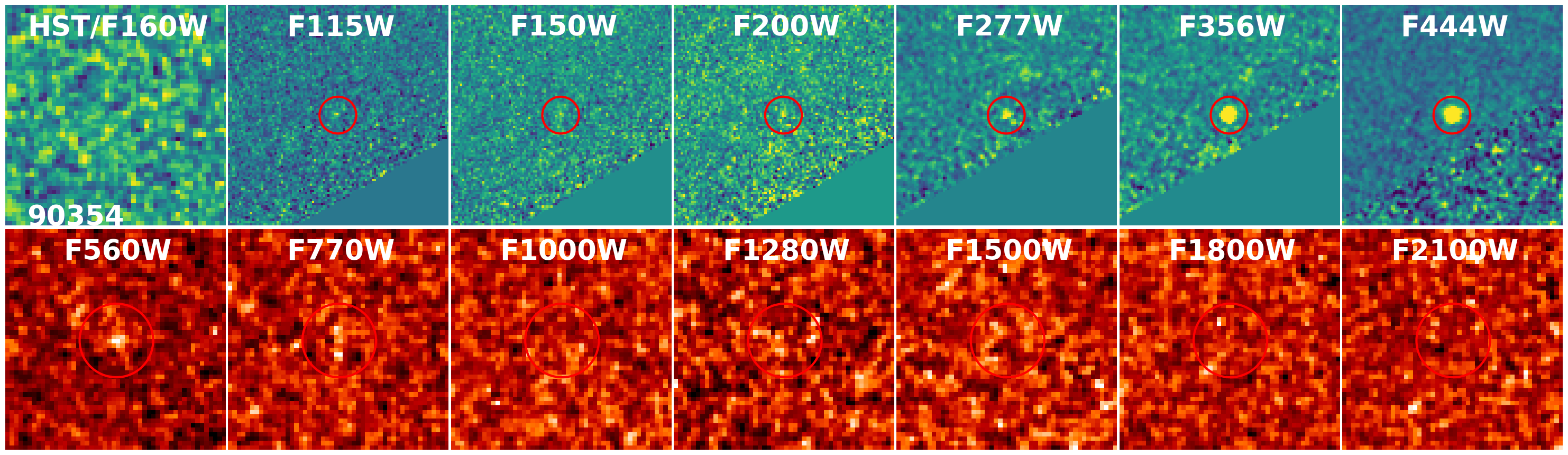}
\includegraphics[width=1\textwidth,trim=10 0 10 10, clip]{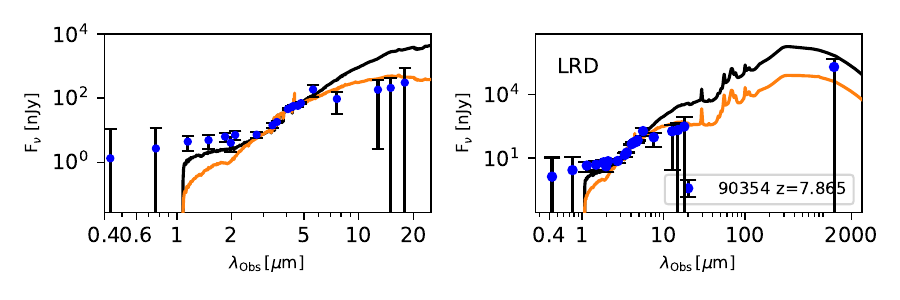}
\end{figure*}

\begin{figure*}[t]
\includegraphics[width=1\textwidth]{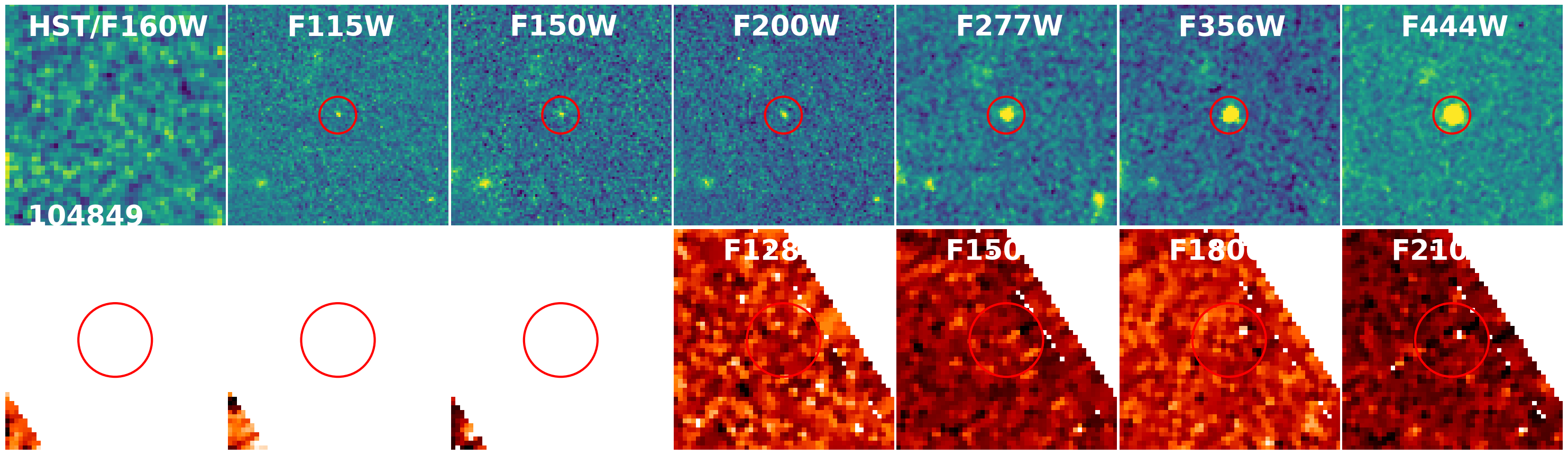}
\includegraphics[width=1\textwidth]{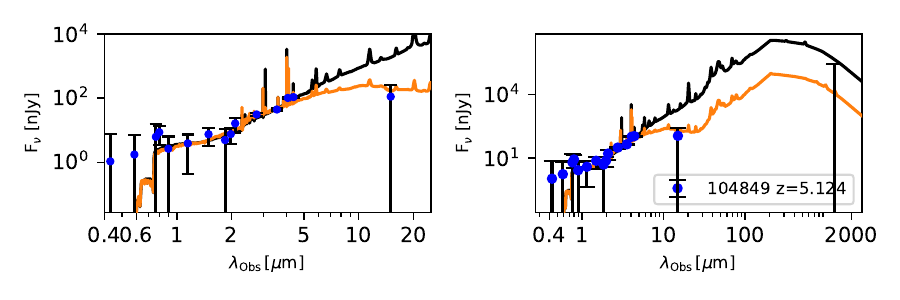}
\includegraphics[width=1\textwidth]{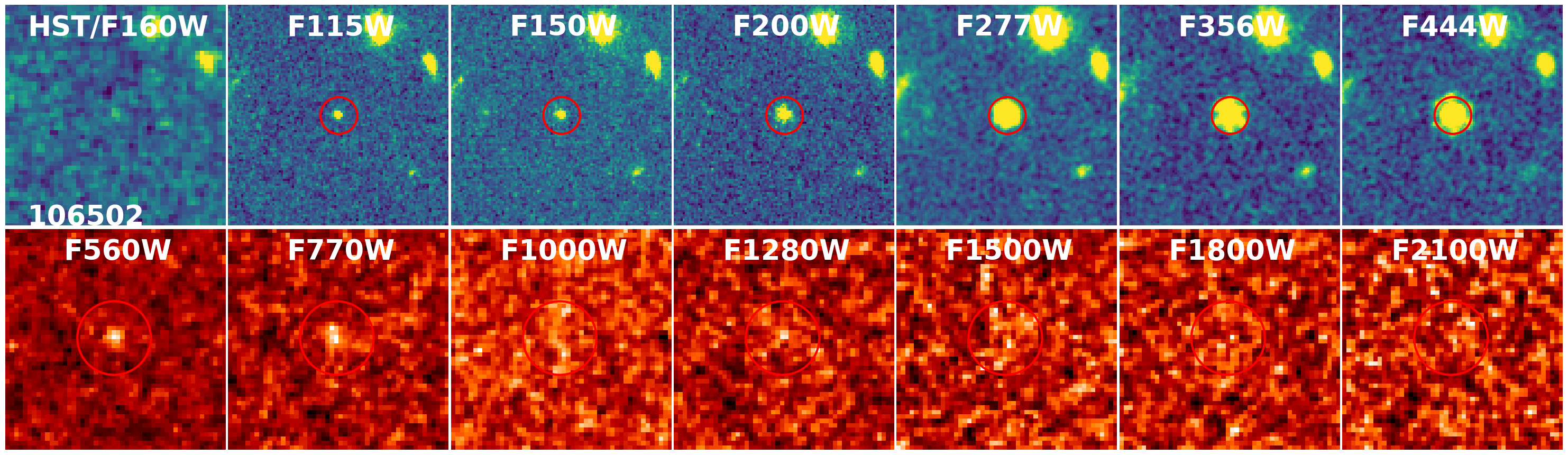}
\includegraphics[width=1\textwidth]{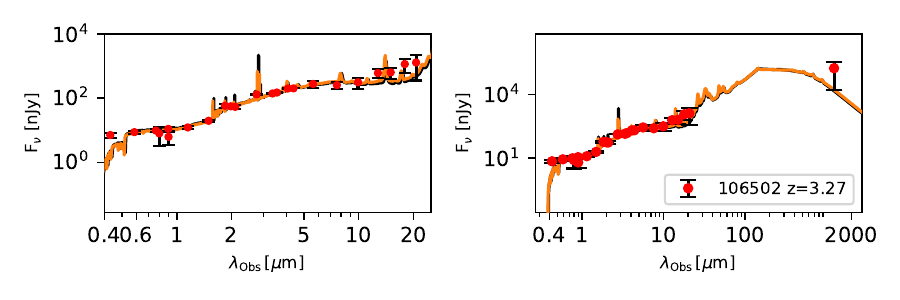}
\end{figure*}
\begin{figure*}[t]
\includegraphics[width=1\textwidth]{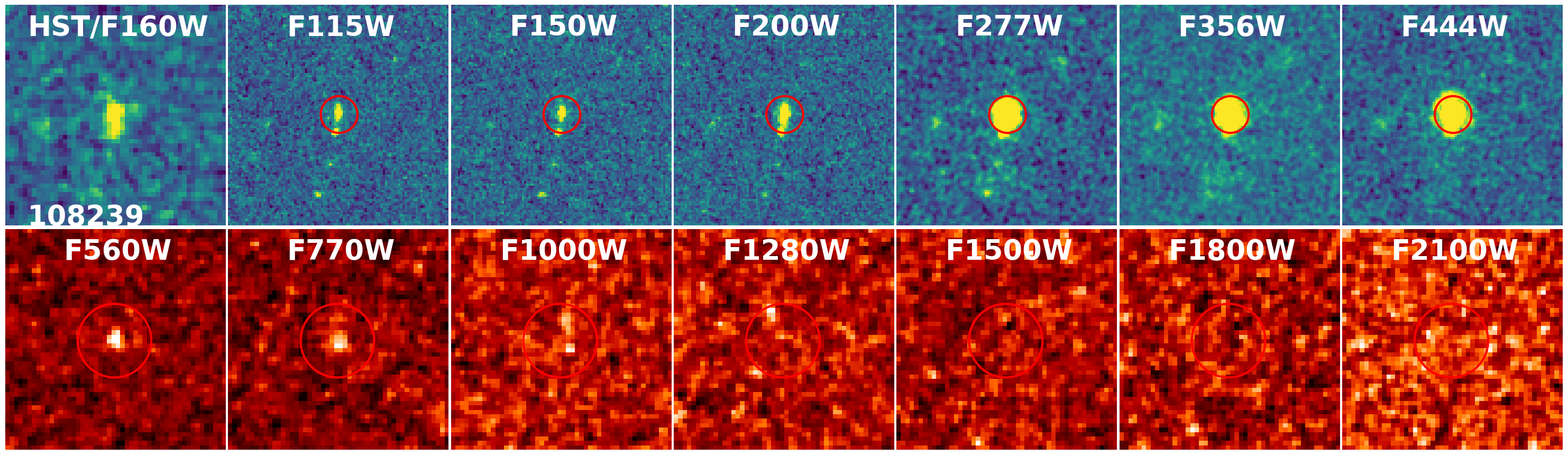}
\includegraphics[width=1\textwidth]{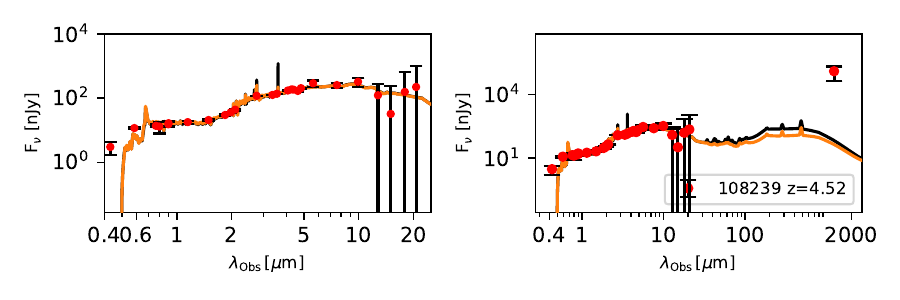}
\end{figure*}
\begin{figure*}[t]
\includegraphics[width=1\textwidth]{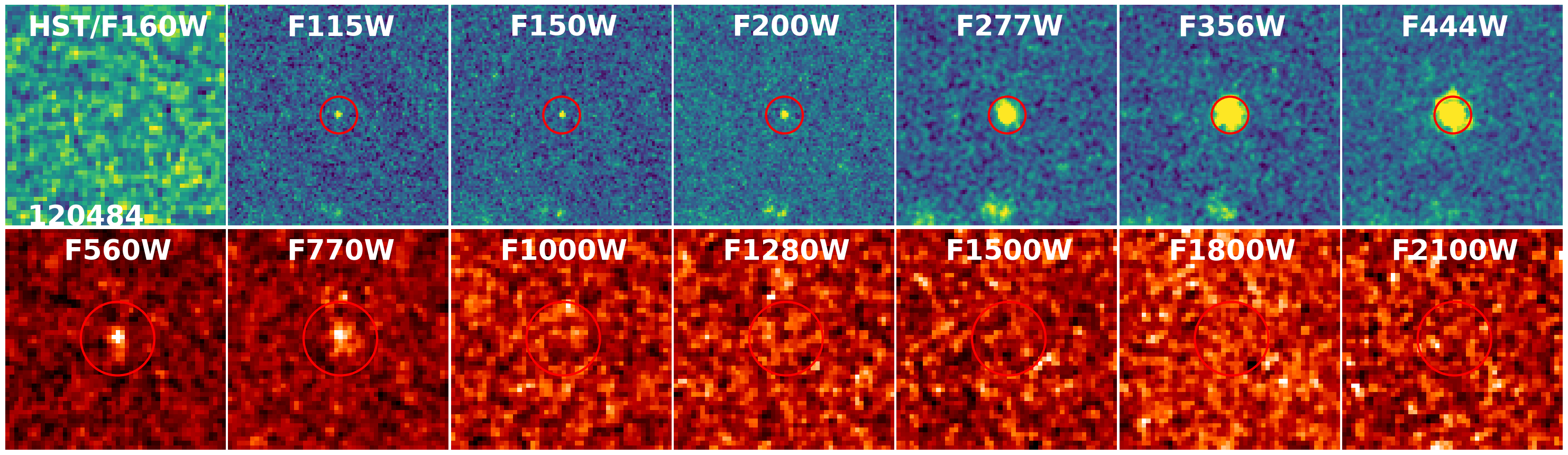}
\includegraphics[width=1\textwidth]{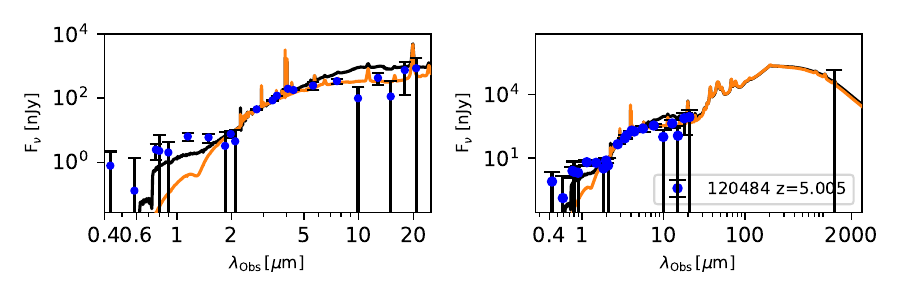}
\includegraphics[width=1\textwidth]{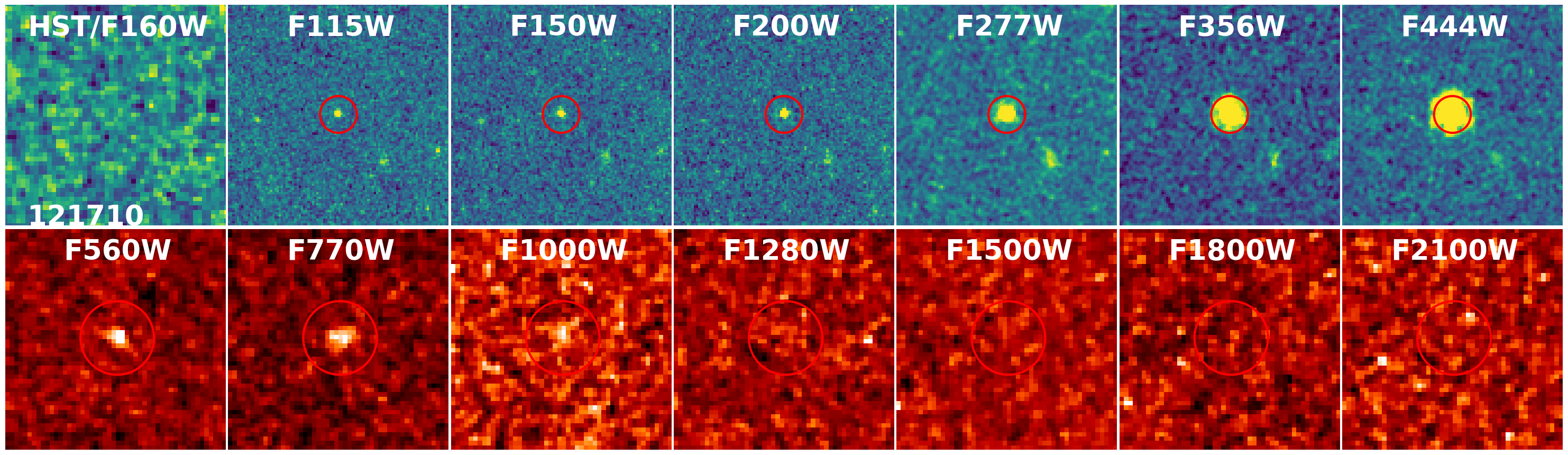}
\includegraphics[width=1\textwidth]{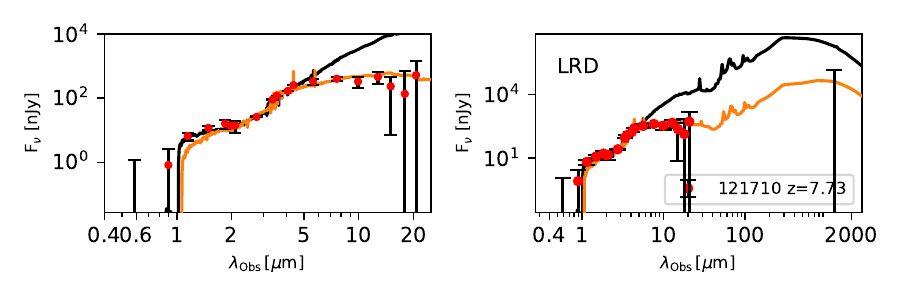}
\end{figure*}
\begin{figure*}[t]
\includegraphics[width=1\textwidth,trim=10 35 10 10, clip]{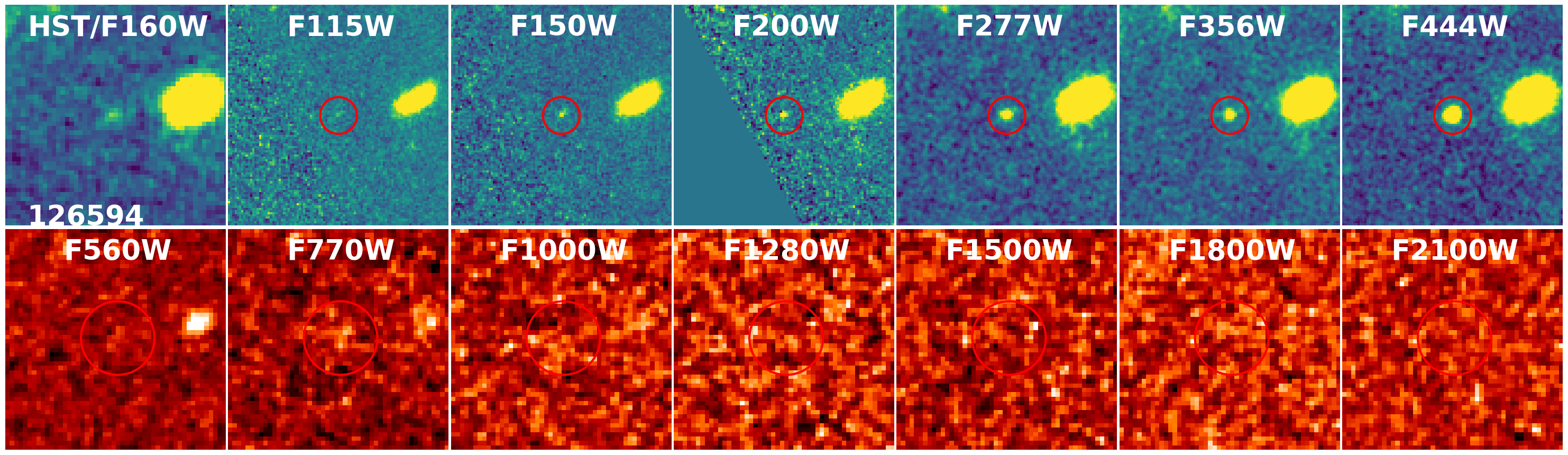}
\includegraphics[width=1\textwidth,trim=10 0 10 10, clip]{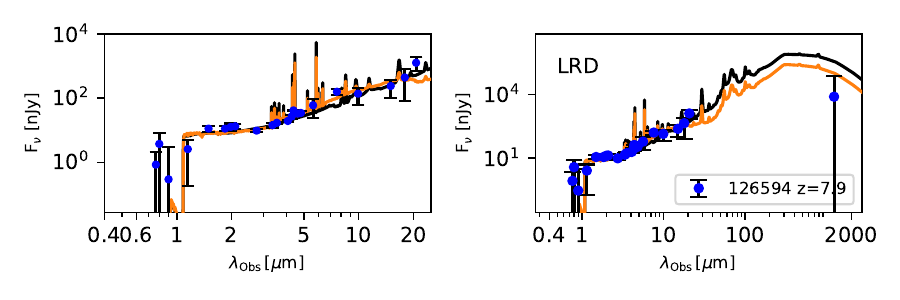}
\includegraphics[width=1\textwidth,trim=10 35 10 10, clip]{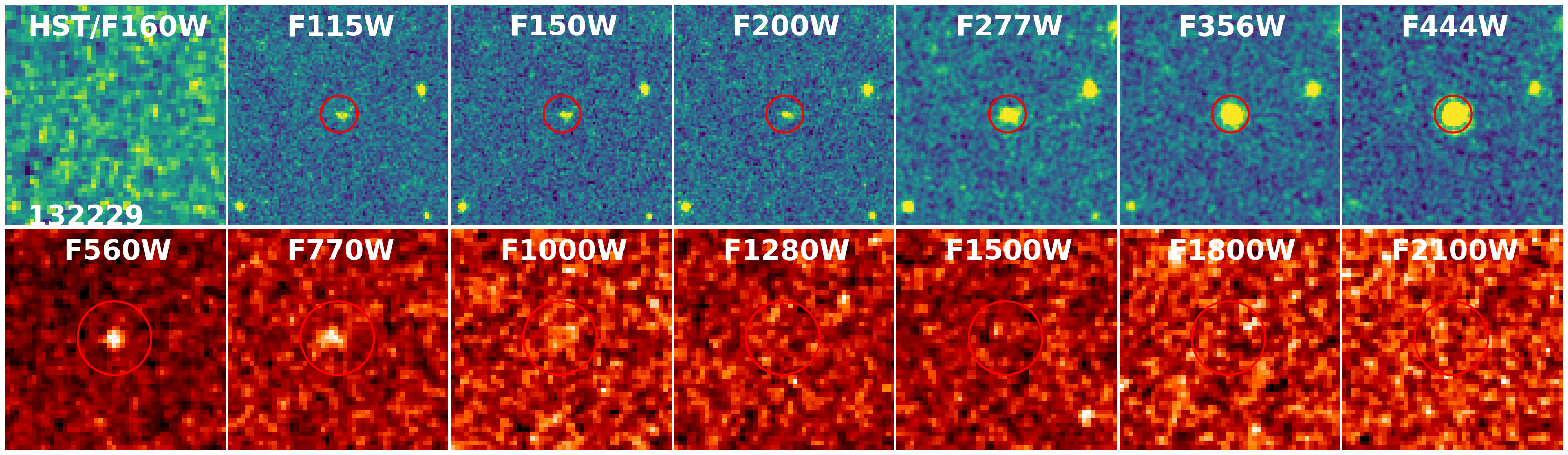}
\includegraphics[width=1\textwidth,trim=10 0 10 10, clip]{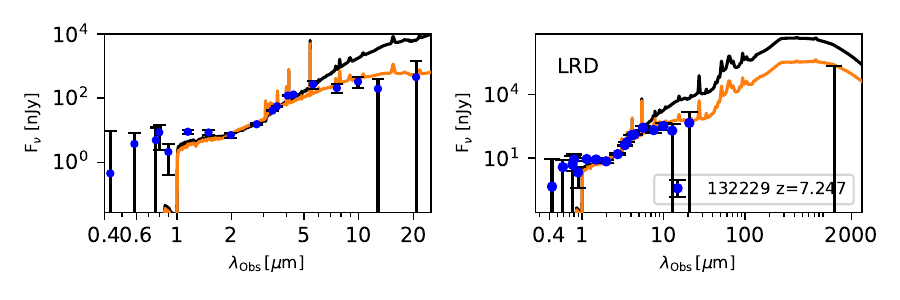}
\end{figure*}
\begin{figure*}[t]
\includegraphics[width=1\textwidth,trim=10 35 10 10, clip]{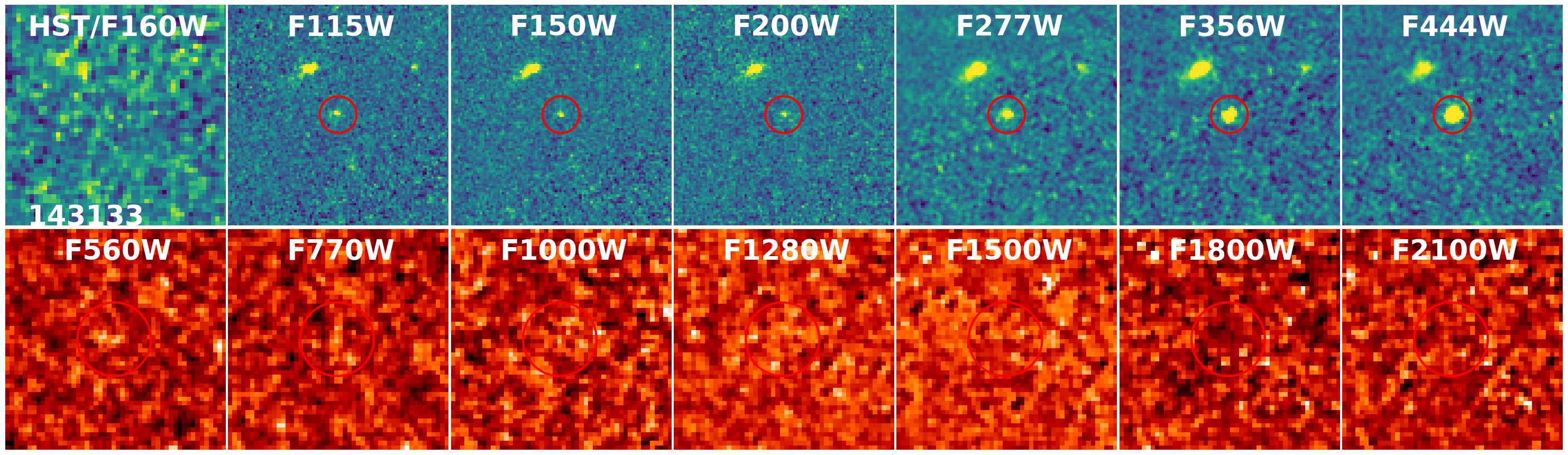}
\includegraphics[width=1\textwidth,trim=10 0 10 10, clip]{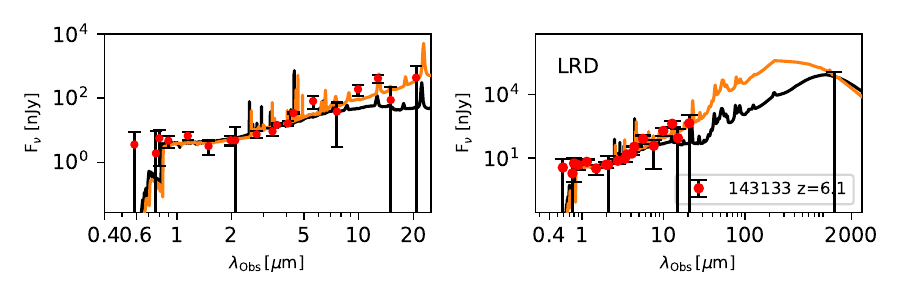}
\includegraphics[width=1\textwidth,trim=10 35 10 10, clip]{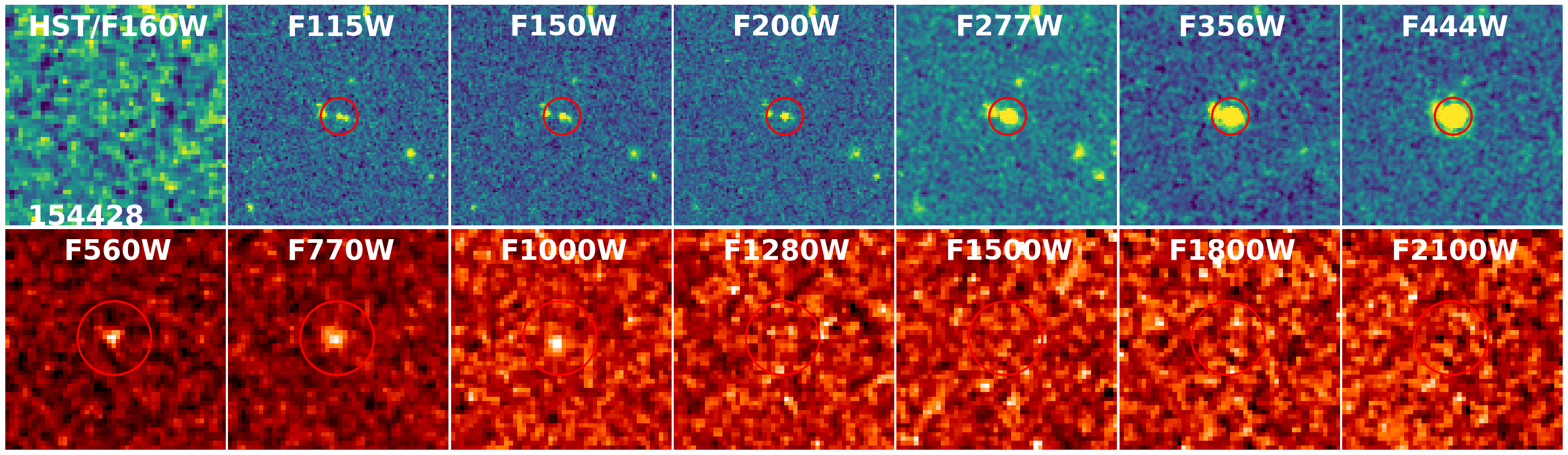}
\includegraphics[width=1\textwidth,trim=10 0 10 10, clip]{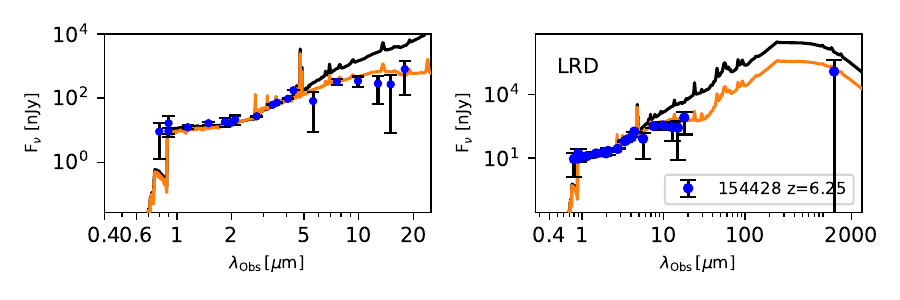}
\end{figure*}
\begin{figure*}[t]
\includegraphics[width=1\textwidth,trim=10 35 10 10, clip]{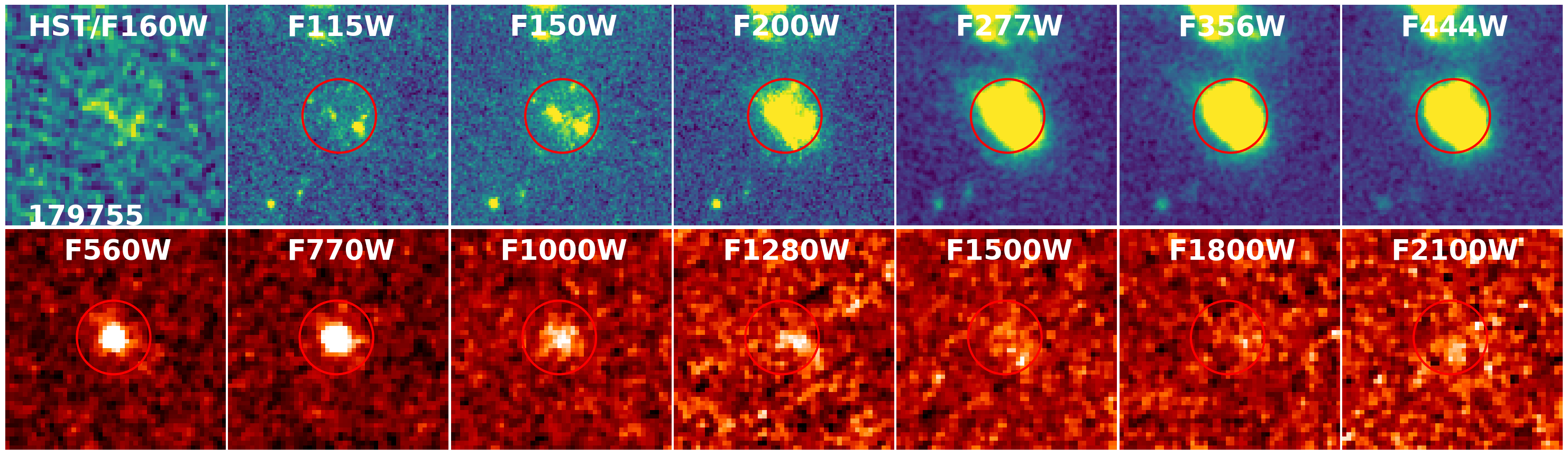}
\includegraphics[width=1\textwidth,trim=10 0 10 10, clip]{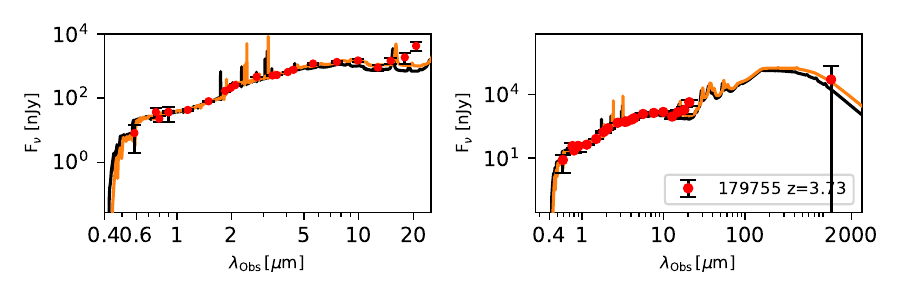}
\includegraphics[width=1\textwidth,trim=10 35 10 10, clip]{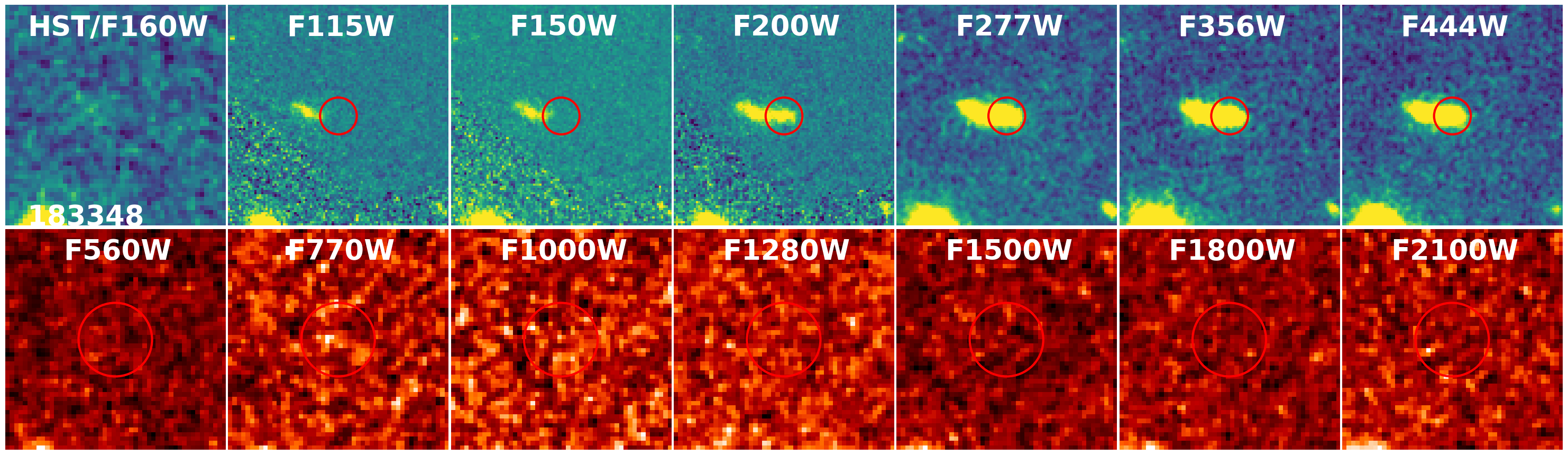}
\includegraphics[width=1\textwidth,trim=10 0 10 10, clip]{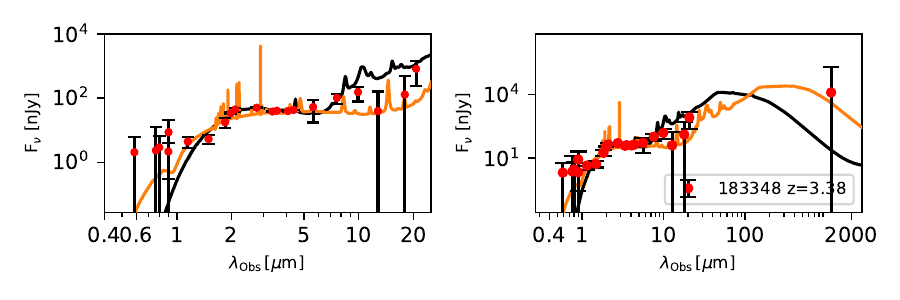}
\end{figure*}
\begin{figure*}[t]
\includegraphics[width=1\textwidth,trim=10 35 10 10, clip]{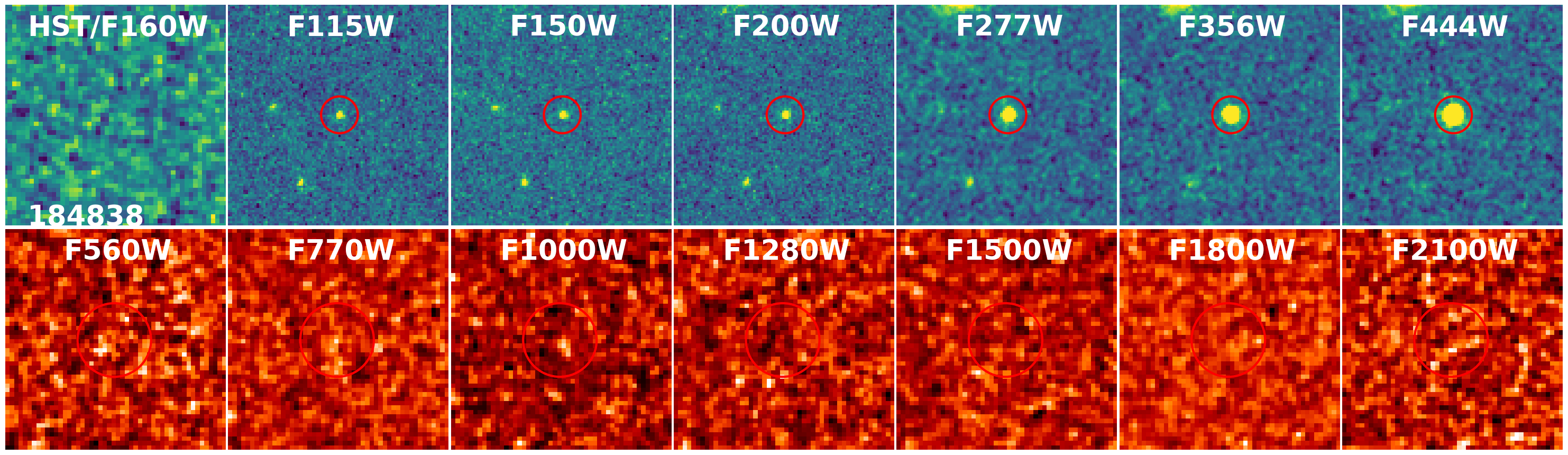}
\includegraphics[width=1\textwidth,trim=10 0 10 10, clip]{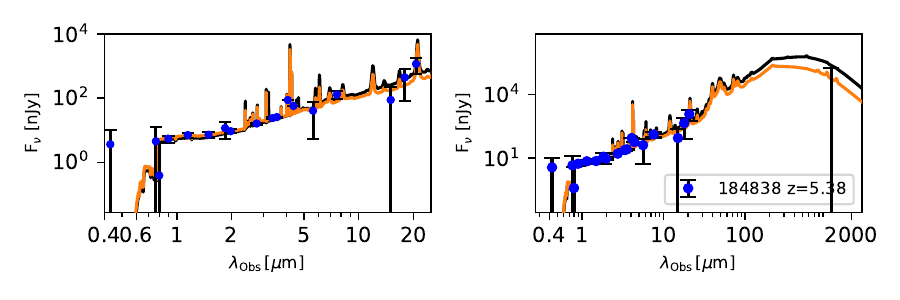}
\includegraphics[width=1\textwidth,trim=10 35 10 10, clip]{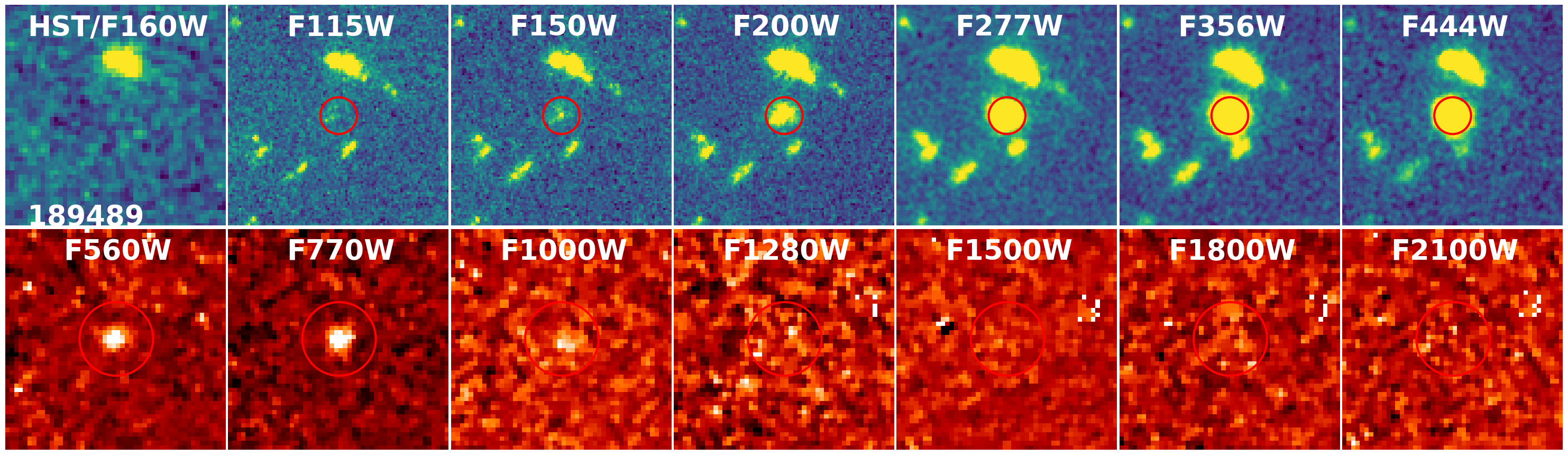}
\includegraphics[width=1\textwidth,trim=10 0 10 10, clip]{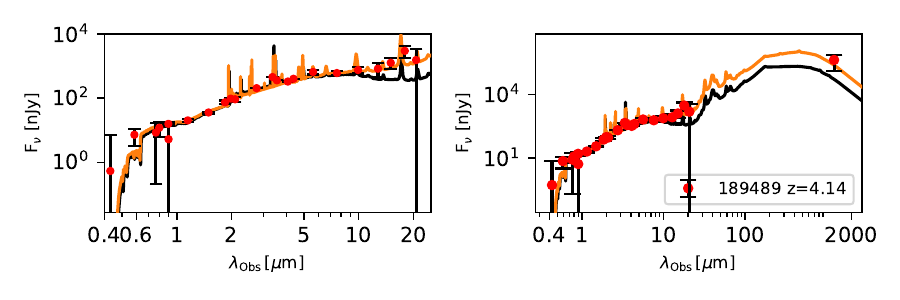}
\end{figure*}
\begin{figure*}[t]
\includegraphics[width=1\textwidth,trim=10 35 10 10, clip]{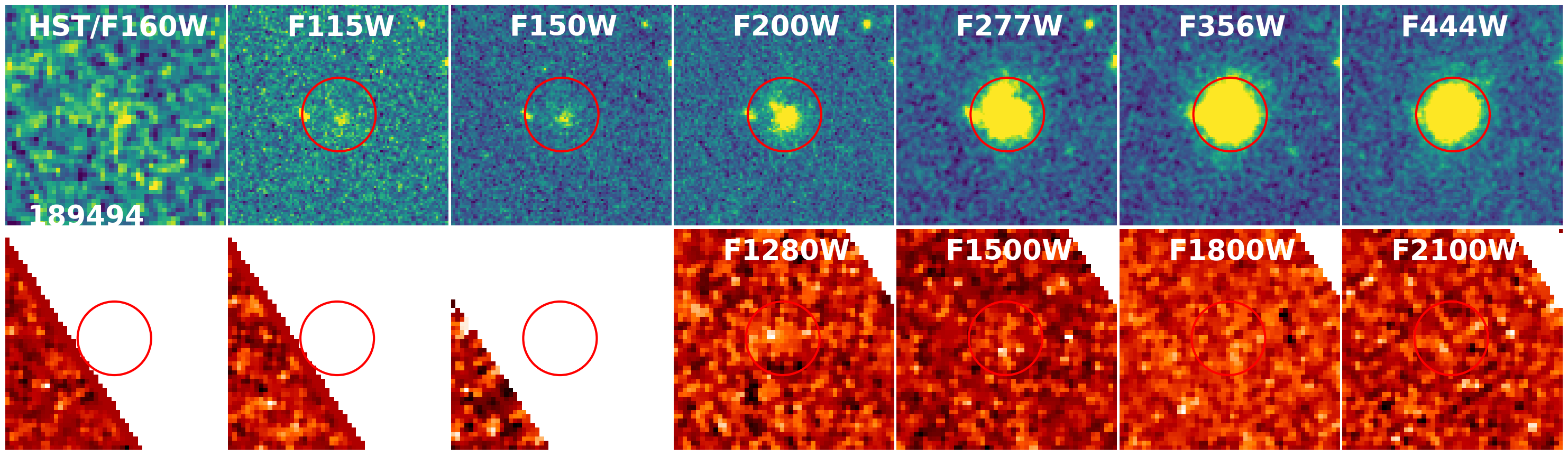}
\includegraphics[width=1\textwidth,trim=10 0 10 10, clip]{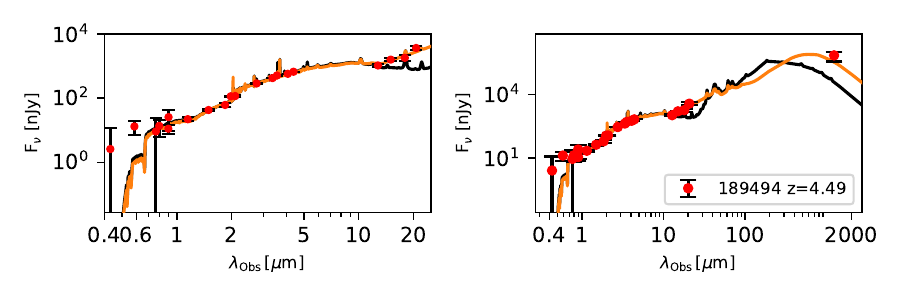}
\includegraphics[width=1\textwidth,trim=10 35 10 10, clip]{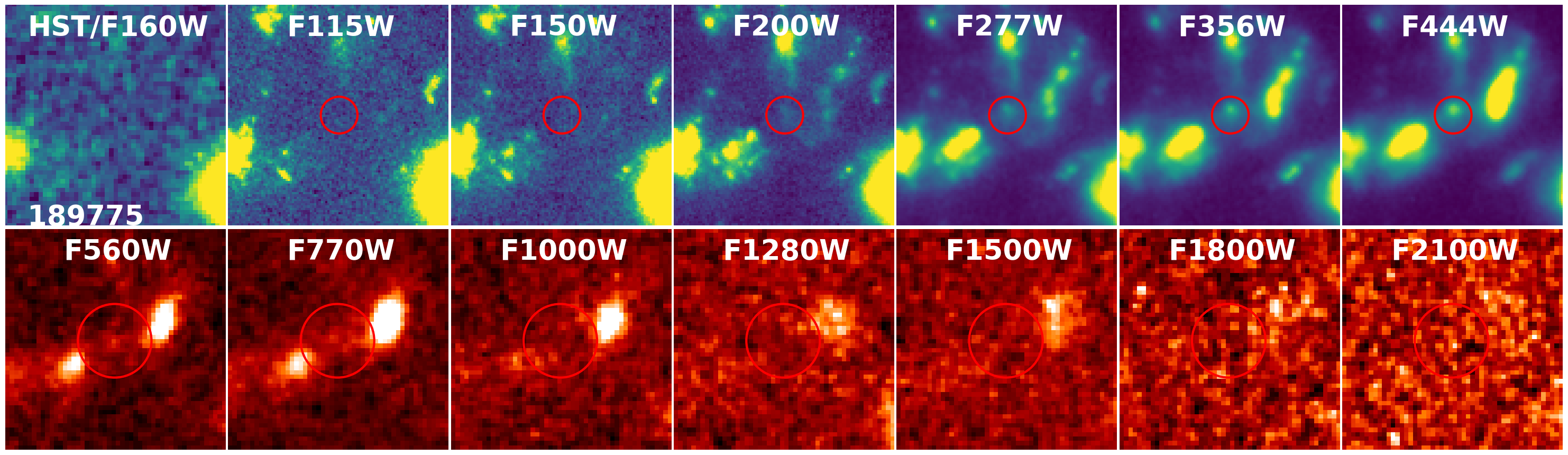}
\includegraphics[width=1\textwidth,trim=10 0 10 10, clip]{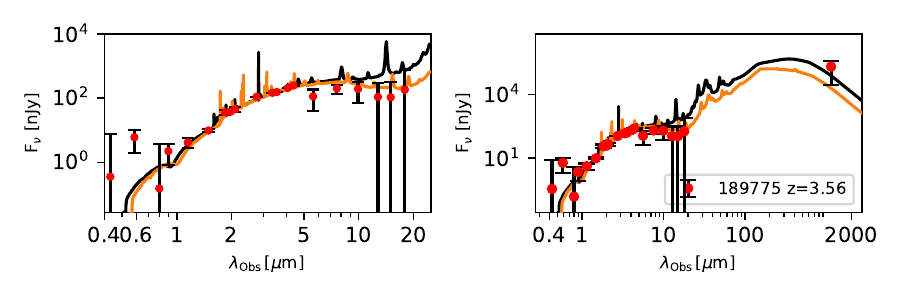}
\end{figure*}
\begin{figure*}[t]
\includegraphics[width=1\textwidth,trim=10 35 10 10, clip]{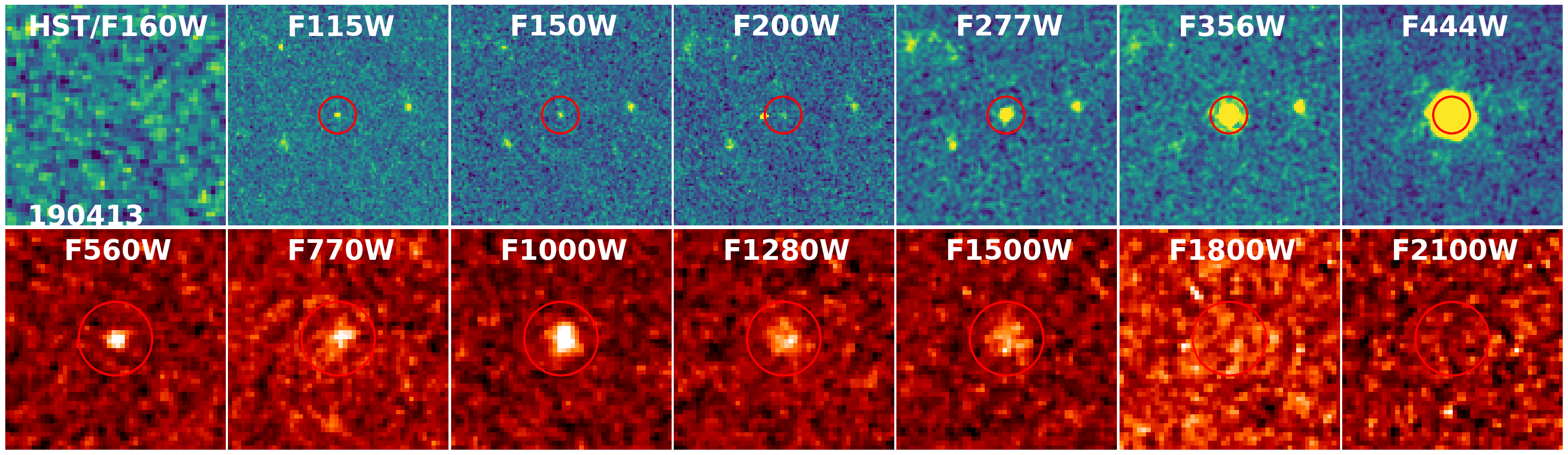}
\includegraphics[width=1\textwidth,trim=10 0 10 10, clip]{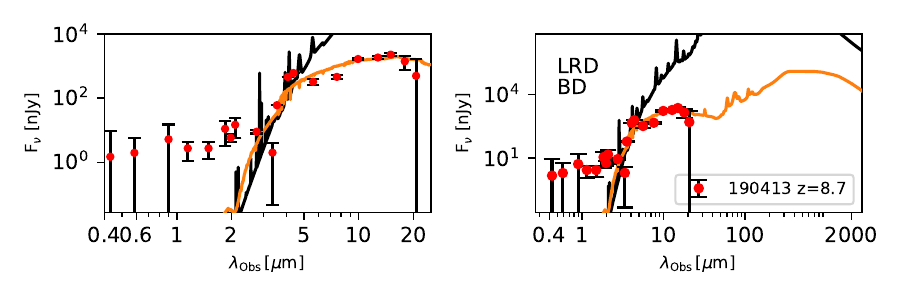}
\includegraphics[width=1\textwidth,trim=10 35 10 10, clip]{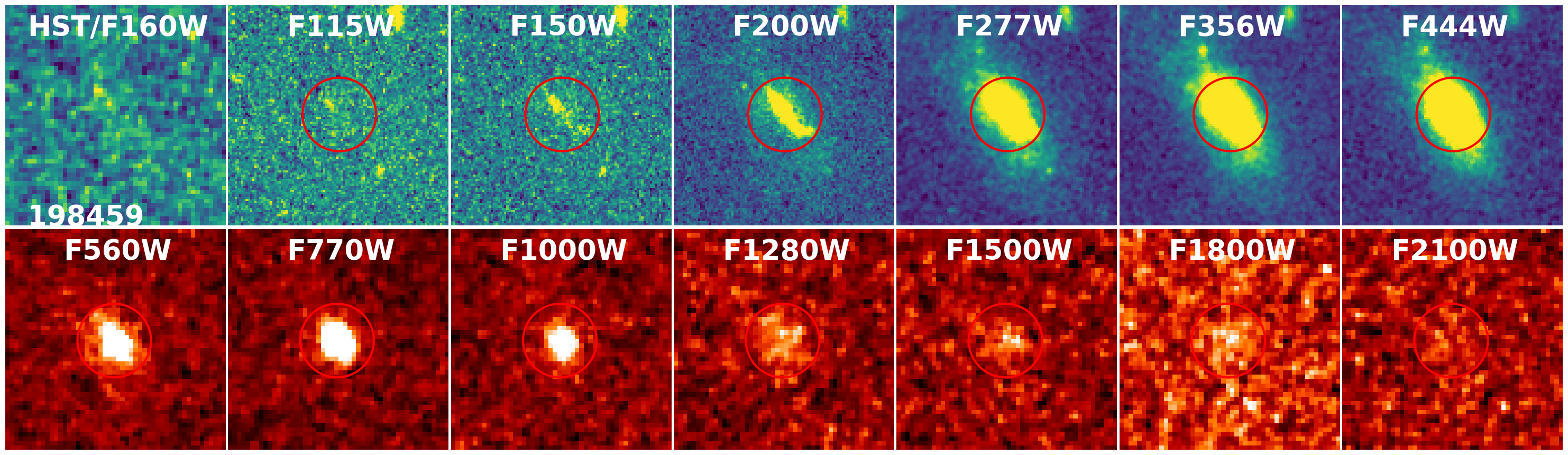}
\includegraphics[width=1\textwidth,trim=10 0 10 10, clip]{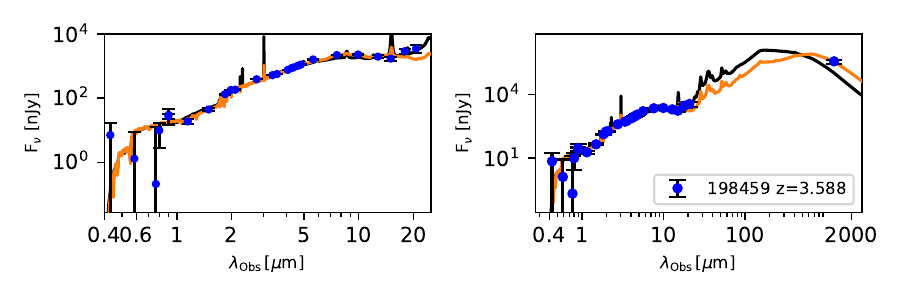}
\end{figure*}
\begin{figure*}[t]
\includegraphics[width=1\textwidth,trim=10 35 10 10, clip]{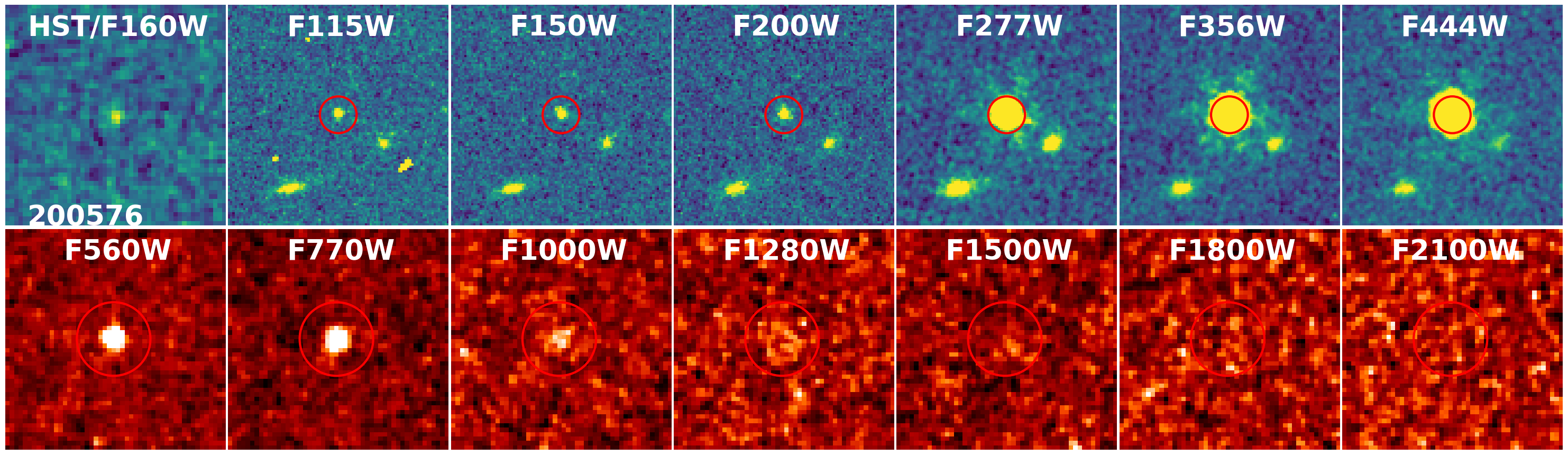}
\includegraphics[width=1\textwidth,trim=10 0 10 10, clip]{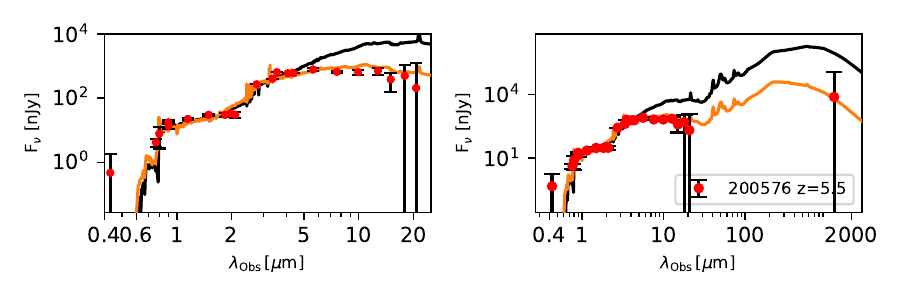}
\includegraphics[width=1\textwidth,trim=10 35 10 10, clip]{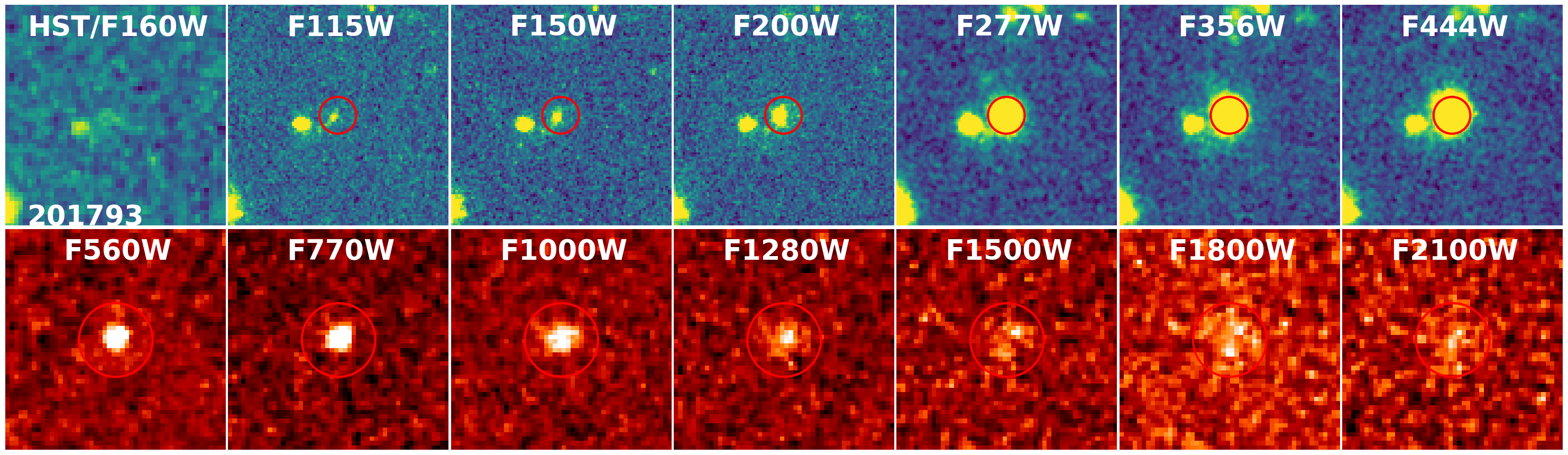}
\includegraphics[width=1\textwidth,trim=10 0 10 10, clip]{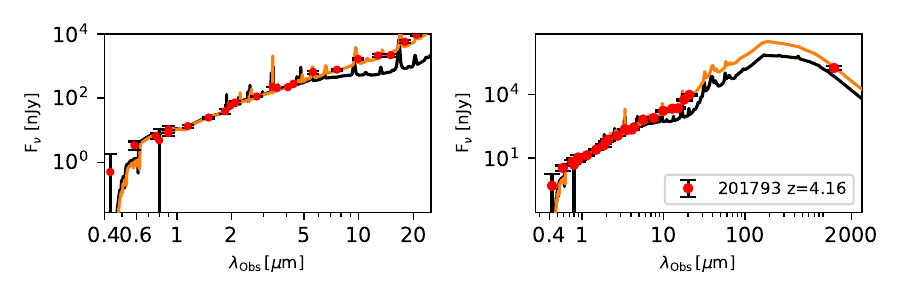}
\end{figure*}
\begin{figure*}[t]
\includegraphics[width=1\textwidth,trim=10 35 10 10, clip]{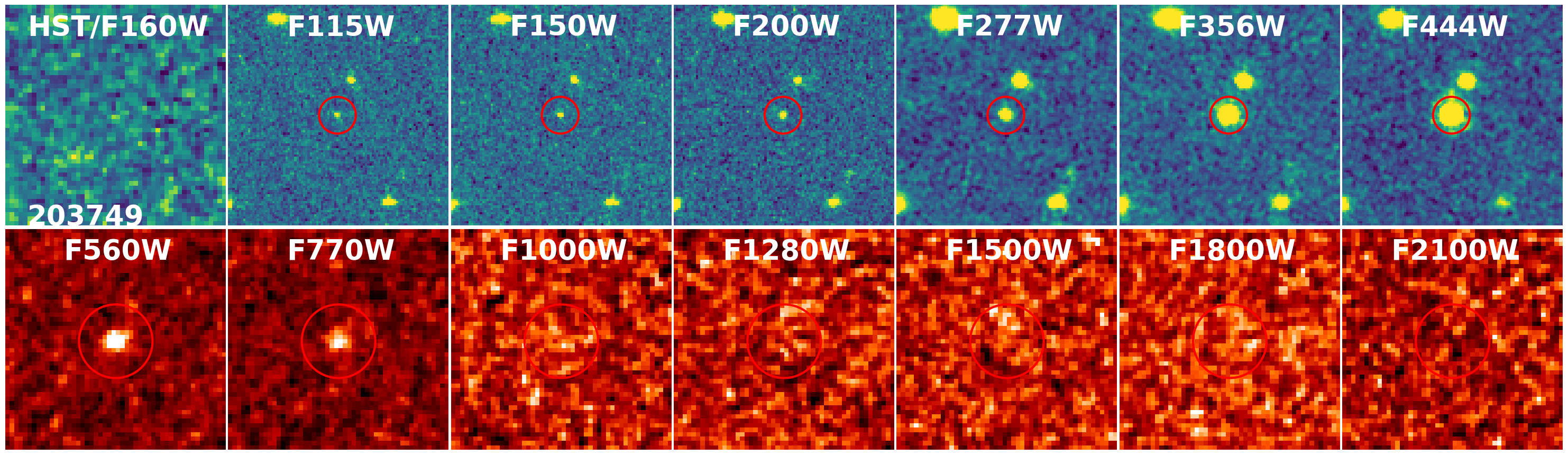}
\includegraphics[width=1\textwidth,trim=10 0 10 10, clip]{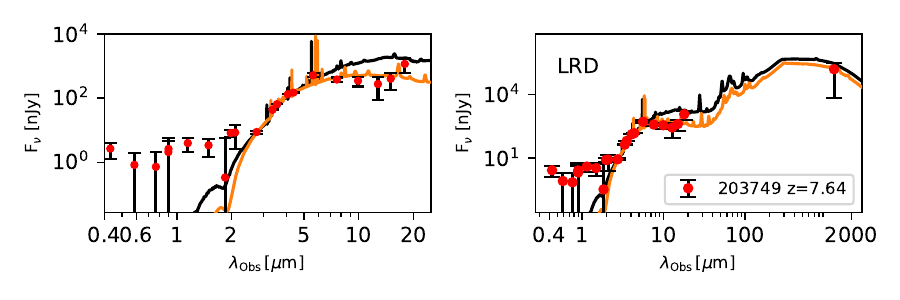}
\includegraphics[width=1\textwidth,trim=10 35 10 10, clip]{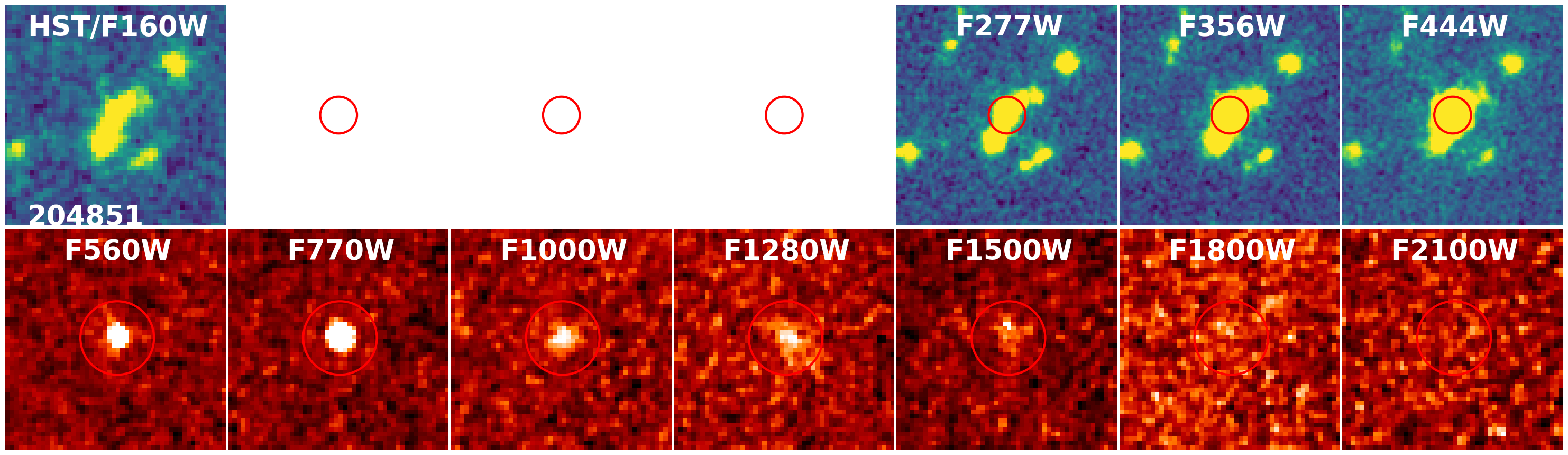}
\includegraphics[width=1\textwidth,trim=10 0 10 10, clip]{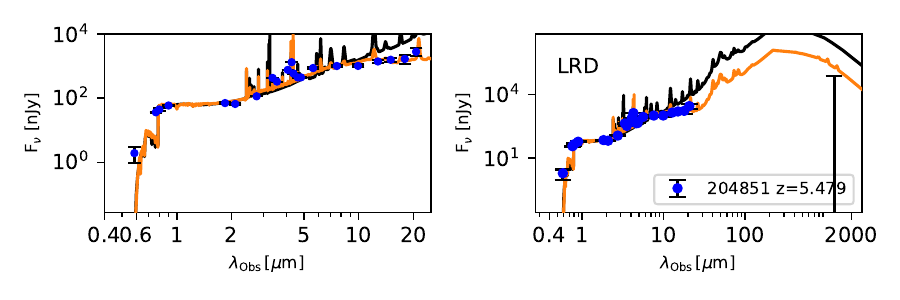}
\end{figure*}
\begin{figure*}[t]
\includegraphics[width=1\textwidth,trim=10 35 10 10, clip]{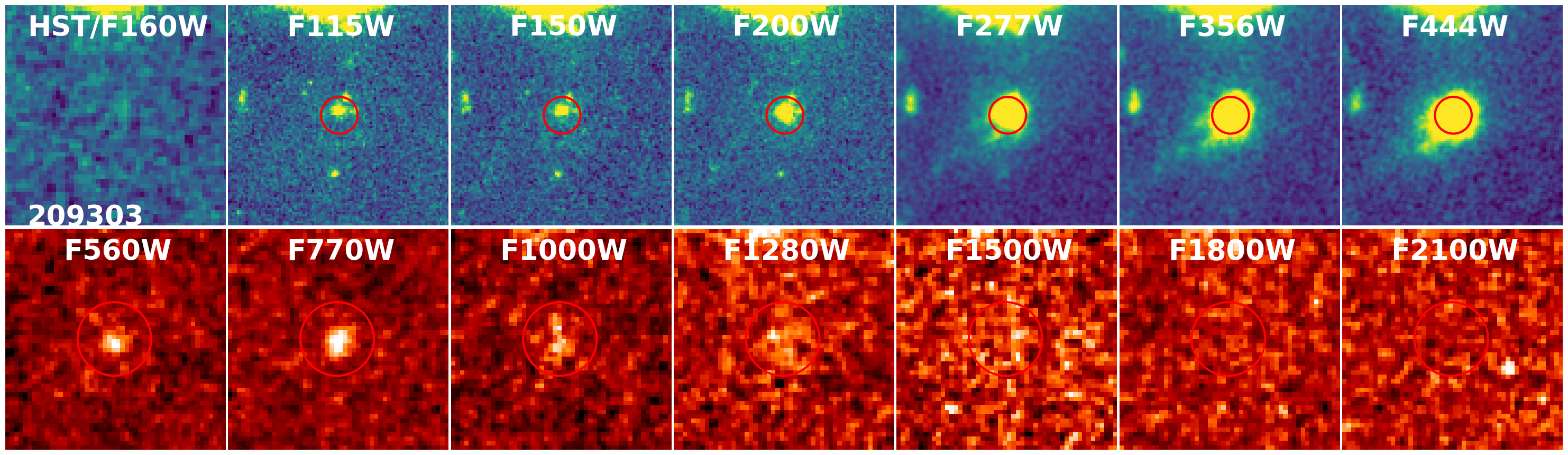}
\includegraphics[width=1\textwidth,trim=10 0 10 10, clip]{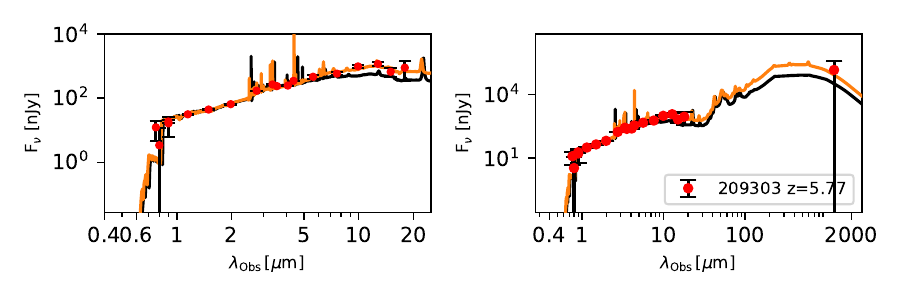}
\includegraphics[width=1\textwidth,trim=10 35 10 10, clip]{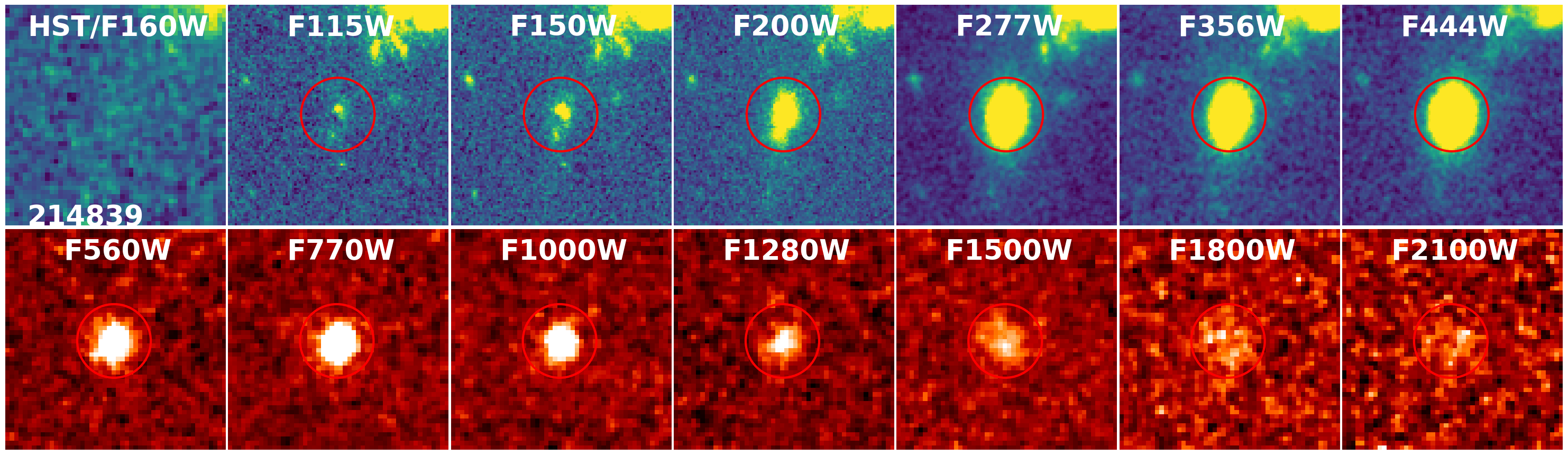}
\includegraphics[width=1\textwidth,trim=10 0 10 10, clip]{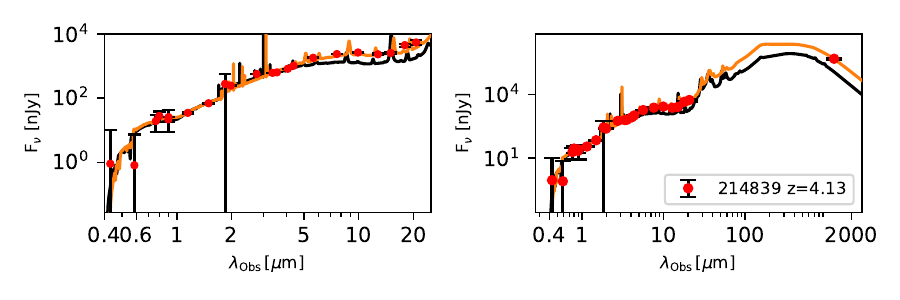}
\end{figure*}
\begin{figure*}[t]
\includegraphics[width=1\textwidth,trim=10 35 10 10, clip]{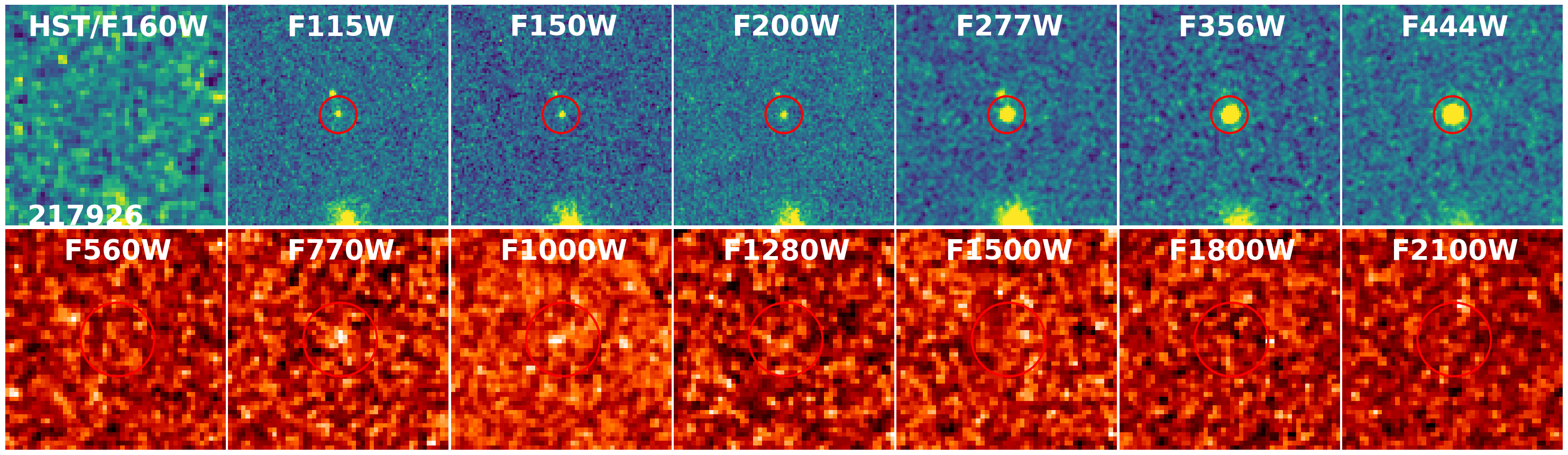}
\includegraphics[width=1\textwidth,trim=10 0 10 10, clip]{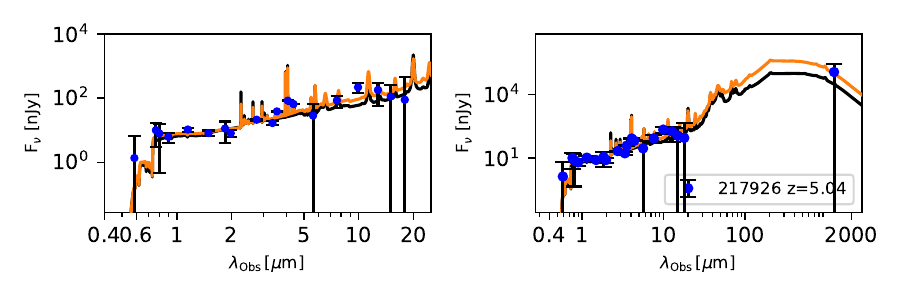}
\includegraphics[width=1\textwidth,trim=10 35 10 10, clip]{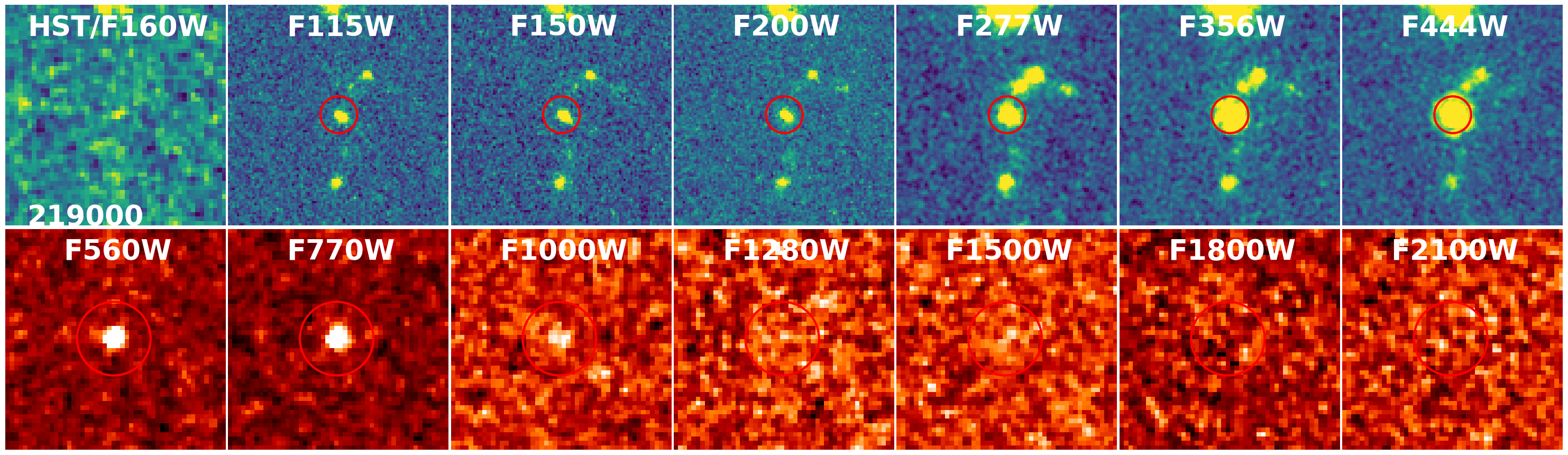}
\includegraphics[width=1\textwidth,trim=10 0 10 10, clip]{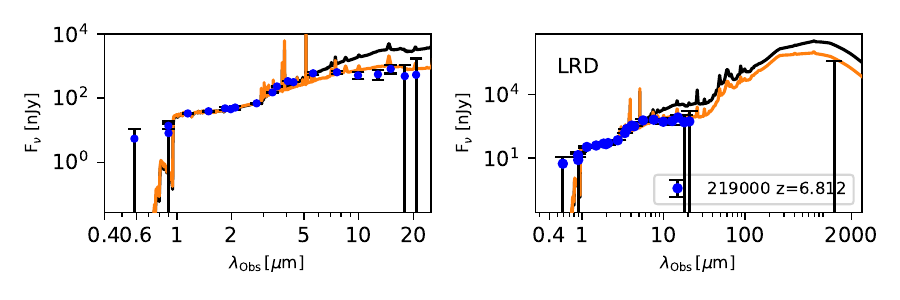}
\end{figure*}
\begin{figure*}[t]
\includegraphics[width=1\textwidth,trim=10 35 10 10, clip]{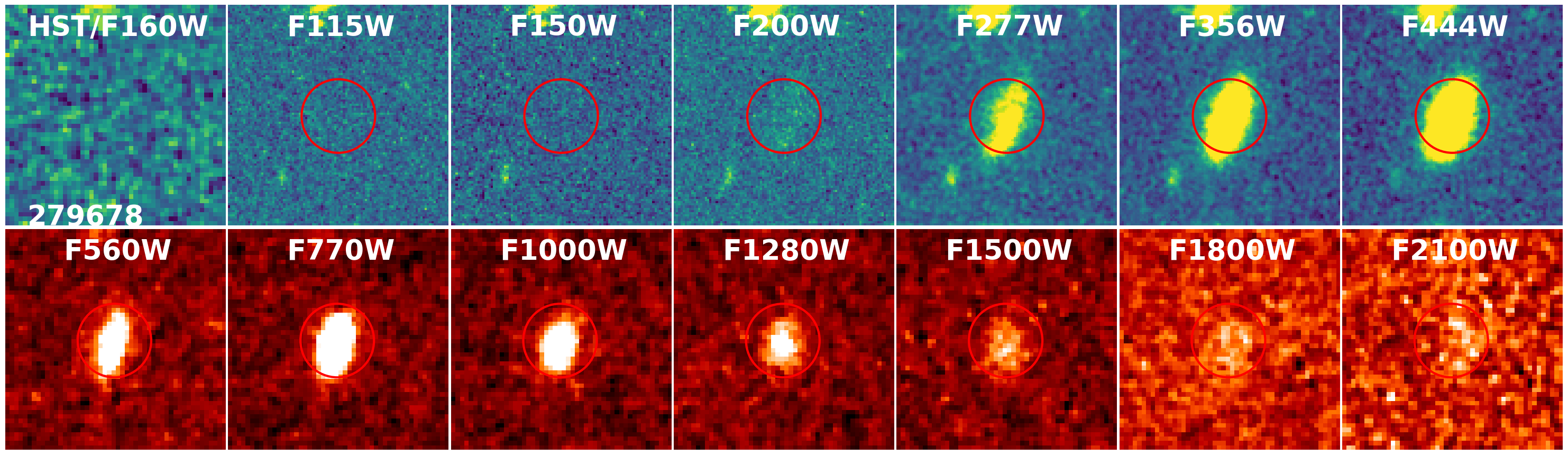}
\includegraphics[width=1\textwidth,trim=10 0 10 10, clip]{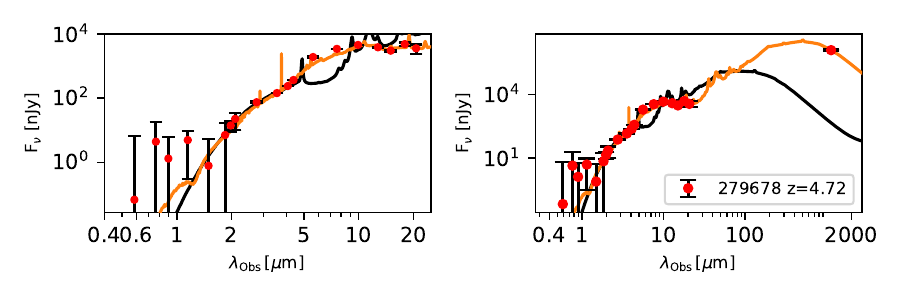}
\caption{Top panels: cutouts for sources that lie inside our MIRI footprint (NIRCam cutouts in blue-green, MIRI cutouts in red). Bottom panel shows photometry (points) and best SED model  fit to only HST+NIRCam (black) and best SED model  fit to HST+NIRCam+MIRI+ALMA (orange). Left is zoomed in on the restframe optical and near-infrared SEDs. Photometry for sources with spectroscopic redshfits are shown in blue (for spectroscopically confirmed objects). Sources which meet the ``Little Red Dot" (LRD) AGN selection are flagged. The brown dwarf candidate based on colors and proper motion is also flagged (BD). }\label{fig:miricutouts}
\end{figure*}

\begin{deluxetable*}{llllllllllll}[!htbp]
\caption{Properties of H-dropouts without MIRI data.   }\label{tab:NIRCAM}
\tablehead{\colhead{ID$^{a}$}   & \colhead{RA} &  \colhead{Dec }& \colhead{F444W} & \colhead{Photo-z$^{b}$} & \colhead{Spec-z}  & \colhead{Stellar Mass$^{c}$} & \colhead{SFR}$^{d}$ & \colhead{Av$^{e}$}& \colhead{Age$^{f}$} &\colhead{f$_{\rm AGN}$$^{g}$} & \colhead{LRD$^{h}$}  }
\startdata
5070 & 53.092006 & -27.903137 & 24.9 & 4.59 & \nodata & 10.0  $^{+ 0.0 }_{- 0.0 }$ & 1.43  $^{+ 0.2 }_{- 0.2 }$ & 0.02  $^{+ 0.03 }_{- 0.01 }$ & 0.96  $^{+ 0.07 }_{- 0.08 }$ & 0.0  $^{+ 0.0 }_{- 0.0 }$ & 0 \\ 
5756 & 53.062410 & -27.901838 & 25.6 & 3.53 & \nodata & 9.6  $^{+ 0.3 }_{- 0.5 }$ & 14.45  $^{+ 11.5 }_{- 4.7 }$ & 1.94  $^{+ 0.54 }_{- 0.51 }$ & 0.57  $^{+ 0.37 }_{- 0.35 }$ & 0.0  $^{+ 0.9 }_{- 0.0 }$ & 0 \\ 
11004 & 53.059444 & -27.893633 & 28.9 & 5.07 & \nodata & 8.7  $^{+ 0.9 }_{- 0.8 }$ & 2.26  $^{+ 12.4 }_{- 1.4 }$ & 3.18  $^{+ 1.45 }_{- 1.35 }$ & 0.12  $^{+ 0.09 }_{- 0.06 }$ & 0.0  $^{+ 0.6 }_{- 0.0 }$ & 0 \\ 
11786 & 53.119075 & -27.892556 & 26.3 & 7.28 & \nodata & 10.8  $^{+ 0.3 }_{- 0.3 }$ & 1242.03  $^{+ 876.4 }_{- 459.4 }$ & 4.30  $^{+ 0.57 }_{- 0.62 }$ & 0.10  $^{+ 0.11 }_{- 0.06 }$ & 0.0  $^{+ 0.2 }_{- 0.0 }$ & 1 \\ 
35203 & 53.140041 & -27.874603 & 27.2 & 4.20 & \nodata & 9.0  $^{+ 0.1 }_{- 0.1 }$ & 0.11  $^{+ 0.4 }_{- 0.1 }$ & 0.51  $^{+ 0.43 }_{- 0.30 }$ & 0.79  $^{+ 0.20 }_{- 0.26 }$ & 0.0  $^{+ 0.5 }_{- 0.0 }$ & 0 \\ 
35453 & 53.057030 & -27.874375 & 25.1 & 5.09 & \nodata & 10.2  $^{+ 0.1 }_{- 0.1 }$ & 1.83  $^{+ 1.7 }_{- 1.0 }$ & 1.01  $^{+ 0.27 }_{- 0.26 }$ & 0.65  $^{+ 0.11 }_{- 0.13 }$ & 0.0  $^{+ 0.3 }_{- 0.0 }$ & 0 \\ 
39005 & 53.080165 & -27.871306 & 28.5 & 0.88 & \nodata & 7.2  $^{+ 0.2 }_{- 0.2 }$ & 0.01  $^{+ 0.0 }_{- 0.0 }$ & 6.82  $^{+ 2.03 }_{- 1.81 }$ & 0.17  $^{+ 0.04 }_{- 0.05 }$ & 0.0  $^{+ 0.0 }_{- 0.0 }$ & 0 \\ 
39376 & 53.064113 & -27.870933 & 25.8 & 7.36 & \nodata & 11.0  $^{+ 0.3 }_{- 0.2 }$ & 1665.58  $^{+ 1148.3 }_{- 565.0 }$ & 4.24  $^{+ 0.66 }_{- 0.55 }$ & 0.12  $^{+ 0.16 }_{- 0.07 }$ & 0.0  $^{+ 0.1 }_{- 0.0 }$ & 1 \\ 
40983 & 53.046050 & -27.869633 & 28.8 & 3.52 & \nodata & 8.0  $^{+ 0.1 }_{- 0.1 }$ & 0.01  $^{+ 0.0 }_{- 0.0 }$ & 0.27  $^{+ 0.31 }_{- 0.17 }$ & 0.24  $^{+ 0.03 }_{- 0.03 }$ & 0.0  $^{+ 0.2 }_{- 0.0 }$ & 0 \\ 
45564 & 53.055347 & -27.866132 & 25.1 & 3.56 & \nodata & 9.7  $^{+ 0.2 }_{- 0.2 }$ & 11.96  $^{+ 6.9 }_{- 3.7 }$ & 1.56  $^{+ 0.41 }_{- 0.59 }$ & 0.56  $^{+ 0.24 }_{- 0.26 }$ & 0.0  $^{+ 0.3 }_{- 0.0 }$ & 0 \\ 
65559 & 53.041051 & -27.854478 & 25.2 & 4.46 & \nodata & 9.8  $^{+ 0.0 }_{- 0.0 }$ & 1.28  $^{+ 0.1 }_{- 0.1 }$ & 0.00  $^{+ 0.00 }_{- 0.00 }$ & 1.27  $^{+ 0.03 }_{- 0.07 }$ & 0.0  $^{+ 0.0 }_{- 0.0 }$ & 0 \\ 
75446 & 53.036392 & -27.846575 & 25.3 & 6.36 & \nodata & 10.2  $^{+ 0.2 }_{- 0.0 }$ & 1.89  $^{+ 27.8 }_{- 0.8 }$ & 0.16  $^{+ 1.13 }_{- 0.14 }$ & 0.62  $^{+ 0.07 }_{- 0.13 }$ & 0.0  $^{+ 0.2 }_{- 0.0 }$ & 0 \\ 
79086 & 53.044180 & -27.842933 & 27.2 & 4.68 & \nodata & 8.9  $^{+ 0.1 }_{- 0.3 }$ & 0.15  $^{+ 1.1 }_{- 0.1 }$ & 0.21  $^{+ 0.34 }_{- 0.15 }$ & 0.55  $^{+ 0.19 }_{- 0.27 }$ & 0.0  $^{+ 0.2 }_{- 0.0 }$ & 0 \\ 
82667 & 53.067296 & -27.838335 & 26.2 & 6.29 & \nodata & 10.5  $^{+ 0.4 }_{- 0.4 }$ & 333.17  $^{+ 415.8 }_{- 138.5 }$ & 3.82  $^{+ 0.95 }_{- 0.86 }$ & 0.21  $^{+ 0.14 }_{- 0.12 }$ & 0.0  $^{+ 0.1 }_{- 0.0 }$ & 1 \\ 
152330 & 53.102742 & -27.744467 & 25.8 & 5.07 & \nodata & 9.9  $^{+ 0.2 }_{- 0.2 }$ & 189.12  $^{+ 33.8 }_{- 29.0 }$ & 3.09  $^{+ 0.22 }_{- 0.29 }$ & 0.16  $^{+ 0.15 }_{- 0.12 }$ & 0.0  $^{+ 0.2 }_{- 0.0 }$ & 0 \\ 
161143 & 53.024862 & -27.894925 & 27.3 & 3.39 & \nodata & 8.7  $^{+ 0.1 }_{- 0.1 }$ & 0.53  $^{+ 0.2 }_{- 0.2 }$ & 0.77  $^{+ 0.21 }_{- 0.29 }$ & 0.75  $^{+ 0.23 }_{- 0.27 }$ & 0.0  $^{+ 0.2 }_{- 0.0 }$ & 0 \\ 
161967 & 53.036504 & -27.894148 & 25.6 & 3.00 & \nodata & 9.7  $^{+ 0.1 }_{- 0.1 }$ & 67.87  $^{+ 13.8 }_{- 12.6 }$ & 3.47  $^{+ 0.35 }_{- 0.31 }$ & 0.28  $^{+ 0.19 }_{- 0.16 }$ & 0.0  $^{+ 0.0 }_{- 0.0 }$ & 0 \\ 
168364 & 53.144844 & -27.879612 & 27.5 & 5.93 & \nodata & 9.5  $^{+ 0.4 }_{- 0.4 }$ & 27.47  $^{+ 62.6 }_{- 12.4 }$ & 2.67  $^{+ 0.95 }_{- 0.82 }$ & 0.20  $^{+ 0.16 }_{- 0.12 }$ & 0.0  $^{+ 0.4 }_{- 0.0 }$ & 0 \\ 
170384 & 53.074866 & -27.875907 & 24.2 & 5.62 & \nodata & 11.0  $^{+ 0.2 }_{- 0.2 }$ & 581.71  $^{+ 159.6 }_{- 135.7 }$ & 2.61  $^{+ 0.32 }_{- 0.26 }$ & 0.20  $^{+ 0.12 }_{- 0.10 }$ & 0.0  $^{+ 0.4 }_{- 0.0 }$ & 0 \\ 
171209 & 53.043730 & -27.874185 & 24.8 & 3.75 & \nodata & 9.9  $^{+ 0.2 }_{- 0.1 }$ & 6.32  $^{+ 9.5 }_{- 3.6 }$ & 1.32  $^{+ 0.48 }_{- 0.34 }$ & 0.59  $^{+ 0.21 }_{- 0.21 }$ & 0.1  $^{+ 1.2 }_{- 0.0 }$ & 0 \\ 
171973 & 53.086837 & -27.873047 & 23.0 & 3.42 & \nodata & 11.5  $^{+ 0.1 }_{- 0.1 }$ & 3843.17  $^{+ 1012.1 }_{- 687.9 }$ & 5.36  $^{+ 0.19 }_{- 0.23 }$ & 0.25  $^{+ 0.18 }_{- 0.13 }$ & 0.0  $^{+ 0.1 }_{- 0.0 }$ & 0 \\ 
172813 & 53.047201 & -27.870031 & 23.4 & 4.20 & \nodata & 11.1  $^{+ 0.2 }_{- 0.2 }$ & 704.25  $^{+ 213.4 }_{- 155.5 }$ & 3.88  $^{+ 0.42 }_{- 0.28 }$ & 0.13  $^{+ 0.05 }_{- 0.05 }$ & 0.0  $^{+ 0.0 }_{- 0.0 }$ & 0 \\ 
173706 & 53.080377 & -27.869467 & 24.1 & 3.55 & \nodata & 10.4  $^{+ 0.2 }_{- 0.3 }$ & 231.42  $^{+ 59.3 }_{- 47.4 }$ & 3.10  $^{+ 0.24 }_{- 0.26 }$ & 0.44  $^{+ 0.22 }_{- 0.23 }$ & 0.0  $^{+ 1.1 }_{- 0.0 }$ & 0 \\ 
177680 & 53.064791 & -27.862624 & 23.3 & 1.33 & \nodata & 10.0  $^{+ 0.9 }_{- 0.3 }$ & 95.58  $^{+ 1517.9 }_{- 21.0 }$ & 5.27  $^{+ 0.55 }_{- 1.05 }$ & 0.11  $^{+ 0.13 }_{- 0.05 }$ & 0.0  $^{+ 0.0 }_{- 0.0 }$ & 0 \\ 
189276 & 53.042071 & -27.842660 & 24.8 & 5.89 & \nodata & 10.3  $^{+ 0.5 }_{- 0.3 }$ & 58.38  $^{+ 73.8 }_{- 23.0 }$ & 1.12  $^{+ 1.90 }_{- 0.43 }$ & 0.22  $^{+ 0.26 }_{- 0.11 }$ & 0.0  $^{+ 0.8 }_{- 0.0 }$ & 0 \\ 
189494 & 53.162966 & -27.841946 & 24.5 & 4.50 & \nodata & 10.4  $^{+ 0.1 }_{- 0.3 }$ & 19.31  $^{+ 10.3 }_{- 6.0 }$ & 1.49  $^{+ 0.24 }_{- 0.27 }$ & 0.65  $^{+ 0.28 }_{- 0.28 }$ & 0.2  $^{+ 0.4 }_{- 0.1 }$ & 0 \\ 
189508 & 53.042426 & -27.841815 & 23.8 & 4.80 & \nodata & 11.7  $^{+ 0.1 }_{- 0.2 }$ & 1417.83  $^{+ 646.2 }_{- 688.5 }$ & 4.46  $^{+ 0.47 }_{- 0.84 }$ & 0.66  $^{+ 0.17 }_{- 0.15 }$ & 0.0  $^{+ 0.1 }_{- 0.0 }$ & 0 \\ 
212950 & 53.132665 & -27.765488 & 24.0 & 4.32 & \nodata & 10.4  $^{+ 0.0 }_{- 0.0 }$ & 0.56  $^{+ 0.1 }_{- 0.1 }$ & 0.04  $^{+ 0.06 }_{- 0.03 }$ & 1.01  $^{+ 0.05 }_{- 0.05 }$ & 0.0  $^{+ 0.1 }_{- 0.0 }$ & 0 \\ 
236329 & 53.089004 & -27.738202 & 24.8 & 5.69 & \nodata & 11.1  $^{+ 0.1 }_{- 0.1 }$ & 41.50  $^{+ 23.1 }_{- 15.5 }$ & 2.56  $^{+ 0.25 }_{- 0.24 }$ & 0.53  $^{+ 0.07 }_{- 0.08 }$ & 0.0  $^{+ 0.0 }_{- 0.0 }$ & 0 \\ 
237934 & 53.055599 & -27.725428 & 27.9 & 5.87 & \nodata & 8.4  $^{+ 0.3 }_{- 0.5 }$ & 2.91  $^{+ 2.2 }_{- 1.1 }$ & 1.13  $^{+ 1.43 }_{- 0.42 }$ & 0.07  $^{+ 0.03 }_{- 0.03 }$ & 0.0  $^{+ 0.2 }_{- 0.0 }$ & 0 \\ 
242342 & 53.054126 & -27.694044 & 25.5 & 2.65 & \nodata & 9.4  $^{+ 0.1 }_{- 0.1 }$ & 88.36  $^{+ 19.1 }_{- 11.6 }$ & 3.14  $^{+ 0.27 }_{- 0.21 }$ & 0.03  $^{+ 0.10 }_{- 0.02 }$ & 0.0  $^{+ 0.0 }_{- 0.0 }$ & 0 \\ 
247084 & 53.060897 & -27.718430 & 24.1 & 2.74 & \nodata & 9.6  $^{+ 0.2 }_{- 1.4 }$ & 28.05  $^{+ 12.7 }_{- 18.9 }$ & 2.56  $^{+ 4.13 }_{- 0.30 }$ & 0.12  $^{+ 0.08 }_{- 0.04 }$ & 0.8  $^{+ 2.3 }_{- 0.0 }$ & 0 \\ 
283711 & 53.082649 & -27.864812 & 28.8 & 3.66 & \nodata & 7.9  $^{+ 0.2 }_{- 0.3 }$ & 0.74  $^{+ 0.5 }_{- 0.3 }$ & 1.21  $^{+ 0.53 }_{- 0.62 }$ & 0.09  $^{+ 0.06 }_{- 0.04 }$ & 0.0  $^{+ 0.2 }_{- 0.0 }$ & 0 \\ 
284487 & 53.078688 & -27.839396 & 26.8 & 3.71 & \nodata & 9.1  $^{+ 0.1 }_{- 0.1 }$ & 0.07  $^{+ 0.4 }_{- 0.0 }$ & 0.27  $^{+ 0.42 }_{- 0.19 }$ & 0.86  $^{+ 0.21 }_{- 0.28 }$ & 0.0  $^{+ 0.4 }_{- 0.0 }$ & 0 \\ 
284527 & 53.041065 & -27.837737 & 23.3 & 0.45 & \nodata & 8.0  $^{+ 0.4 }_{- 0.2 }$ & 0.13  $^{+ 0.1 }_{- 0.1 }$ & 7.50  $^{+ 0.70 }_{- 1.34 }$ & 0.23  $^{+ 0.07 }_{- 0.05 }$ & 0.4  $^{+ 2.1 }_{- 0.2 }$ & 0 \\ 
284756 & 53.073995 & -27.828662 & 24.4 & 3.57 & \nodata & 10.1  $^{+ 0.2 }_{- 0.2 }$ & 58.65  $^{+ 21.9 }_{- 18.9 }$ & 2.14  $^{+ 0.31 }_{- 0.39 }$ & 0.46  $^{+ 0.32 }_{- 0.25 }$ & 0.0  $^{+ 0.4 }_{- 0.0 }$ & 0 \\ 
286677 & 53.155126 & -27.727785 & 24.9 & 3.57 & \nodata & 9.7  $^{+ 0.2 }_{- 0.3 }$ & 82.16  $^{+ 19.8 }_{- 22.7 }$ & 2.39  $^{+ 0.25 }_{- 0.28 }$ & 0.17  $^{+ 0.19 }_{- 0.11 }$ & 0.0  $^{+ 0.8 }_{- 0.0 }$ & 0 \\ 
\enddata
\tablecomments{Properties of our H-dropout sample, measured including MIRI+ALMA data: (a) JADES DR1 ID (b) Photometric redshift measured by {\tt prospector}, or from EAZY in the case of a spectroscopic redshift  (c) Log$_{10}$ of stellar mass  in units M$_{\odot}$ (d) SFR measured by {\tt prospector} using the most recent 30 Myr time bin of the SFH in units M$_{\odot}$/yr (e) V-band Attenuation (f) Mass weighted age in units Gyr (g) the ratio of bolometric luminosity from the galaxy divided by that from the AGN (h) flag (1) indicating the source meets our LRD color selection. }
\end{deluxetable*}

\begin{acknowledgements}
We thank Carlos Gomez-Guijarro and David Elbaz for sharing the GOODS-ALMA imaging. We thank Ivo Labbe and Jenny Greene for sharing their insight and for immensely helpful discussions. The authors acknowledge the FRESCO team led by PI Pascal Oesch for developing their observing program with a zero-exclusive-access period. This work is based in part on observations made with the NASA/ESA/CSA James Webb Space Telescope.  The data were obtained from the Mikulski Archive for Space Telescopes at the Space Telescope Science Institute, which is operated
by the Association of Universities for Research in Astronomy, Inc., under NASA contract NAS 5-03127 for
JWST. These observations are associated with JWST
Cycle 1 GO program \#1180, 1181, 1207, 1210, 1286, 1895, 1963. Data used in this work are available via MAST at \dataset[DOI:10.17909/fsc4-dt61]{https://doi.org/10.17909/fsc4-dt61}, \dataset[DOI: 10.17909/z2gw-mk31]{https://doi.org/10.17909/z2gw-mk31} and \dataset[DOI: 10.17909/T91019]{https://doi.org/10.17909/T91019}. Support for program JWST-GO-1963 was provided by NASA through a grant from the Space Telescope Science Institute, which is operated by the Association of Universities for Research in Astronomy, Inc., under NASA contract NAS 5-03127. 

 The work of CCW is supported by NOIRLab, which is managed by the Association of Universities for Research in Astronomy (AURA) under a cooperative agreement with the National Science Foundation. SA acknowledges support from the JWST Mid-Infrared Instrument (MIRI) Science Team Lead, grant 80NSSC18K0555, from NASA Goddard Space Flight Center to the University of Arizona. MR, EE, DJE, BDJ, BR, GR, FS, and CNAW acknowledge support from the NIRCam Science Team contract to the University of Arizona, NAS5-02015.  DJE is further supported as a Simons Investigator. AJB has received funding from the European Research Council (ERC) under the European Union’s Horizon 2020 Advanced Grant 789056 “First Galaxies”.  ECL acknowledges support of an STFC Webb Fellowship (ST/W001438/1).  RM, WB acknowledge support by the Science and Technology Facilities Council (STFC) and by the ERC through Advanced Grant 695671 "QUENCH". This material is based upon High Performance Computing (HPC) resources supported by the University of Arizona TRIF, UITS, and Research, Innovation, and Impact (RII) and maintained by the UArizona Research Technologies department. 

\end{acknowledgements}
\software{astropy \citep{Astropy2013,Astropy2022},  
          Cloudy \citep{Ferland2013}, 
          photutils \citep{photutils}
          WebbPSF \citep{Perrin2015ascl.soft04007P}
          }

\bibliography{manu}{}
\bibliographystyle{aasjournal}

\end{document}